\documentclass[11pt,a4paper]{article}
\pdfoutput=1
\usepackage{jheppub}
\usepackage{mathptmx} 
\usepackage{bbm,latexsym,url,graphicx}
\usepackage{amssymb,amsmath}
\usepackage[utf8]{inputenc}
\usepackage{amsfonts}
\usepackage{amsthm,array,booktabs,dsfont,mathtools,multirow,placeins,rotating,stmaryrd,tensor,verbatim,subfig}
\usepackage[all,color]{xy}

\newcommand{\xtexttt}[1]{}

\newcommand{\sst}[1]{{\scriptscriptstyle #1}}

\def\0{{\sst{(0)}}}
\def\1{{\sst{(1)}}}
\def\2{{\sst{(2)}}}
\def\3{{\sst{(3)}}}
\def\4{{\sst{(4)}}}
\def\5{{\sst{(5)}}}
\def\6{{\sst{(6)}}}
\def\7{{\sst{(7)}}}

\newcommand{\be}{\begin{equation}}
\newcommand{\ee}{\end{equation}}
\def\ba{\begin{array}}
\def\ea{\end{array}}

\newtheorem*{theorem*}{Theorem}

\numberwithin{equation}{section}

\newcommand{\bea}{\begin{eqnarray}}
\newcommand{\eea}{\end{eqnarray}}

\DeclareMathOperator{\Aut}{Aut}

\DeclareMathOperator{\Tr}{Tr} 

\DeclareMathOperator{\Iso}{Iso}
\DeclareMathOperator{\Hom}{Hom}
\DeclareMathOperator{\SO}{SO}

\DeclareMathOperator{\Span}{span}
\DeclareMathOperator{\sgn}{sgn}

\newcommand{\J}{\mathfrak{J}}

\newcommand{\F}{\mathds{F}}
\newcommand{\R}{\mathds{R}}
\newcommand{\C}{\mathds{C}}

\newcommand{\Z}{\mathds{Z}}

\newcommand{\N}{\mathcal{N}}

\newcommand{\FTS}{\mbox{$\mathfrak{F}$}}

\newcommand{\autFF}{\mbox{$\mathfrak{aut}(\mathfrak{F})$}}
\newcommand{\fx}{\mbox{$\FTS_x$}}
\newcommand{\fxp}{\mbox{$\FTS_{x}^{\perp}$}}
\newcommand{\tx}{\mbox{$\tilde{x}$}}
\newcommand{\FT}{\mbox{$\mathfrak{T}$}}

\newcommand{\txx}{\mbox{$\tilde{x}$}}
\newcommand{\xdot}{\mbox{$\dot{x}$}}

\newcommand{\hS}{\mbox{$ \hat{S} $}}
\newcommand{\sdx}{\mbox{$\sqrt{\mid\Delta(x)\mid\ }$}}
\newcommand{\sdxx}[1]{\mbox{$\sqrt{\mid\Delta(x_{#1})\mid\ }$}}
\newcommand{\spdx}{\mbox{$\sqrt{\Delta(x)}\ $}}

\newcommand{\ooUpsilon}{ \mbox{$\overline{\overline{\Upsilon}}$}}

\newcommand{\ups}[1]{ \mbox{$\Upsilon_x(#1)$}}
\newcommand{\fxy}{ \mbox{$\FTS_{y\perp x}$}}
\newcommand{\fxt}{ \mbox{$\FTS_{x}^\perp$}}
\newcommand{\ff}{ \mbox{$F\ $}}

\newcommand{\beq}{\begin{eqnarray}}
\newcommand{\eeq}{\end{eqnarray}}

\newcommand{\M}{\mathcal{M}}

\newcommand{\xdodo}{\mbox{$x_{ D0D6}$}}

\newcommand{\xdodc}{\mbox{$x_{ D0D4}$}}

\renewcommand{\blacksquare}{ \qed}

\newcommand{\pmtwo}[4]{\begin{pmatrix}#1 & #2 \\ #3 & #4 \end{pmatrix}}

\begin{document}
\title{Black holes and general Freudenthal transformations}

\author[a]{ L.~Borsten,}
\author[b,c,d]{ M.~J.~Duff,}
\author[e]{ J.~J.~Fern\'andez-Melgarejo,}
\author[f,g]{  A.~Marrani,}
\author[e]{ E.~Torrente-Lujan}

\affiliation[a]{School of Theoretical Physics, Dublin Institute for Advanced Studies, \\ 10 Burlington Road, Dublin 4, Ireland}
\affiliation[b]{Theoretical Physics, Blackett Laboratory, Imperial College London, London SW7 2AZ, United Kingdom}
\affiliation[c]{Mathematical Institute, University of Oxford, Andrew Wiles Building, \\Woodstock Road, Radcliffe Observatory Quarter,
Oxford, OX2 6GG, United Kingdom}
\affiliation[d]{Institute for Quantum Science and Engineering and Hagler Institute for Advanced Study, Texas A\&M University, College Station, TX, 77840, USA}
\affiliation[e]{Fisica Teorica, Dep. de F\'{\i}sica, Universidad de Murcia, Campus de Espinardo, E-30100 Murcia, Spain}
\affiliation[f]{Museo Storico della Fisica e Centro Studi e Ricerche “Enrico Fermi”, \\via Panisperna 89A, I-00184, Roma, Italy}
\affiliation[g]{Dipartimento di Fisica e Astronomia “Galileo Galilei”, Università di Padova, and INFN, sezione di Padova, via Marzolo 8, I-35131 Padova, Italy}

\emailAdd{leron@stp.dias.ie}
\emailAdd{m.duff@imperial.ac.uk}
\emailAdd{jj.fernandezmelgarejo@um.es}
\emailAdd{alessio.marrani@pd.infn.it}
\emailAdd{torrente@cern.ch}
\preprint{
\vspace{-0.8cm}
{\flushright{DIAS-STP-19-03,\\
 Imperial-TP-2019-MJD-01,\\
FISPAC-TH/19-31415,  \\
UQBAR-TH/78-27182.\\
}}
}
\keywords{Gravity, Supergravity, Black Holes, Dualities, Freudenthal triple systems}

\abstract{
We study  General Freudenthal Transformations (GFT)
on black hole solutions in Einstein-Maxwell-Scalar (super)gravity theories with global symmetry of type  $E_7$.
GFT can be considered as a  2-parameter,  $a, b\in \R$, generalisation of Freudenthal duality: $x\mapsto x_F= a x+b\tilde{x}$,
where   $x$  is the vector of  the electromagnetic charges, an element of a Freudenthal triple system (FTS), carried by a large black hole  and  $ \tilde{x}$ is its Freudenthal dual.
These transformations leave  the Bekenstein-Hawking
entropy invariant up to  a scalar factor given by
$a^2\pm b^2$. For any $x$ there exists  a one parameter subset of GFT  that leave the entropy invariant,   $a^2\pm b^2=1$,  defining the subgroup of Freudenthal rotations. The Freudenthal plane defined by $\Span_\R\{x, \tilde{x}\}$ is closed under GFT and is foliated by the orbits of the Freudenthal rotations.
Having introduced the basic definitions and presented their properties in detail,
we consider the relation of GFT to the global symmetries or U-dualities in the context of supergravity.
We consider explicit examples in pure supergravity, axion-dilaton theories
 and $N=2,D=4$ supergravities
 obtained from $D=5$ by dimensional reductions  associated
to  (non-degenerate) reduced FTS's descending from cubic Jordan Algebras.}

\maketitle

\section{Introduction}

Recent  observations,  consistent with expectations for the shadow of a Kerr black hole (BH)  as predicted by general relativity,
have been, for the first time, presented \cite{Akiyama:2019cqa,Akiyama:2019eap}. This 
result demonstrates  yet again 
the 
effectiveness of general relativity, but also serves to emphasise  the need to address the long-standing 
puzzles presented by its  BH solutions. A classical stationary BH solution is  characterised by its mass $M$, angular momentum $J$ and charge $Q$ alone. In particular, its horizon area  is a simple function of these three quantities.   Identifying the horizon area as an 
entropy (determined up to a numerical constant of proportionality),  the classical  mechanics of BHs obeys a set of laws are  directly analogous to those of  thermodynamics  \cite{B1, B2, BCH}. 
Hawking's  
prediction
 \cite{H1, H2} that BHs quantum mechanically emit thermal radiation at the semiclassical level  fixes the Bekenstein-Hawking  
 area/entropy relation to be precisely (where the usual constants $c=\hbar=G=1$),
\be
S_{\text{BH}} = \frac{A_{\text{horizon}}}{4}
\ee 
and suggests that the  thermodynamic interpretation of BH mechanics  is more than a mere analogy. However, it also presents an immediate question.  A large BH carries a huge entropy, yet  is classically characterised entirely by $M, J$ and $Q$. Where, then, are the microscopic degrees of freedom underpinning the entropy?.

Any complete theory of quantum gravity should address this challenge in some way or at least advance in this direction. 
String/M-theory provides an answer for a very special class of extremal dyonic BHs, where the calculations are made tractable by 
the presence of some preserved supersymmetries \cite{SV}. 
This result and its generalisations depend on a range of mathematical and theoretical insights. 
In particular, symmetries, duality transformations and the mathematical structures  upon  which they are realised,  constitute  important
tools in the study of black hole solutions in general relativity and its
supersymmetric extension, supergravity, which provides the low-energy effective field theory limit of string/M-theory.  
For instance, the non-compact global symmetries of supergravity theories \cite{Cremmer:1979up}, or U-dualities in
the context of M-theory \cite{Duff:1990hn, Hull:1994ys},  have played a particularly crucial role,
starting with the  work of \cite{Breitenlohner:1987dg}.
For a large class of $\N\geq 2$ Poincar\'e supergravity theories with symmetric scalar manifolds\footnote{For a survey of symmetric spaces in supergravity  see \protect\cite{Ferrara:2008de}.}
the U-duality groups   are  of ``type $E_7$'', a class of groups sharing the same algebraic structure as the second largest exceptional Lie group $E_7$ \cite{Brown:1969}.
%
%
Groups of type $E_7$ are axiomatically characterised by Freudenthal triple systems (FTS)
\cite{Freudenthal:1954,Freudenthal:1959, Brown:1969}.
An FTS is a vector space $\FTS$ with, in particular,
  a symmetric four-linear form
 $\Delta(x,y,z,w)$ (see \autoref{sec:fts} for full details).
The automorphism group $\Aut(\FTS)$ of the FTS is the U-duality group $G_4$ of
the associated 4d supergravity. The electromagnetic  charges carried by the static extremal
  black hole solutions in such theories  correspond to elements $x\in\FTS$ and fall into linear representations of the associated U-duality groups. For such theories the leading-order Bekenstein-Hawking black hole entropy  is given by
\be
S_{\text{BH}}=\pi\sqrt{|\Delta(x)|},
\ee
 where $\Delta(x):=\Delta(x,x,x,x)$ is the
unique U-duality invariant quartic polynomial of the BH charges.

In \cite{Duff-FD} it was shown that when the U-duality group is
 of type $E_{7}$ \cite{Cartan, Brown:1969},
these black hole solutions enjoy a  nonlinear symmetry, named \textit{Freudenthal
duality}, acting on their associated charge vectors $x$. This holds for instance in all $\mathcal{N}>2$-extended, $D=4$
supergravities, as well as in all $\mathcal{N}=2$ supergravities coupled to
vector multiplets with symmetric scalar manifolds. However, supersymmetry is
not a necessary ingredient (\textit{e.g.} in the case
of $\mathfrak{F}(J_{3}^{\mathds{C}_{s}})$ and
$\mathfrak{F}(J_{3}^{\mathds{H}_{s}})$; \textit{cfr.}  \autoref{tt1}).

In \cite{FMY-FD}
Freudenthal duality was then generalised to a symmetry not only of the
Bekenstein-Hawking black hole entropy $S_{\text{BH}}$, but also of the
critical points of the black hole effective potential $V_{\text{BH}}$:
regardless of supersymmetry, such a formulation of Freudenthal duality
actually holds for\textit{\ any} Maxwell-Einstein system coupled to a
non-linear sigma model of scalar fields, in four dimensions.

The role of Freudenthal duality in the structure of extremal black hole
solutions was investigated in \cite{Ortin:2012gg}, in the framework of
ungauged $\mathcal{N}=8$, $D=4$ maximal supergravity. In particular,  the most
general solution to the supersymmetric stabilisation equations where shown to
be given by  the F-dual of a suitably defined real $56$-dimensional
vector, whose components are real harmonic functions in $%
\R ^{3}$ transverse space. Then, in \cite{Galli:2012ji} Freudenthal
duality was also shown to be an \textit{on-shell} symmetry of the effective,
one-dimensional action describing the dynamics of scalar fields in the
background of a static, spherically symmetric and asymptotically flat black
hole in $\mathcal{N}=2$, $D=4$ supergravity. In \cite{F-Dual-L} it was  shown that the generalised,
scalar-dependent Freudenthal duality introduced in \cite{FMY-FD}
 actually is a
symmetry of the equations of motion of the full theory, and is not
restricted to the extremal black hole solutions or their effective action.
Remarkably, in \cite{F-Dual-L} Freudenthal duality was also applied to
world-sheet actions, such as the Nambu-Goto world-sheet action in any $(t,s)$%
-signature spacetime, then allowing for an F-dual formulation of
Gaillard-Zumino duality \cite{Gaillard-Zumino,Cecotti:1988zz, Duff:1989tf}
on the world-sheet.

It is also here worth remarking that, in recent years, groups of type $E_{7}$,
Freudenthal triple systems, and Freudenthal duality have also appeared in
several indirectly related contexts, such as the relation to minimal
coupling of vectors and scalars in cosmology and supergravity
\cite{FK,FKM-Deg}, Freudenthal gauge theory (in which the
scalar fields are $\mathfrak{F}$-valued) \cite{Marrani:2012uu}, multi-centered BPS black holes
\cite{Fernandez-Melgarejo:2013ksa}, conformal isometries \cite{Borsten:2018djw}, Hitchin functionals and entanglement in quantum information theory
\cite{Borsten:2008,levay-2008,Borsten:2009yb, Levay:2012tg, Ashmore:2018npi}\footnote{Freudenthal duality in the context of entanglement and Hitchin functionals can actually be related back to its application to black holes via the \textit{%
black-hole/qubit correspondence}
\cite{Borsten:2008wd, 4-qubits, rev-1,Borsten:2012fx, Levay:2012tg}}.

 Our focus here is on the notion of general Freudenthal
 transformations (GFT), introduced  in \cite{Fernandez-Melgarejo:2013ksa}. In this  work it was shown
that  F-duality can be generalised to an Abelian group of transformations
\be
x\mapsto x_F= a x+b\tilde{x}.
\ee
The GFT leave the quartic form invariant up to a scalar factor
$\lambda^2=a^2\pm b^2$,
\be\label{GFTdelta}
\Delta(x_F) = \lambda^4 \Delta(x)
\ee
The  entropy, ADM mass and,
 for multicenter solutions in some specific models,
the  inter-centre distances scale as
\be
S_{BH}\to\lambda^2S_{BH},\quad M_{ADM}\to\lambda M_{ADM},\quad r_{ab}\to\lambda r_{ab},
\ee
while the scalars on the horizon  and  at infinity are left invariant.

The properties of GFT, in particular the  properties
of the quartic FTS invariant (or the Bekenstein-Hawking entropy in physical terms), can be traced back to the existence
and properties of  Freudenthal planes in $\FTS$.
This notion  first appears  in the mathematical literature in \cite{Brown:1969}.
 Given an F-dual pair $x$ and $\tilde{x}$ we define the Freudenthal plane $\fx\subset \FTS$ as the
set of all elements
\be
y_x=a x+ b\tx, \quad a,b\in \R.
\ee
The $\FTS_x$-plane is closed under GFT. From \eqref{GFTdelta} we see that the quartic form and, thus, the Bekenstein-Hawking entropy, is invariant under the special set of GFT with $\lambda=\pm1$. In particular, for any $x, \Delta(x)\not=0$ there exists a one-parameter subgroup of $\Aut(\FTS)$ that preserves the $\FTS_x$-plane and the Bekenstein-Hawking entropy. These will be referred to as  Freudenthal rotations. Although GFT are non-linear, there  always exists a linearly acting ``gauged''
 U-duality\footnote{Note, here we are considering continuous charges; when they are quantised this no longer holds.}  transformation that sends $x$ to $x_F$.

In the present work we introduce in detail these constructions and develop their applications to  black holes in supergravity, as summarised here.
 An extended treatment of GFT is presented in
\autoref{sec:2}-\autoref{sec:5}.
In the following sections these mathematical tools will be
 applied to the physics of black holes solutions in supergravity. First, in \autoref{sec:pure}  we will study the entropy properties of
 $\N=2$, $D=4$ pure supergravity from the point of view of  the FTS formalism. This provides an example of a degenerate FTS, where the quartic invariant  is a positive definite perfect square. As a consequence the Freudenthal plane in this case coincides with the entire FTS and the GFT are transitive on the space of charges. The Freudenthal rotations correspond precisely to familiar electromagnetic duality. To go beyond electromagnetic duality we consider in  \autoref{sec:axion}  the axion-dilaton model, an $\N=2, d=4$ supergravity
 minimally coupled to
one vector multiplet, which can be considered a consistent truncation of $\N=4$ supergravity.
Again, this model is degenerate and  cannot be uplifted to $D=5$. This
 is reflected in the non-reduced character of the FTS; it is not built from an underlying cubic Jordan algebra.
In \autoref{sec:reduced}
we proceed to the analysis of $\N=2,D=4$ supergravities admitting a  $D=5$ origin.  The mathematical
structure of these models is that of a reduced FTS, which may be derived from a cubic Jordan Algebra, $J_3$, so that $\FTS\cong \FTS(J_3)$.
In first place we study the  $T^3$ model,  or in Freudenthal terminology  $\FTS(\R)$.

In \autoref{sec:84} we study the question of
  orbit
stratification of the $\Delta >0$ locus of $\mathfrak{F}(\R)$ and its preservation by  GFT.
In \autoref{sec:90} we show, in different examples,
how the action of GFT, and, in particular,
Freudenthal duality can be
realised by U-duality
transformations that are ``gauged" in the sense that they depend on the element of
$\mathfrak{F}$ to which they are applied.
Finally in \autoref{sec9} we present some further physical discussion, summary and conclusions.
We  study the properties of asymptotically small interacting black holes.
In the different appendices we present a summary of formulae  used throughout the work and further technical
details.

\section{Freudenthal triple systems:   definitions and properties}

\label{sec:fts}
\label{sec:2}

In 1954 Freudenthal \cite{Freudenthal:1954, Freudenthal:1959} constructed
the exceptional Lie group $E_{7}$ (of dimension $133$) as the automorphism
group of a structure based on the smallest, non-trivial $E_{7}$ irrepr. $%
\mathbf{56}$, in turn related to the exceptional Jordan algebra $J_{3}^{%
\mathds{O}}$ of $3\times 3$ Hermitian octonionic matrices (also referred to as the  Albert
algebra) \cite{JVNW}.  Freudenthal's
aforementioned construction is often referred to as a \textit{Freudenthal triple system}
(FTS) for reasons that shall become clear shortly.

At the end of 60's, Meyberg \cite{Meyberg:1968} and Brown \cite{Brown:1969}
elaborated the axioms on which the, completely symmetric, ternary structure underlying an FTS is
based; in fact, the $E_{7}$ irrepr. $\mathbf{56}$ is just an example of a
class of modules, characterising certain Lie groups as of groups \textit{%
\textquotedblleft of type }$E_{7}$\textit{\textquotedblright }.
The role of
the FTS's in $D=4$ Maxwell-Einstein (super)gravity theories was discovered
later \cite{GST, GST-2, FG-1} to be related to the representation of the
electric-magnetic (dyonic) charges of black hole solutions.

A  FTS is  defined \cite{Brown:1969} as
a finite dimensional vector space $\mathfrak{F}$ over a
field $\mathds{F}$
\footnote{In the following treatment, we will consider
$\F=\R$ (classical/(super)gravity level).
The (quantum/Dirac-Schwinger-Zwanzinger-quantized) case
(and further extensions thereof) will be investigated elsewhere.
The complex case $\F =\mathds{C}$ is relevant for  quantum qubit entanglement applications.}
 (not of characteristic 2 or 3), such that:

\begin{enumerate}
\item $\mathfrak{F}$ possesses a non-degenerate antisymmetric bilinear form $\{x, y\}.$

\item $\mathfrak{F}$ possesses a completely symmetric four-linear form
$\Delta(x,y,z,w)$
which is not identically zero.
This quartic linear form induces a
 ternary product $T(x,y,z)$ defined on $\mathfrak{F}$ by
$$\{T(x,y,z), w\}=2\Delta(x, y, z, w).$$

\item For the ternary product $T(x,y,z)$
 it is required that
\begin{equation}
\label{eq:def}
3\{T(x, x, y), T(y,y,y)\}=2\{x, y\}\Delta(x, y, y, y).
\end{equation}

\end{enumerate}

In our case of interest, the semi-classical supergravity limit, the  physical vector of charges $x$
is to be  regarded as continuous and the associated FTS is taken to be   over $\R$ or $\C$.

The \emph{automorphism} group of an FTS is defined as the set of invertible $\mathds{F}$-linear transformations preserving the quartic and quadratic
forms:
\begin{equation}
\Aut(\mathfrak{F})\equiv\{\sigma\in\Iso_\mathds{F}(\mathfrak{F})|\{\sigma x,
\sigma y\}=\{x, y\},\;\Delta(\sigma x)=\Delta(x)\} .
\label{eq:brownfts}
\end{equation}

An important operation in what follows is the  \FT-dual\footnote{Not to be confused with T-duality in string theory.}, $':\mathfrak{F}\rightarrow \mathfrak{F}$, defined by
\begin{equation}
\qquad x\mapsto {x'}:= T(x,x,x)\equiv T(x).
\label{ee0021}
\end{equation}
Note that, the conditions $\{\sigma x, \sigma y\}=\{x, y\}$ and $\Delta(\sigma x)=\Delta(x)$
immediately imply the homogeneity of the \FT-map
\begin{equation}
T(\sigma x)=\sigma T(x).
\label{e8801}
\end{equation}
Hence $\Aut(\mathfrak{F})$ is the set of automorphisms in the conventional sense. 

The Lie algebra $\autFF$ of $\Aut(\mathfrak{F})$ is
given by
\begin{equation}
\autFF=\{\phi \in \Hom_{\mathds{F}}(\mathfrak{F}%
)|\Delta(\phi x,x,x,x)=0,\{\phi x,y\}+\{x,\phi y\}=0,\;\forall x,y\in \mathfrak{F}%
\},
\end{equation}%
as is easily verified \cite{Borsten:2011nq}.  The first of the conditions can be
restated as $\{\phi x,x'\}=0$.

The $\mathds{F}$-linear map $\Upsilon_x:\mathfrak{F}\rightarrow \mathfrak{F}$
defined by
\begin{equation}
 \label{Ipsilon}
\Upsilon_x(y)=3T(x,x,y)+\{x,y\}x
\end{equation}
is in $\autFF$. This is a direct consequence of axiom III (\autoref{Ipsilon}). In fact
note that \autoref{eq:def} can be reexpressed as
$$\{\Upsilon_{x}(y),y'\}=0.$$
Note that, in particular,
$$\Upsilon_x(x)= 3 x',$$
but $'$ is not in $\autFF$.

The linear map $\Upsilon_x$ was introduced in this $\Aut(\mathfrak{F})$-covariant form in \cite{Krutelevich:2004}.
For $\Delta(x)\not=0$ we may also define the normalized map
\be\label{eq:defY}
\overline{\Upsilon }_{x}\equiv \frac{1}{3\sqrt{|\Delta (x)|}}\Upsilon _{x}.
\ee
Linearizing  \autoref{Ipsilon} with
respect to $x$ implies that $\Upsilon_{x,y}:\mathfrak{F}\rightarrow
\mathfrak{F}$ defined by
\begin{equation}
\Upsilon_{x,y}(z)=6T(x,y,z)+\{x,z\}y+\{y, z\}x
\end{equation}
is also in $\autFF$ \cite{Yokota:2009}
(see also Eq.11.b in \cite{Brown:1969}). In particular the following relation holds
$\{\Upsilon_{x,y}(z),z'\}=0$. We have also  $\Upsilon_{x, x}(z)= 2 \Upsilon_x(z)$
and, (see \cite{Brown:1969}), for any $y,z$, $\Upsilon_{y,y'}(z)=0$


Following \cite{Krutelevich:2004} an FTS element may be assigned a
manifestly $\Aut(\mathfrak{F})$ invariant \emph{rank}, an integer function between 1 and 4 defined
by the relations:
\begin{equation}  \label{eq:FTSranks}
\begin{split}
\text{{Rank}} (x) =1& \Leftrightarrow \Upsilon_x(y)=0\; \forall y,\; x\not=0;
\\
\text{{Rank}} (x) =2& \Leftrightarrow T(x)=0,\;\exists y \;\text{s.t.}%
\;\Upsilon_x(y)\not=0; \\
\text{{Rank}} (x) =3& \Leftrightarrow \Delta(x)=0,\;T(x)\not=0; \\
\text{{Rank}} (x) =4& \Leftrightarrow \Delta(x)\not=0.\\
\end{split}%
\end{equation}
The Rank 1 conditions appeared
before in \cite{Freudenthal:1954}.
We define the sets of elements of a given rank
$\mathfrak{F}_{(k)}\equiv \{x\in \mathfrak{F}\;|\;Rank(x)=k\}$.
The rank of a element can be related to the degree
of supersymmetry preserved by the solution (see \cite{Borsten:2009zy} and references therein).

\subsubsection*{Supergravity and the classification of FTS: An outline}

An FTS is said to be reduced if it contains a strictly regular element: $%
\exists \;u\in \mathfrak{F}$ such that $T(u)=0$ and $u\in \text{ Range }%
L_{u,u}$ where $L_{x,y}:\mathfrak{F}\rightarrow \mathfrak{F};\quad
L_{x,y}(z)\equiv T(x,y,z)$.
It can be proved \cite{Brown:1969,Ferrar:1972} that every simple \textit{reduced}
FTS $\mathfrak{F}$ is isomorphic to an FTS $\mathfrak{F}(J_{3})$,
where
\begin{equation}
\mathfrak{F}(J_{3})\equiv \F \oplus \F \oplus J_{3}\oplus \overline{%
J}_{3},
\end{equation}%
with $J_{3}$ denoting a rank-$3$ Jordan algebra. All algebraic structures in
$\mathfrak{F}(J_{3})$ can be defined in terms of the basic Jordan algebra
operations \cite{Brown:1969, Ferrar:1972} (also \textit{cfr.} \cite%
{small-orbits-maths} and Refs. therein). In a Maxwell-Einstein physical
framework, the presence of an underlying Jordan algebra $J_{3}$ corresponds to the fact
that the $D=4$ Maxwell-Einstein (super)gravity theories can be obtained by
dimensional reduction of a $D=5$ theory, whose electric-magnetic (U-)duality\footnote{%
Here U-duality is referred to as the \textquotedblleft
continuous\textquotedblright\ symmetries of \cite{CJ-1}. Their discrete
versions are the U-duality non-perturbative string theory symmetries
introduced by Hull and Townsend \cite{HT-1}.} is nothing but the reduced
structure group of $J_{3}$ itself.

For $\mathfrak{F}(J_{3}^{\mathds{A}})$, the automorphism group has a two
element centre, and its quotient yields the simple groups listed \textit{e.g.%
} in Table 1 of \cite{F-Dual-L}, whereas for
$\mathfrak{F}(\R \oplus {\Gamma }_{m,n})$
one obtains the semi-simple groups $SL(2,%
\R )\times SO(m+1,n+1)$ \cite{Brown:1969,Krutelevich:2004,
Gunaydin:2009zza}. In all cases, $\mathfrak{F}$ fits into a symplectic
representation of $\Aut(\mathfrak{F})$, with dimensions listed \textit{e.g.}
in the rightmost column of Table 1 of \cite{F-Dual-L}.

By confining ourselves to reduced FTS's $\mathfrak{F}(J_{3})$ related to
\textit{simple} or \textit{semi-simple} rank-$3$ Jordan algebras $J_{3}$,
one can exploit the Jordan-Von Neumann-Wigner classification \cite{JVNW},
and enumerate the possible FTS's, depending on their dimension dim$\mathfrak{%
F}=2N$.
\footnote{Reduced FTS's have \textit{at least} dimension $2N=4$, namely they contain
\textit{at least} $N=2$ Abelian vectors in $D=4$. Within the $\mathcal{N}=2$
interpretation, they are the $5D\rightarrow 4D$ Kaluza-Klein (KK) vector
(\textit{aka} the $D=4$ graviphoton) and the $D=5$ graviphoton (which becomes
a matter photon in $D=4$).}
  A summary of this classification is presented in \autoref{tt1}.

Various $D=4$ supergravities are listed in \autoref{tt1}: the semi-simple
cases $\mathfrak{F}\left(\R \oplus {\Gamma}%
_{1,n-1}\right) $ and $\mathfrak{F}\left(\R \oplus {%
\Gamma }_{5,n-1}\right) $ correspond to $\mathcal{N}=2$ resp. $4$
Maxwell-Einstein supergravity, while $\mathfrak{F}\left(J_{3}^{\mathds{A}}\right) \equiv \mathfrak{F}^{\mathds{A}%
}$ correspond to the so-called $\mathcal{N}=2$
\textquotedblleft magic\textquotedblright\ Maxwell-Einstein supergravities\footnote{%
The theories based on Lorentzian cubic Jordan algebras $J_{2,1}^{\mathds{A}}$
and $J_{2,1}^{\mathds{A}_{s}}$ correspond to certain classes of $\mathcal{N}%
=2$ supergravities with non-homogeneous vector multiplets' scalar manifolds
(cfr. e.g. \cite{GZ}, \cite{Squaring-Magic}).}\cite{GST}. Moreover, $\mathfrak{F}^{\mathds{O}_{s}}\equiv \mathfrak{F}\left(
J_{3}^{\mathds{O}_{s}}\right) $ pertains to maximal $\mathcal{N}=8$
supergravity, and the simplest reduced FTS is $\mathfrak{F}\left(\R \right) $, related to the so-called $T^{3}$ model of $\mathcal{N}%
=2 $, $D=4$ supergravity (treated in \autoref{sec:T^3}).

As evident from \autoref{tt1}, there are two (for $\mathds{A}=%
\R $) or three (for $\mathds{A}=\C,\mathds{H},\mathds{O%
}$)  possible FTS
structures for $N=3q+4$, where $q=$dim$\left( \R ,\mathds{C},\mathds{H%
},\mathds{O}\right) =1,2,4,8$, corresponding to $\mathfrak{F}^{\mathds{A}}$ \cite{GST, GST-2} and $%
\mathfrak{F}^{\mathds{A}_{s}}$\cite{Magic-Non-Susy,CJ-1}.

Finally, an FTS is said to be degenerate if its quartic form is identically proportional
 to the square of a quadratic polynomial.
Note that FTS on  ``degenerate" groups
of type $E_{7}$ (as defined in \cite{FKM-Deg}, and Refs. therein) are not
reduced and hence cannot be written as $\mathfrak{F}(J_{3})$; they
correspond to theories which cannot be uplifted to $D=5$ dimensions
consistently reflecting the lack of an underlying rank-$3$ Jordan algebra $%
J_{3}$.
\begin{table}[h]
\begin{tabular}{|l|l|l|}
\hline
$N$ 
& $J_{3}$\qquad & $D=4$ Maxwell-Einstein~theory \\    \hline
2 &   $ \R $\qquad\qquad & $\mathcal{N}=2$ ${T}^{3}$\\
3 & $\R \oplus \R $ & $\mathcal{N}=2$, ${ST}^{2}$  \\
4  & $\R \oplus \R \oplus \R $ & $\mathcal{N}=2$, ${STU}$\\
5-6& $ \R \oplus {\Gamma}_{n,3-n},~~\R \oplus {\Gamma}_{n,4-n}$ & $n=1: \mathcal{N}=2$, $n_{V}=4,5$\\
7  & $\left\{
\begin{array}{l}
\R \oplus \Gamma _{n,5-n} \\
J_{3}^{\R }%
\end{array}%
\right. $ & $\left\{
\begin{array}{l}
n=1:\mathcal{N}=2,n_{V}=5,~n=5:\mathcal{N}=4,n_{V}=1 \\
\mathcal{N}=2~\text{magic}~\R %
\end{array}%
\right. $ \\
8-9 & $ \R \oplus {\Gamma}_{n,6-n},~\R \oplus {\Gamma}_{n,7-n}$ &$n=1:\mathcal{N}=2,n_{V}=7,8$,~$n=5:\mathcal{N}=4,n_{V}=2,3$,\\
10 & $\left\{
\begin{array}{l}
\R \oplus \Gamma _{n,8-n} \\
J_{3}^{\mathds{C}},~~J_{3}^{\mathds{C}_{s}}%
\end{array}%
\right. $ &$\left\{
\begin{array}{l}
n=1:\mathcal{N}=2,n_{V}=9,~n=5:\mathcal{N}=4,n_{V}=3 \\
\mathcal{N}=2~\text{magic}~\mathds{C},~~\mathcal{N}=0~\text{magic}~\mathds{C}_{s}
\end{array}%
\right. $ \\
11-15  & $\R \oplus {\Gamma}_{n,9-n},...,~\R \oplus {\Gamma}_{n,13-n}$&$n=1:\mathcal{N}=2,n_{V}=10-14$,~$n=5:\mathcal{N}=4,n_{V}=5-9$,  \\
16  & $\left\{
\begin{array}{l}
\R \oplus \Gamma _{n,14-n} \\
J_{3}^{\mathds{H}},~~J_{3}^{\mathds{H}_{s}}%
\end{array}%
\right. $ &
$\left\{
\begin{array}{l}
n=1:\mathcal{N}=2,n_{V}=15,~n=5:\mathcal{N}=4,n_{V}=10 \\
\mathcal{N}=2~\text{magic}~\mathds{H},~~\mathcal{N}=0~\text{magic}~\mathds{H}_{s}
\end{array}%
\right. $ \\
17-27 & $\R \oplus {\Gamma}_{n,15-n},...,~\R \oplus {\Gamma}_{n,25-n} $ & $n=1:\mathcal{N}=2,n_{V}=16-26$,~$n=5:\mathcal{N}=4,n_{V}=11-21$\\
28 &$\left\{
\begin{array}{l}
\R \oplus \Gamma _{n,26-n} \\
J_{3}^{\mathds{O}},~~J_{3}^{\mathds{O}_{s}}%
\end{array}%
\right. $ &
$\left\{
\begin{array}{l}
n=1:\mathcal{N}=2,n_{V}=27,~n=5:\mathcal{N}=4,n_{V}=22 \\
\mathcal{N}=2~\text{magic}~\mathds{O},~~\mathcal{N}=8\text{~~}%
\end{array}%
\right. $ \\
$>28$ &  $\R \oplus {\Gamma}_{n,N-2-n} $ & $n=1:\mathcal{N}=2,n_{V}>27$,~$n=5:\mathcal{N}=4,n_{V}>22$\\ \hline
\end{tabular}
\centering
\caption{Classification of Freudenthal triple systems (see text for explanation).}
\label{tt1}
\end{table}

\section{Freudenthal dualities and  planes}
\label{sec3}

\subsubsection*{ F-duality}

We have defined already the  transformation $x'=T(x)$, valid for a vector of any rank (see \autoref{ee0021}).
For rank-4 charge vectors $x\in \FTS_{(4)}$,
the black hole charge
Freudenthal duality is defined by ($\epsilon\equiv\epsilon(x)\equiv\sgn \Delta(x)$)
\begin{equation}
\tilde{}:\mathfrak{F}_{(4)}\rightarrow \mathfrak{F}_{(4)},\qquad
 x\mapsto \tilde{ x}\equiv \epsilon(x)\frac{T(x)}{\sqrt{|\Delta (x)|}}.
\label{ee0029}
\end{equation}%
The Freudenthal duality has the following elementary properties \cite{Borsten:2009zy}:
\begin{itemize}
\item It preserves the quartic norm  $\Delta (\tilde{x})=\Delta (x)$;
obviously
$\Delta(\tilde{x},x,x,x)=\{\tilde{x},x'\}=0.$

\item It is an anti-involution: $\tilde{\tilde{x}}=-x$;

\item It is not a U-duality,  since it is non-linear and  generically $\{\tilde{x},\tilde{y}\}\not=\{x,y\}$.
Also, in general, $\{\tilde{x},y\}+\{x,\tilde{y}\}\not=0$.
\end{itemize}

Note that, although the map $x\to \tilde x$ is not a U-duality,   the map
$\Upsilon_x$ (or $\overline{\Upsilon}$), for $x$ fixed, is indeed:
\begin{itemize}
\item[a)] linear,

\item[b)] $\Upsilon_x\in \autFF $ and, finally

\item[c)]  $\overline{\Upsilon}_x(x)= \epsilon(x)\tilde{x}$.
\end{itemize}
It follows from  (c), that for $\Delta(x)>0$ the Freudenthal map $x\mapsto \tilde{x}$ can be considered as an ``$x$-dependent'' U-duality.

\subsubsection*{The \FT- and $\FTS$-planes: Definition and general properties}

For  a general  element in $\FTS$, respectively an
element $x\in \mathfrak{F}_{(4)}$, we define
the associated \FT- and $\FTS$-planes, respectively denoted $\FT_x,\FTS_x$, as
the $\R$-linear spans of $x, x'$ or $x, \tilde{x}$. In each case:
\begin{eqnarray}
\FT_x&\equiv&\{y \in \FTS | y=ax+b x',~a,b\in \mathds{R}\},\\
\FTS_x&\equiv&\{y \in \FTS_{(4)} | y=ax+b\tilde{x},~a,b\in \mathds{R}\}.
\end{eqnarray}

Naturally the $\FTS$-plane $\FTS_x$ is only defined  as long as
$\Delta(x)\neq 0$ (maximal rank elements), while
$\FT_x$ is defined for any $x$, although it degenerates to a \FT-line for rank $x<3$ elements.
If they both exist, $\FT_x$  and $\FTS_x$ are the same space.
It is advantageous to study the properties of the \FT-planes, and when needed, to specialise to
$\FTS$-planes. We will follow this strategy in what follows.

\subsubsection*{Linearity of \FT-transformations on the \FT-plane}

We first  show the linearity of the \FT-dual on the \FT-plane: \FT-planes are closed
under \FT-transformations.
For any linear combination, one has,
because of the multi-linearity of $T$,  ($a,b$ constants, $\Delta_x=\Delta(x)$),
\beq
T(a x+b x')&=& a^3 T(x)+ b^3 T(x')+3 a^2 b T(x,x,x')+3 a b^2 T(x,x',x') \nonumber \\
           &=& a^3 x' +a b^2\Delta_x x'-b^3\Delta_x^2 x -a^2 b \Delta_x x \nonumber\\
           &=& (a^2+ b^2\Delta_x) (-b\Delta_x x+a x').
\label{ee001}
\eeq
Where we have used the properties (\cite{Brown:1969}, lemma 11.(abcf)):
\beq
T(x,x,x')&=& -1/3 \Delta_x x ,\label{ee002a}\\
T(x,x',x')&=& 1/3 \Delta_x x' ,\label{ee002b}\\
T(x',x',x') &=& -\Delta^2_x x.
\label{ee002}
\eeq
We can see that \autoref{ee001} is equivalent to, or simply  summarizes, the relations
 \autoref{ee002a}-\autoref{ee002}.

Using  \autoref{ee001} we can compute the map $\Delta$ for any element on the \FT-plane. After a short explicit computation we have
(using $2\Delta(x)=\{x',x\}$)
\beq
\Delta(a x+b x')&=& \frac{1}{2}\{ T(a x+b x'),a x+b x'\}\\
                 &=&\left (a^2+b^2\Delta(x)\right)^2  \Delta(x).
\label{ee003}
\eeq
The sign of $\Delta$ on the \FT-plane is constant. Hence, 
in any \FTS-plane, there is  an element $y\in \FTS_x$ such $\Delta(y)=0$
 if and only if  $\Delta(x)$ is negative.

 Similarly, we have the following expressions describing the behaviour of the map $\Upsilon$ on the \FT-plane:
 \beq
  \Upsilon_x(  x')&=& 3 T(x,x,x')+\{x,x'\} x= -\Delta_x x - 2 \Delta_x x= -3 \Delta_x x,\\
 \Upsilon_x( a x+b x') &=& -3 b \Delta(x) x + 3 a x' = 3 T(a x+b x')/ (a^2+b^2\Delta_x).
 \label{ee00801}
 \eeq
 Further mathematical properties of the \FT-planes are shown in  \autoref{sec:properties} and \autoref{secab}.

 \subsubsection*{\FT-  and $\FTS$-transformations on the planes}

For $\tilde{x}$ well-defined, we rewrite   \autoref{ee001},  \autoref{ee003}, by a redefinition of the parameters
 $a,b$ in terms of the F-dual   \autoref{ee0029} as ($\epsilon=\sgn \Delta(x)$)
\beq
T(a x+b \txx) &=& \epsilon \left(a^2+\epsilon b^2\right)\sdx \left(- \epsilon b x+ a \txx\right),\label{e9001}\\
\Delta(a x+b \txx) &=& \left(a^2+\epsilon b^2\right)^2\Delta(x).
\label{ee003b}
\eeq
It is clear from this last expression that $a^{2}+\epsilon b^{2}=\pm 1$ defines a subset of elements in $\fx$ with fixed entropy: an $SO(2)$ or $SO(1,1)$ symmetry.
Moreover, $\sgn \Delta(x)=\sgn \Delta(ax+b \txx)=\epsilon$, unless $a^{2}+\epsilon b^{2}=0$.
In particular
\beq
\Delta(x\pm \txx) &=& (1+\epsilon)^2\Delta(x).
\eeq

As consequence of \autoref{e9001} and \autoref{ee003b}, the F-dual of a linear combination is given by
($\Delta(x)\neq 0,\Delta( ax+b\txx)\neq 0$, $\sgn \Delta(x)=\sgn \Delta(ax+b \txx)=\epsilon$)
\beq
\widetilde{ax+b \tilde{x}} &=&\frac{\epsilon T(a x+b \txx)}{\sqrt{\mid \Delta (a x+b\txx)\mid}}
\eeq
and, finally,
\beq
\widetilde{ax+b \tilde{x}} &=& \eta \left(-\epsilon b x +a\txx\right )=\begin{cases}
-b x+a\txx , & (\Delta(x)>0),\\
 \eta \left(b x+a\txx\right) ,&(\Delta(x)<0),
 \end{cases}
 \label{e8802}
\eeq
where $\epsilon=\sgn \Delta(x), \eta=\sgn (a^2+b^2\epsilon)$.
Clearly, if $\epsilon=1$ then $\eta=1$
%



As discussed in  appendix \autoref{ss3a}  the $\FTS$-plane is a, quadratic, two dimensional, sub-FTS system
with suitably restricted operations $\Delta_{\FTS},T_{\FT},\{,\}_{\FT}$ and $I_2$.

\label{ss331}


On the full FTS, for maximal rank elements,  one can define an ($
\Aut(\mathfrak{F})$-invariant) ``metric''  by the (non quadratic) expression
\begin{equation}
(u,v)\equiv \frac{1}{4}[\{\tilde{u},v\}+\{\tilde{v},u\}].  \label{dd-1}
\end{equation}
and a ``pseudo-norm'' by
\begin{equation}
||u||=(u,u)=\frac{1}{2}\{\tilde{u},u\}=\epsilon \sqrt{|\Delta (u)|}.
\end{equation}%

If we  fix $x$, and restrict ourselves to the $\FTS_x$ plane  we can use the expressions in
  \autoref{ss3a}  (see also further properties in Ref.\cite{torrentemelgarejoquad}) and
connect $||u||$ with $I_2(u)$:
\begin{eqnarray}
(u,v)&\equiv&\frac{1}{4}\left [\{\tilde{u},v\}+\{\tilde{v},u\}\right]\\
        &=&  \frac{1}{4} \left [\eta(u)\{ \hS u,v\}+\eta(v)\{\hS v,u\}\right ] \\
&=&\frac{1}{4} (\eta(u)+\eta(v))\{\hS v,u\}\\
&=&\frac{1}{4} \epsilon (\eta(u)+\eta(v)) I_2( v,u)
\end{eqnarray}
where $\epsilon=\sgn \Delta(x), \eta(x)=\sgn I_2(x)$ and $\hat S$ is a linear map
given in \autoref{secab}. In particular
\begin{eqnarray}
||u||&=&\frac{1}{2} \epsilon \eta(u) I_2( u)=\frac{1}{2} \epsilon \mid I_2( u)\mid
\end{eqnarray}

We arrive at the same conclusions as for the \FT-plane.
 For $\epsilon =1$, the pseudo-norm $(\cdot ,\cdot )$ (or $I_2(u)$) is positive definite and the norm-preserving group is $\SO%
(2)$; thus, the $\FTS$-plane $\mathfrak{F}_{x}$ undergoes a \textit{\textquotedblleft spherical foliation"}. On the other hand,
for $\epsilon =-1$, the norm is positive semi-definite and the norm-preserving group is $\SO(1,1)$; thus, the $\FTS$-plane $\mathfrak{F}_{x}$ undergoes an \textit{\textquotedblleft hyperboloid-like foliation"}.
While the norm is timelike or
null, the vector $u$  can be timelike, spacelike and null according  to $\eta(x)$,
the sign of $I_2(x)$.

As for the \FT-operation,  the F-duality $x\to \tilde{x}$,  change the character of the vector.
The vectors $x,\txx$  are   ``$I_2$-orthogonal", by \autoref{eq31b},
$(x,\txx)=I_2(x,\txx)=0$.
 $\txx$ is timelike (resp. spacelike) if $x$ is spacelike (resp. timelike):
\begin{eqnarray}
x: lightlike & \longleftrightarrow & \txx: lightlike,\quad \txx=\pm x,\\
x: timelike (spacelike)& \longleftrightarrow & \txx: spacelike (timelike).
\end{eqnarray}

It is noted that, although the metric $I_2(x,y)$ is defined only inside a concrete $\FTS$-plane, the character
null, time or spacelike of a vector is an intrinsic property, as any given element belongs to one and only
one $\FTS$-plane, ``its'' plane, from the disjointness of the $\FTS$-planes (see \autoref{ss32}).

\section{The orthogonal space $\FTS_x^\perp$ and the orthogonal  plane $\fxy$}
\label{sec4}

In general $\left\{ x,\tilde{x}\right\} =-2
\sqrt{\left\vert \Delta \left( x\right)\right\vert }\neq 0$.
The bilinear form  $\{\cdot ,\cdot \}$ is non-degenerate on $\FTS_x$ by construction, since $x$ is neccesarily of
maximal rank ($\Delta(x)\not = 0$).
Consequently, for a given $x\in \mathfrak{F}_{(4)}$, the FTS $\mathfrak{F}$
may be decomposed as
\be
\FTS = \FTS_x\oplus\FTS_{x}^{\perp},
\ee
where $\mathfrak{F}_{x}$ is the $2$-dimensional $\FTS$-plane and $\fxp$ is its $\left(\dim_{\F} \FTS-2\right) $-dimensional orthogonal complement
w.r.t. the bilinear form $\left\{ \cdot ,\cdot \right\}$:
\beq
\FTS_x^\bot&=&\left\{y\in \FTS : \{ x, y \} = \{ x', y \}= 0\right\}\,.
\label{ee00351}
\eeq
Hence, for a given $x\in \mathfrak{F}$, any
element $y\in \mathfrak{F}$ enjoys the decomposition%
\be
y =y_{\parallel x}+y_{\perp x },
\ee
where $y_{\parallel x}\in
\mathfrak{F}_{x}$ and $y_{\perp x}\in \fxp$. Note that $y_{\parallel x},y_{\perp x}$ (also denoted $y_\parallel,y_\perp$ if there is no risk of confusion) are uniquely determined by $x$.
The coordinates of $y_{\parallel x}=a x + b\tilde{x}$ are uniquely determined by the expressions,
\begin{eqnarray}
\left\{ x,y_{\perp }\right\} &=&0\Leftrightarrow b=
\frac{-1}{2\sqrt{\left\vert \Delta \left( x\right) \right\vert }}\left\{x,y\right\} , \\
\left\{ \tilde{x},y_{\perp }\right\} &=&0\Leftrightarrow a=
\frac{1
}{2\sqrt{\left\vert \Delta \left( x\right) \right\vert }%
}\left\{ \tilde{x},y\right\} .
\end{eqnarray}%
Or, in compact notation (with respect a fixed element $x$),
\beq
y_{\parallel} &=&  \frac{1}{2 \surd \mid \Delta(x)\mid} \left | \begin{array}{cc} x & \tilde{x} \\  \{x  ,y  \}&\{\tilde{x}  ,y  \}  \\\end{array} \right|,\label{ee0046}\\
&=&  \frac{1}{\{x'  ,x  \}} \left | \begin{array}{cc} x & x' \\  \{x  ,y  \}&\{x'  ,y  \}  \\\end{array} \right|,\\
y_\perp &=& y-y_\parallel .
\eeq
The following properties hold (for the parallel component with respect a fixed $x$):
\beq
\Delta(y_{\parallel } )&=& \frac{1}{16\Delta(x)} \left( \epsilon\{\txx,y\}^2+\{x,y\}^2 \right)^2 \\
\left (a y+b \tilde{y}\right)_\parallel &=& a y_\parallel+ b (\tilde{y})_\parallel \\
\widetilde{(y_\parallel)} &=& \frac{\eta\epsilon}{2\surd\mid\Delta(x)\mid} \left( \epsilon\{\txx,y\}\txx+\{x,y\} x \right) .
\label{ee0048}
\eeq
with $\eta=\sgn ( \epsilon\{\txx,y\}^2+\{x,y\}^2 )$. In particular, note the distributivity of $\parallel$ (second expression).

Note, however that%
\begin{equation}
\widetilde{y_{\parallel }}\neq \left( \tilde{y}\right) _{\parallel }=\frac{%
\left( \left\{ \tilde{x},\tilde{y}\right\} x-\left\{ x,\tilde{y}\right\}
\tilde{x}\right) }{2\sqrt{\left\vert \Delta _{x}\right\vert }}.
\end{equation}

%

Obviously, a similar construction can be performed for the $\FT$-plane (see \cite{Brown:1969}, from pg. 89 on,   where such a
space is used to build a Jordan Algebra for reduced FTSs).

This decomposition into ``parallel" and ``orthogonal" spaces and the further decomposition of the
orthogonal space in orthogonal planes (to be defined in the next section) will be important in  what follows.

\subsubsection*{The  $\fxy$ plane}

Consider an arbitrary reference vector $x$ (of maximal rank for simplicity) and 
a perpendicular vector $y\in \FTS_x^\perp$. We define the
space, $\fxy:=\Span\{y,\Upsilon_x(y)\}$. That is
\beq
\fxy\ &:=&\{a y +b\Upsilon_x(y)~ | ~~y\in\FTS_x^\perp \quad a,b\in \R\}.
\eeq
We will show that the ``planes'' $\FTS_x$ and $\fxy$ are $\{,\}$-orthogonal:
$$\fxy\subseteq \FTS_x^\perp.$$
For any $y\in \FT_x^\perp$, we can  show  that also
$\Upsilon_x(y)\in \FT_x^\perp$.
%
We have indeed (using Equation (11c) in \cite{Brown:1969} in the first line and axiom 3 in the second line)
\beq
\{x',\Upsilon_x(y)\}&=&3 \{ T(x,x,x), T(x,x,y)\}=-\{y,x'\}\Delta(x)=0,\label{ee116}\\
\{x,\Upsilon_x(y)\}&=& \{ x,3 T(x,x,y)\}=3\{y,T(x,x,x)\}=3\{y,x'\}=0.
\label{ee117}
\eeq
%
which implies
$\Upsilon_x(y)\in \FTS_x^\perp.$
Obviously, the equality $\fxy\ = \FTS_x^\perp$ is only possible for $dim(\FTS)=4$,
as $dim (\FTS_x^\perp)=dim (\FTS)-2$.

We  show next that, in the same case, successive powers of  $\Upsilon_x$ acting on $y$  belong to the orthogonal plane.
In fact,  $\Upsilon_x^n(y)$ is proportional to $y$ or $\Upsilon_x(y)$.
 We have for example (as for  any $y\in \FTS_x^\perp$) 
\beq
\Upsilon^2_x(y)\equiv \Upsilon_x(\Upsilon_x(y))&=& 9 T(x,x,T(x,x,y))\nonumber\\
&=& -2\Delta(x,x,x,y) x-\Delta(x) y-\{y,x\}x' =-\{x',y\}- \Delta(x) y
\label{ee119}\\
&=&-\Delta(x) y.
\label{ee120}
\eeq
In the second line we have used the Lemma 1(11e)  in \cite{Brown:1969}.
In general for any $n$, we have, by using induction (for any $y\in \FTS_x^\perp$), for $n\geq 1$,
\beq
\Upsilon_x^{2n}(y)&=&  (-1)^n \ \Delta(x)^n y,\label{ee900}\\
 \Upsilon_x^{2n+1}(y)&=&  (-1)^n\ \Delta(x)^n \Upsilon_x(y)\label{ee901}.
\eeq


Let us remark that $\fxy$ is
not closed in general under \FT-transformations, it is not a sub-FTS with the operations inherited from the parent FTS.
The plane  $\fxy$ is however closed under the $\Upsilon_x$ map.
For any element belonging to it ($u\in \fxy$, $u=a y +b\ups{y}$),
\beq
\Upsilon_x (u)&=&    3 T(x,x,a y+b\ups{y})+\{x,a y+ b\ups{y}\} x\\
&=&   3 aT(x,x,y)+ 3 b T(x,x,\ups{y})\\
&=& -b \Delta (x) y+ a\ups{y},  
\eeq
where in the second line we have used the equality  expressed by
\autoref{ee119} (see also \cite{Brown:1969}).
According to this $\Upsilon_x(u)\in \fxt$.

\subsection{Behaviour of $\Delta$ on the \fxy\ plane }
\label{sec:41}
We are interested in the
behaviour of $\Delta$ on the \fxy\ plane.
For any $u\in \fxy, u=a y+b\ups{y}$, we have, by combining
 \autoref{e90101} with \autoref{e90104}
\beq
\Delta(u)&=& \left(a^2+b^2\Delta(x)\right)^2\Delta(y),\label{ee00527}\\
\Delta\left(\Upsilon_x(y)\right)&=&\Delta(x)^2 \Delta(y). \label{ee00527b}
 \eeq
Or, in normalized terms
\beq
\Delta(\tilde\Upsilon_x(y))\equiv
\Delta\left( \frac{\Upsilon_x(y)}{\sqrt{\mid \Delta(x)\mid}}\right)&=& \Delta(y).
\label{ee00527d}
\eeq
The $\tilde\Upsilon_x$ map thus  preserves both the bilinear and quartic invariants in each of
the  \fxy\ planes.
Applying twice \autoref{ee00527d} we arrive to
\beq
\Delta(\tilde\Upsilon_x^2(y)) &=& \Delta(y)
\label{ee00527e}
\eeq
and in general
\beq
\Delta(\tilde\Upsilon_x^n(y)) &=& \Delta(y).
\label{ee00527f}
\eeq

Combining \autoref{ee00527},\autoref{ee00527b} with \autoref{ee003}
we arrive to an expression
\beq
\Delta(a y+b\ups{y})\Delta(x) &=&
\Delta\left(a +bx'\right)\Delta(y).
\label{ee00527c}
\eeq
which relates the behaviour of the quartic invariant $\Delta$ on the $\FT_x$   and \fxy\  planes.

The behaviour of $\Delta$ on the $\FTS_x$ or \fxy\  planes is similar but with some important differences.
In the \fxy\ case it depends on the  signs of both
$\Delta(x)$ and $\Delta(y)$.
The overall sign of all the elements of the \fxy\ plane is the same as   $\Delta(y)$
 excluding the null elements such that
\beq
 a^2+b^2\Delta(x)&=&0.
\label{e3001}
\eeq
For example, any element of the form
\beq
z_\pm\propto&&\sqrt{\mid\Delta(x)\mid} y \pm \Upsilon_x(y)
\label{ee00525}
\eeq
is null, $\Delta(z_\pm)=0$   (for $\Delta(x)<0, y\in \fxt$).
We observe that the null elements of $\FTS_x$ and any \fxy\ are aligned, they are given by the same \autoref{e3001} which
it is independent of $y$.


%
%
%


\section{Freudenthal rotations: The $\Upsilon$ map and its exponential}
\label{sec5}
\label{sec:5}

The closure of $\fx$ under F-duality implies the existence of a one-parameter family of U-duality transformations  
stabilising $\fx$ as it will be shown in this section.

Recall, for any fixed $x$,   $\Upsilon_x$  is in
 $\autFF$.
In particular, the
 normalised version,  $\overline \Upsilon_x$,  given
in   \autoref{eq:defY}, acting on $x$ itself
maps $x$ into its F-dual,
\be
\overline{\Upsilon}_x(x)=\epsilon \tx.
\ee
Note, we also have the relation ($\epsilon=\sgn \Delta(x)$)
\begin{eqnarray}
\Upsilon_{\tx} (x)&=&3 T(\tx,\tx,x)+\{\tx,x\}\tx \\
       &=&\frac{1}{\mid \Delta(x)\mid }\left (3 T(x',x',x)+\{x',x\}x' \right ), \\
       &=&\frac{1}{\mid\Delta(x)\mid}\left (\Delta(x) x'+2\Delta(x)x' \right )
 = 3\epsilon x',
\end{eqnarray}
then
\begin{eqnarray}
\overline{\Upsilon}_{\tx} (x)&=&\frac{\epsilon}{\sqrt{\Delta(\tx)|}} x'
                                           =\tx.
\end{eqnarray}
It is obvious that  $\Upsilon_x$ (as well as $\Upsilon_{a x+b \tilde{x}}$) is a $\FTS_x\to\FTS_x$ map.
Furthermore, the set of maps
$\{\Upsilon_{a x+b \tilde{x}}\}_{a,b\in \R}$ for a fixed $x$ forms an  two-parametric automorphism subalgebra.

We are interested here in the action of the $\Upsilon_x$ map and the computation of  its exponential.
For this purpose, it is convenient to distinguish the action of any
 $\Upsilon_x$ on its particular associated $\FTS_x$ plane and on
the respective orthogonal complement $\FTS_x^\perp$
\footnote{For convenience, we work indistinctly on the $\FTS_x,\FTS_x^\perp$ or
on $\FT_x,\FT_x^\perp$. They are equivalent as long $\Delta(x)\neq 0$.}
.

\subsection{The exponential map on the $\FTS_x$-plane}\label{f-rot}

The action of the exponential of the (normalized)  map $\overline{\Upsilon}_x$ reads as follows.
For any rank-4, fixed, $x\in \mathfrak{F}$, ($\theta\in \F$, $\epsilon(x)= \sgn \Delta(x)$)
\begin{eqnarray}
\exp \left( \theta \bar{\Upsilon}_{x}\right) (x) &=&\cos \left( \sqrt{%
\varepsilon }\theta \right) x+\sqrt{\varepsilon }\sin \left( \sqrt{%
\varepsilon }\theta \right) \tilde{x}, \label{eq:Frot1}\\
\exp \left( \theta \bar{\Upsilon}_{x}\right) (\tilde{x}) &=&-\varepsilon
\sqrt{\varepsilon }\sin \left( \sqrt{\varepsilon }\theta \right) x+\cos
\left( \sqrt{\varepsilon }\theta \right) \tilde{x}, \label{ee115}\\
\exp \left( \theta \bar{\Upsilon}_{\tilde{x}}\right) (x) &=&\cos \left(
\sqrt{\varepsilon }\theta \right) x+\varepsilon \sqrt{\varepsilon }\sin
\left( \sqrt{\varepsilon }\theta \right) \tilde{x}, \\
\exp \left( \theta \bar{\Upsilon}_{\tilde{x}}\right) (\tilde{x}) &=&-\sqrt{%
\varepsilon }\sin \left( \sqrt{\varepsilon }\theta \right) x+\cos \left(
\sqrt{\varepsilon }\theta \right) \tilde{x}, \label{ee115-bis}
\end{eqnarray}%
where $x\in \mathfrak{F}$ and
$\exp$ is defined by the usual infinite series.

The proof of \autoref{eq:Frot1} - \autoref{ee115-bis} is based in the following properties:
\begin{eqnarray}
\Upsilon _{x}^{2n}(x) &=&(-1)^{n}(3)^{2n}\Delta(x)^n x,\label{Y-1} \\
\Upsilon _{x}^{2n+1}(x) &=&(-1)^{n}(3)^{2n+1}\Delta(x)^n \ x',\label{Y-2} \\
\Upsilon _{\tilde{x}}^{2n}(x) &=&\Upsilon _{x}^{2n}(x), \\
\Upsilon _{\tilde{x}}^{2n+1}(x) &=&\varepsilon \Upsilon _{x}^{2n+1}(x),
\end{eqnarray}
which are obtained by induction starting on with
$\Upsilon_x(x)=3 x'$, $\Upsilon_x^2(x)=9T(x,x,x')=-3\Delta (x)x$.
By linearity $\exp \theta \Upsilon_x$ can be extended to the full $\FTS_x$ plane.

Explicitly, for $\Delta _{x}=\Delta _{\tilde{x}}>0\Leftrightarrow
\varepsilon =1$, it holds that%
\begin{eqnarray}
\exp \left( \theta \bar{\Upsilon}_{x}\right) (x) &=&\cos \left( \theta
\right) x+\sin \left( \theta \right) \tilde{x}, \\
\exp \left( \theta \bar{\Upsilon}_{x}\right) (\tilde{x}) &=&-\sin \left(
\theta \right) x+\cos \left( \theta \right) \tilde{x}, \\
\exp \left( \theta \bar{\Upsilon}_{\tilde{x}}\right) (x) &=&\cos \left(
\theta \right) x+\sin \left( \theta \right) \tilde{x}=\exp \left( \theta
\bar{\Upsilon}_{x}\right) (x), \\
\exp \left( \theta \bar{\Upsilon}_{\tilde{x}}\right) (\tilde{x}) &=&-\sin
\left( \theta \right) x+\cos \left( \theta \right) \tilde{x}=\exp \left(
\theta \bar{\Upsilon}_{x}\right) (\tilde{x}),
\end{eqnarray}%
whereas for $\Delta _{x}=\Delta _{\tilde{x}}<0\Leftrightarrow \varepsilon =-1
$, it holds that%
\begin{eqnarray}
\exp \left( \theta \bar{\Upsilon}_{x}\right) (x) &=&\cosh \left( \theta
\right) x-\sinh \left( \theta \right) \tilde{x}, \\
\exp \left( \theta \bar{\Upsilon}_{x}\right) (\tilde{x}) &=&-\sinh \left(
\theta \right) x+\cosh \left( \theta \right) \tilde{x}, \\
\exp \left( \theta \bar{\Upsilon}_{\tilde{x}}\right) (x) &=&\cosh \left(
\theta \right) x+\sinh \left( \theta \right) \tilde{x}, \\
\exp \left( \theta \bar{\Upsilon}_{\tilde{x}}\right) (\tilde{x}) &=&\sinh
\left( \theta \right) x+\cosh \left( \theta \right) \tilde{x}.
\end{eqnarray}
Hence, the set of transformations $\exp \theta \Upsilon_x$ form an
automorphism subgroup $\Aut(\FTS_x)\subseteq \Aut(\FTS)$ preserving the $\FTS_x$ plane.

To summarise, as a consequence
of \autoref{eq:Frot1} and \autoref{ee115},  for any rank-4 $x\in \mathfrak{F}$,
there exists a monoparametric subgroup
$\sigma_x(\theta)  \in \Aut(\FTS_x)$ which is made
of  \textit{\textquotedblleft rotations"} in
$\mathfrak{F}_{x}$ and whose generator is $\Upsilon_x$:
$$\sigma_x(\theta) \equiv e^{\theta\overline{\Upsilon }_{x}}.$$



Let us study the details of the automorphism subgroup $\sigma_x(\theta)$ depending on the sign of $\Delta(x)$.
For $\Delta(x)>0, (\epsilon=1)$ the subgroup $\sigma_x(\theta)$ is $SO(2)$. The Freudenthal rotation with $\theta=\pi/2$
is the U-duality transformation relating $x$ to its F-dual.
For $\mathcal{N}=8$ black holes with $G_4=E_{7(7)}$ the
existence of  a U-duality  connecting $x$ and $\tx$ was
guaranteed since all $x$ with  the same $\Delta(x)>0$ belong
to the same $E_{7(7)}$ orbit. For $\mathcal{N}<8$  not all $x$
with  the same $\Delta(x)>0$ necessarily lie in the same
U-duality orbit; the orbits are split by further U-duality invariant
conditions. See \cite{Borsten:2011nq, Borsten:2011ai} and the references
therein. Nonetheless, for $\epsilon=1$ \label{eq:rot1} the Freudenthal rotation given by
\autoref{eq:Frot1} with $\theta=\pi/2$ implies that  $x$ and $\tx$ are  in the same U-duality orbits for all FTS.

  On the other hand, for $\Delta(x)<0, (\epsilon=-1)$ the subgroup $\sigma_x(\theta)$ is
  $SO_0(1,1)$ which has  three different kinds of orbits: the origin (a group fixed point), the four rays $\{(\pm t,\pm t),t>0\}$,
and the hyperbolae  $a^2-b^2=\pm r^2$.
  The  Freudenthal rotation cannot relate $x$ to its F-dual
(by  inspection of  \autoref{ee115}, the orbits of the exponential
of the $\Upsilon$ are hyperbolic). Therefore $x,\tilde{x}$ lie in different branches.
However, for any FTS, all $x$ with the same $\Delta(x)<0$ lie in the
same  U-duality orbit  \cite{Borsten:2011nq, Borsten:2011ai}.
Hence, there exists a U-duality transformation, which is determined by $x$, connecting $x$ and $\tx$ for $\Delta(x)<0$.
But, this U-duality transformation is not represented by any member of $\sigma_x(\theta)$.
In fact, as we shall see there is a one-parameter family of U-dualities which connects  $x$ and $\tilde{x}$ but  does not preserve the $\FTS_x$-plane.
We will return to this question in  the next sections.

In summary, putting together the previous comments, we arrive to the
conclusion that

\begin{itemize}
\item[a)] For all supergravities with $E_7$-type duality group of any $\N$, large BH have charges $x$ and $\tx$   in the same U-duality orbit,
irrespective of the sign of $\Delta(x)$.
\item[b)] For $\Delta(x)>0$ the orbit of $\sigma_x(\theta)$, which relates  the F-dual BHs, is
contained in the $\FTS_x$-plane. For $\Delta(x)<0$, the orbit of of the one-parameter subgroup, introduced later, connecting $x$ and $\tilde{x}$ does not preserve  the $\FTS_x$-plane. It would perhaps  be ``natural''  to conjecture that this orbit only intersects the $\FTS_x$-plane only at $x$ and $\tilde{x}$. We will come back to this point later on.
\end{itemize}

Note, a similar treatment can be performed   for the case of small BHs, $\Delta(x)=0$. In this case
 the group generated by $\Upsilon_x$  has orbits corresponding  to  null rays.

\section{Pure $\mathcal{N}=2,D=4$ supergravity and degenerate FTS}
\label{sec:pure}

The simplest example of a FTS (which is, being two-dimensional, a Freudenthal plane
with $\Delta \left( x\right) >0$) in supergravity
is provided by the one associated to \textquotedblleft pure" $\mathcal{N}=2$%
, $D=4$ supergravity, whose purely bosonic sector is the simplest
(scalarless) instance of Maxwell-Einstein gravity. In such a theory, the
asymptotically flat, spherically symmetric, dyonic extremal Reissner-Nordstr%
\"{o}m (RN) black hole (BH) solution has Bekenstein-Hawking entropy%
\begin{equation}
\frac{S_{RN}}{\pi }=\frac{1}{2}\left( p^{2}+q^{2}\right) ,
\end{equation}%
where $p$ and $q$ are the magnetic resp. electric fluxes associated to the
unique Abelian vector field (which, in the $\mathcal{N}=2$ supersymmetric
interpretation, is the so-called graviphoton).

In this case, the associated  FTS $\mathfrak{F}_{\mathcal{N}=2~\text{\textquotedblleft
pure"}}$ has dim$=2$ (\textit{i.e.}, it has $N=1$, within the previous
treatment); it is immediate to realize that this \textit{cannot} be a
reduced FTS, because\footnote{%
The case $\Delta =0$ corresponds to the uncharged limit $p=0=q$.}%
defining
\beq
x&=&\left( p,q\right) ^{T}\
\eeq
then the associated quartic invariant $\Delta(x)$ is defined by
\begin{equation}
\Delta(x) =\frac{1}{4}\left( p^{2}+q^{2}\right) ^{2}=\frac{S_{RN}^{2}}{\pi ^{2}%
}\geqslant 0
\label{deg}
\end{equation}%
for any choice of $p$ and $q$.

This system can be considered a
BPS ($\Delta(x)>0$)  prototype.
Let us start by doing some, simple, explicit computations. For this purpose
let us  choose (without any loss of generality) a vector given by
\begin{equation}
x=\left( p,0\right) ^{T},
\end{equation}
which corresponds  to a purely magnetic extremal RN BH.
For this configuration,
\beq
 \Delta \left( x\right) &=&\frac{1}{4}p^{4},\\
 \frac{S_{RN}\left( x\right) }{\pi }&=&\frac{1}{2}p^{2}.
\label{late-2}
\eeq%
 Introducing a basis\footnote{For $\dim \FTS=2N$, Latin capital indices are symplectic, and take values over $0,1,...,N-1$
contravariant and $0,1,...,N-1$ covariant indices.} $\{e_M\}_{\M=1}^{\dim \FTS}$, the Freudenthal dual
 $\widetilde{x}$ of $x$ can be computed \cite{Duff-FD, FMY-FD}
 by using
 \begin{equation}
\widetilde{x}^{M}=\Omega ^{MN}\frac{1}{\pi}\frac{\partial S(x)}{\partial x^{N}}=\Omega ^{MN}\frac{\partial \sqrt{\left\vert \Delta (x)\right\vert }}{\partial x^{N}}=\frac{\epsilon }{2\sqrt{\epsilon \Delta(x)%
}}\Omega ^{MN}\frac{\partial \Delta\left( x\right) }{\partial x^{N}},
\label{ee00122}
 \end{equation}
 where we recall that $\epsilon \equiv \sgn \Delta(x)$. Note, we have introduce here   the $\dim \FTS \times \dim \FTS=2N\times 2N$ symplectic matrix $\Omega$, defined by \begin{equation}
\left\{ x,y\right\}\equiv x^T \Omega y  
\label{sympl-prod-z}
\end{equation}
For a generic FTS,  we can choose  a basis such that $\Omega$ is realized as follows:
\begin{equation}
\Omega =\left (\Omega^{MN}\right ) =\left(
\begin{array}{cc}
\mathbf{0} & -\mathds{1} \\
\mathds{1} & \mathbf{0}%
\end{array}%
\right) ,  \label{Omega-z}
\end{equation}%
where $\mathbf{0}$ and $\mathds{1}$ denote the $N\times N$ zero and identity
matrices, respectively.

In the present case 
\be
x^T \Omega y  =  qp'-pq', \qquad x=(p,q)^T, y=(p', q')^T
\ee
and from \autoref{deg} we find%
\begin{equation}
\widetilde{x}\equiv (\tilde p, \tilde q)^T=(0, p)^{T}.
 \label{late-3}
\end{equation}%

A purely electric extremal RN BH is nothing else as the Freudenthal dual of purely magnetic
extremal RN BH.
The whole FTS $%
\mathfrak{F}_{\mathcal{N}=2~\text{\textquotedblleft pure"}}$ \textit{%
coincides} with the Freudenthal plane $\mathfrak{F}_{x}$ associated to $x$ :%
\begin{equation}
\mathfrak{F}_{\mathcal{N}=2~\text{\textquotedblleft pure"}}=\mathfrak{F}%
_{x}=\left\{ ax+b\widetilde{x},~a,b\in  \R %
\right\} \
\end{equation}
and transverse space is obviously empty
$
\mathfrak{F}_{x}^{\perp }=\varnothing.
$
A general Freudenthal transformation (GFT) depending on the real
parameters $a,b$ is given by
\beq
x\to x_F&=& \ ax+b\widetilde{x}
\eeq
or, in this case
\begin{equation}
x_F=(q_F,p_F)^T=(a p ,b p)^{T},
\end{equation}%
with
\beq
\Delta(x_F)\equiv \Delta \left( ax+b\widetilde{x}\right)
&=&\frac{1}{4}\left(a^{2}p^{2}+b^{2}q^{2}\right) ^{2}\\
&=&\frac{1}{4}\left(a^{2}+b^{2}\right ) p^{4} \geqslant 0.
\eeq%
The corresponding extremal RN BH is supersymmetric and $\frac{1}{2}$-BPS (in
absence of scalar fields, supersymmetry implies extremality). For $a^2+b^2=1$,
the general Freudenthal transformation leaves invariant the entropy of the black hole. In this context
a GFT is nothing else as an instance of EM duality.
Automorphism algebra and group element as $\Upsilon $ and $ \exp (\theta \Upsilon)$ can be explicitly and easily
computed. $\Upsilon_x( a x+b \txx)\propto \widetilde{ a x+b \txx}$.

$\mathfrak{F}_{\mathcal{N}=2~\text{\textquotedblleft pure"}}$ provides the
simplest case of \textit{degenerate} FTS
, in which $2\Delta $ is the square of a quadratic
polynomial $I_{2}$ :%
\begin{equation}
\Delta =I_{2}^{2},  \label{degeneracy}
\end{equation}%
and thus it is always positive. In fact, \textquotedblleft pure" $\mathcal{N}%
=2$, $D=4$ supergravity is the $n_{V}=0$ limit of the sequence of $\mathcal{N%
}=2$, $D=4$ supergravity \textit{\textquotedblleft minimally coupled"} to $%
n_{V}$ vector multiplets\footnote{%
Actually, such a sequence is the unique, \textit{at least} among theories
with homogeneous scalar manifolds, to admit the \textquotedblleft pure"
theory as the limit of $n_{V}=0$ vector multiplets.} \cite{Luciani} (see
also \cite{Gnecchi-1, FMO-mc}), in which the related FTS is degenerate $%
\forall n_{V}\in \mathds{N}\cup \left\{ 0\right\} $; the corresponding
scalar manifold is $\overline{\mathds{CP}}^{n_{V}}$.

In the formalism discussed in \autoref{sec3}, in $\mathfrak{F}_{%
\mathcal{N}=2~\text{\textquotedblleft pure"}}=\mathfrak{F}_{x}$ it holds
that ($I,J=1,2$)
\begin{equation}
I_{2}=\frac{1}{2}\left\Vert \mathbf{x}\right\Vert _{\delta }^{2}=\frac{1}{2}%
\delta _{IJ}x^{I}x^{J},
\end{equation}%
with $x^{1}=x$, $x^{2}=\widetilde{x}$. The Euclidean nature of the metric
structure defined on $\mathfrak{F}_{\mathcal{N}=2~\text{\textquotedblleft
pure"}}=\mathfrak{F}_{x}$ corresponds to a \textit{spherical foliation} of $\mathfrak{F}_{x}$ for $\Delta >0$.

Degenerate FTS's \textit{never} satisfy the reducibility condition \cite%
{Brown:1969}, namely they are globally non-reduced; they have been treated
\textit{e.g.} in \cite{Deg-Garibaldi}, and their application in supergravity
has been discussed in \cite{FKM-Deg} (see also \label{Gnecchi-1} and \cite%
{FMO-mc}). Other (infinite) examples of degenerate FTS's are provided by the
ones related to the $n$-parameterised sequence of $\mathcal{N}=3$, $D=4$
supergravity coupled to $n$ matter (vector) multiplets \cite{N=3,Gnecchi-1,
FKM-Deg}.
On the other hand, $\mathcal{N}=4$ \cite{N=4-pure}and $\mathcal{N}%
=5$
\footnote{A particularly interesting case is provided by $\mathcal{N}=5$, $D=4$
supergravity \cite{N=5}, which is seemingly related to a non-reduced FTS
which is non-degenerate, but also to a triple system denoted by $M_{2,1}(%
\mathds{O})\sim M_{1,2}(\mathds{O})$ \cite{GST, Exc-Reds, ADFMT-1} which deserves a particular study.},
$D=4$ \textquotedblleft pure" supergravities
have FTS's which do not satisfy the degeneracy condition (\autoref{degeneracy})
in all symplectic frames, but rather (\autoref{degeneracy}) is satisfied \textit{%
at least} in the so-called \textquotedblleft
scalar-dressed\textquotedblright\ symplectic frame \cite{Gnecchi-1}.

This FTS cannot be associated to any Jordan Algebra.
Consistently, \textquotedblleft
pure" $\mathcal{N}=2$, $D=4$ supergravity does \textit{not} admit an uplift
to $D=5$, or conversely it cannot be obtained by dimensionally reducing any $%
D=5$ theory down to $D=4$.
In general, degenerate FTS's are \textit{not} built starting from rank-$3$
Jordan algebras, and therefore the corresponding Maxwell-Einstein
(super)gravity models do \textit{not} admit an uplift to $D=5$; rather,
degenerate FTS's are based on Hermitian (Jordan) triple systems (\textit{%
cfr. e.g.} \cite{GST-2, Deg-Garibaldi}, and Refs. therein).

As discussed in Sec. 10 of \cite{FKM-Deg}, \textit{at least} for the
degenerate FTS's relevant to $D=4$ supergravities with symmetric scalar
manifold (\textit{i.e.}, $\mathcal{N}=2$ \textquotedblleft minimally
coupled\textquotedblright\ and $\mathcal{N}=3$ theories\footnote{%
These cases pertain to simple, degenerate FTS's \cite{FKM-Deg}. No examples
of semi-simple or non-semi-simple degenerate FTS's relevant to
(super)gravity ($D=4$) models are known to us.}), Freudenthal duality is
nothing but an anti-involutive U-duality mapping. This can be realized
immediately in the aforementioned case of $\mathcal{N}=2$, $D=4$
\textquotedblleft pure" supergravity; let us  consider ($a=b=1$)%
\begin{equation}
\mathfrak{F}_{\mathcal{N}=2~\text{\textquotedblleft pure"}}\ni y=\left(
p,q\right) ^{T}\Rightarrow \Delta \left( x+\widetilde{x}\right) =\frac{1}{2}%
\left( p^{2}+q^{2}\right) ^{2}.
\end{equation}%
The Freudenthal dual $\widetilde{y}$ of $y$ can be computed (by recalling (\autoref{ee00122}) an using (\autoref{deg})) to read%
\begin{equation}
\widetilde{y}=(-q,p)^{T}=\Omega_0 y,  \label{late-1}
\end{equation}%
where $\Omega_0$ is nothing but the canonical symplectic $2\times 2$ metric $%
\Omega _{2\times 2}$ :%
\begin{equation}
\Omega_0\equiv \left(
\begin{array}{cc}
0 & -1 \\
1 & 0%
\end{array}%
\right) \equiv \Omega _{2\times 2}.  \label{epsilon!}
\end{equation}%
Thus,
Freudenthal duality in $\mathfrak{F}_{\mathcal{N}=2~\text{\textquotedblleft
pure"}}$ is given by the application of the symplectic
metric $\Omega \equiv \Omega_0 $,
and it is thus an \textit{anti-involutive }$U$\textit{%
-duality transformation}.
The relation (\autoref{late-1}) defines a $\mathds{Z}_{4}$
symmetry in the $2$-dim. FTS $\mathfrak{F}_{\mathcal{N}=2~\text{%
\textquotedblleft pure"}}=\mathfrak{F}_{x}$, spanned by $x$ (\autoref{late-2})
and its Freudenthal dual $\widetilde{x}$ (\autoref{late-3}) : in fact, the
iteration of Freudenthal duality yields%
\begin{equation}
\left( p,q\right) ^{T}\overset{\sim }{\rightarrow }\left( -q,p\right) ^{T}%
\overset{\sim }{\rightarrow }-\left( p,q\right) ^{T}\overset{\sim }{%
\rightarrow }\left( q,-p\right) ^{T}\overset{\sim }{\rightarrow }\left(
p,q\right) ^{T}.
\end{equation}%
This provides the realisation  of the $\mathds{Z}_{4}$ in the FTS $\mathfrak{F%
}_{\mathcal{N}=2~\text{\textquotedblleft pure"}}=\mathfrak{F}_{x}$, as a
consequence of the anti-involutivity of Freudenthal duality itself.
The same symmetry will be also  explicitly observed, for example, for the
Freudenthal plane defined by the $D0-D6$ brane charge configuration in
reduced FTS's, to be studied in latter sections.

\section{ The axion-dilaton $\mathcal{N}=2,D=4$ supergravity}
\label{sec:axion}
%

%

Let us consider now
 $\mathcal{N}=2$, $D=4$ supergravity
\textquotedblleft minimally coupled" to one vector multiplet, in the
so-called axion-dilaton (denoted by the subscript \textquotedblleft $ad$")
symplectic frame. Ultimately, this is nothing but the $n_{V}=1$ element of
the sequence of $\overline{\mathds{CP}}^{n_{V}}$ \textquotedblleft minimally
coupled" models \cite{Luciani}
, but in a particular
symplectic frame, which can be obtained as a consistent truncation of
\textquotedblleft pure" $\mathcal{N}=4$ supergravity, in which only two of
the six graviphoton survive (in this frame, the holomorphic prepotential
reads $F(X)=-iX^{0}X^{1}$; \textit{cfr. e.g.} the discussion in \cite%
{Hayakawa}, and Refs. therein).

The purely bosonic sector of such an $\mathcal{N}=2$ theory may be regarded
as the simplest instance of Maxwell-Einstein gravity coupled to one complex
scalar field. In the axion-dilaton symplectic frame, in the particular
charge configuration obtained by setting to zero two charges out of four and
thus having only two non-vanishing charges\footnote{%
In this case, the effective
FTS $\mathfrak{F}_{\mathcal{N}=2~ad}$ given by  the truncation
has dimension $2$.},
namely one magnetic and one electric charge $p$ resp. $q$, the
asymptotically flat, spherically symmetric, dyonic extremal BH solution has
Bekenstein-Hawking entropy%
\begin{equation}
\frac{S_{ad}}{\pi }=\left\vert pq\right\vert ,  \label{ad}
\end{equation}%
and it is non-supersymmetric\footnote{%
Indeed, in presence of scalar fields (in this context stabilized at the
event horizon of the BH by virtue of the attractor mechanism), extremality
does not imply BPS nature, and extremal non-BPS solutions may exist.}
(non-BPS). The expression (\autoref{ad}) is very reminiscent of the
Bekenstein-Hawking entropy of a BH in a reduced FTS in the $D0-D6$ charge
configuration (to be treated later on, we refer to  (\autoref{I4-1})-(\autoref{I4-2}), $I_4\equiv  \Delta(x)$):%
\begin{equation}
\frac{S_{BH}}{\pi }=\sqrt{\left\vert \Delta \right\vert }=\left\vert p^{0}q_{0}\right\vert .
 \label{BH-entropy}
\end{equation}%
However, the $\mathcal{N}=2$ axion-dilaton supergravity model, as the
\textquotedblleft pure" $\mathcal{N}=4$, $D=4$ supergravity from which it
derives, \textit{cannot} be uplifted to $D=5$ (as instead all models related
to reduced FTS's can), consistently with its \textquotedblleft minimally
coupled" nature : in fact, the charges $P$ and $Q$ do \textit{not} have the
interpretation of the magnetic resp. electric charge of the KK vector in the
$D=5\rightarrow 4$ dimensional reduction.


This truncated system can be described by a two dimensional FTS characterized by
a quartic form ($x\equiv(p,q)^T$)
\beq
 \Delta(x)&=&-\frac{1}{2} \left( p q\right)^2.
\eeq
One can recast this expression by defining%
\begin{equation}
p\equiv \frac{1}{\sqrt{2}}\left( P+Q\right) ,~~q\equiv \frac{1}{\sqrt{2}}\left(
P-Q\right) ,  \label{redef}
\end{equation}%
in the following form
\begin{equation}
\frac{S_{ad}}{\pi }=\frac{1}{2}\left\vert P^{2}-Q^{2}\right\vert ,
\end{equation}
Let us start by choosing, without any loss of generality, a  charge configuration given by%
\begin{equation}
x=\left( P,0\right) ^{T}.
\label{ee00899}
\end{equation}
The corresponding entropy is given by
\beq
 \frac{S_{ad}\left( x\right) }{\pi }&=&\frac{1}{2}P^{2}.
  \label{jzz}
\eeq%
By virtue of \autoref{ee00122}, one can compute
 the Freudenthal dual $\widetilde{x}$ of $x$ to read
($\epsilon\equiv \sgn\left( P^{2}-Q^{2}\right)$%
\begin{equation}
\widetilde{x}=(\tilde P, \tilde Q)=(0,\epsilon P)^{T}.
 \label{jzz-2}
\end{equation}%
Thus, one can define a GFT transformations and the $2$-dim. Freudenthal plane $\mathfrak{F}_{x}$
associated to $x$ inside the whole $4$-dim. FTS $\mathfrak{F}_{\mathcal{N}%
=2~ad}$ :%
\begin{equation}
\mathfrak{F}_{\mathcal{N}=2~ad}\supset \mathfrak{F}_{x}=\left\{x_F\equiv  ax+b%
\widetilde{x},~a,b\in \R \right\} ,
\end{equation}%
with%
\begin{equation}
\frac{S_{ad}\left( x_F \right) }{\pi }=\frac{1}{2}\left\vert a^{2}P^{2}-b^{2}Q^{2}\right\vert .
\end{equation}%
The corresponding extremal BH is non-supersymmetric (non-BPS).
In particular  for the \autoref{ee00899} configuration
\begin{equation}
\frac{S_{ad}\left( x_F \right) }{\pi }=\frac{1}{2}\left\vert
a^{2}-Q^{2}\right\vert \frac{S_{ad}\left( x \right) }{\pi }.
\end{equation}%
The entropy is invariant for $a^2-b^2=\pm 1$.

Within the formalism discussed in \autoref{sec3}, in $\mathfrak{F}%
_{x}\subset \mathfrak{F}_{\mathcal{N}=2~ad}$, it holds that
\begin{equation}
I_{2}=\frac{1}{2}\left\Vert \mathbf{x}\right\Vert _{\eta }^{2}=\frac{1}{2}%
\eta _{IJ}x^{I}x^{J},
\end{equation}%
with $x^{1}=x$, $x^{2}=\widetilde{x}$, and $\eta _{IJ}=diag\left(
1,-1\right) $. The Kleinian nature of the metric structure defined on $%
\mathfrak{F}_{x}\subset \mathfrak{F}_{\mathcal{N}=2~ad}$ corresponds to an
\textit{hyperbolic} (\textit{i.e.}, \textit{hyperboloid-like}) \textit{%
foliation} of $\mathfrak{F}_{x}$ for $\Delta <0$. Therefore, notwithstanding the fact that $\mathcal{N}%
=2 $, $D=4$ axion-dilaton supergravity is nothing but the $\overline{\mathds{%
CP}}^{1}$ \textquotedblleft minimally coupled" model in a particular
(non-Fubini-Study) symplectic frame and thus with (\autoref{degeneracy}) holding
true, in the \textit{peculiar} $\left( P,Q\right) $ charge configuration (%
\autoref{redef}), the corresponding $\mathfrak{F}_{x}\subset \mathfrak{F}_{%
\mathcal{N}=2~ad}$ can be considered as a \textit{\textquotedblleft
degenerate" limit} of the $\Delta <0$ prototype of Freudenthal plane
for reduced FTS's.

It is instructive to consider the explicit action of the Freudenthal duality
in the Freudenthal plane $\mathfrak{F}_{x}\subset \mathfrak{F}_{\mathcal{N}%
=2~ad}$. Let us start and consider ($a=b=1$; we disregard the coordinates in
$\mathfrak{F}_{\mathcal{N}=2~ad}$ pertaining to $\mathfrak{F}_{x}^{\perp }=%
\mathfrak{F}_{\mathcal{N}=2~ad}/\mathfrak{F}_{x}$)%
\begin{equation}
\mathfrak{F}_{\mathcal{N}=2~ad}\supset \mathfrak{F}_{x}\ni y=\left(
P,Q\right) ^{T}\Rightarrow \frac{S_{ad}\left( y\right) }{\pi }=\frac{1}{2}%
\left\vert P^{2}-Q^{2}\right\vert .
\end{equation}%
The Freudenthal dual $\widetilde{y}$ of $y$ can be computed
(by recalling (\autoref{ee00122})) to read\footnote{%
By virtue of the discussion made at the end of th previous Subsection (also
\textit{cfr.} Sec. 10 of \cite{FKM-Deg}), $\mathbf{O}$ (\autoref{jz-3}) can be
completed to a $4\times 4$ (consistently anti-involutive; \textit{cfr.}
discussion further below) transformation of the U-duality group $U(1,1)$.}%
\begin{eqnarray}
\widetilde{y} &=&\epsilon (Q,P)^{T}=\epsilon \mathbf{\hat{O}}y,  \label{jz-2}
\eeq
with
\beq
\mathbf{\hat{O}} &\mathbf{:}&=\left(
\begin{array}{cc}
0 & 1 \\
1 & 0%
\end{array}%
\right) .  \label{jz-3}
\end{eqnarray}%
Note that $\mathbf{\hat{O}}$ (\autoref{jz-3}) is involutive:%
\begin{equation}
\mathbf{\hat{O}}^{2}=Id,
\end{equation}%
but since the Freudenthal duality on $\mathfrak{F}_{x}$ exchanges $P$ and $Q$
and thus flips $\epsilon (=\sgn\left( P^{2}-Q^{2}\right)) $, it follows that the correct
iteration of the Freudenthal duality on $\mathfrak{F}_{x}\subset \mathfrak{F}%
_{\mathcal{N}=2~ad}$ is provided by the application of
$\epsilon\mathbf{\hat{O}}$ and then necessarily of
$-\epsilon \mathbf{\hat{O}}$, thus corresponding to $-\mathbf{\hat{O}%
}^{2}=-Id$ acting on $x$, and thus correctly yielding
$$\widetilde{\widetilde{x}}=-x.$$

As at the end of previous Subsection for \textquotedblleft pure" $\mathcal{N}%
=2$, $D=4$ supergravity, in this case
due to the relations
 (\autoref{jz-2})-(\autoref{jz-3}), we can  define
a $\mathds{Z}_{4}$ symmetry in the $2$-dim. Freudenthal plane $\mathfrak{F}%
_{x}\subset \mathfrak{F}_{\mathcal{N}=2~ad}$, spanned by $x$ (\autoref{jzz}) and
its Freudenthal dual $\widetilde{x}$ (\autoref{jzz-2}) : \textit{e.g.}, starting
from $\epsilon =1$, the iteration of Freudenthal duality
yields%
\begin{equation}
\left( P,Q\right) ^{T}\overset{\sim }{\rightarrow }\left( Q,P\right) ^{T}%
\overset{\sim }{\rightarrow }-\left( P,Q\right) ^{T}\overset{\sim }{%
\rightarrow }-\left( Q,P\right) ^{T}\overset{\sim }{\rightarrow }\left(
P,Q\right) ^{T}.
\end{equation}%
This provides the realisation  of the $\mathds{Z}_{4}$ in the Freudenthal
plane $\mathfrak{F}_{x}\subset \mathfrak{F}_{\mathcal{N}=2~ad}$, as a
consequence of the anti-involutivity of Freudenthal duality itself.

\section{$\N=2,D=4$  supergravities from $D=5$: The reduced $\FTS$ case}
\label{secJ3}
\label{sec8}
\label{sec:reduced}


We will now proceed to present an analysis of the (non-degenerate) \textit{%
reduced} FTS's, of the properties of Freudenthal duality defined in them,
and of the corresponding Freudenthal planes.

Unless otherwise noted, we will essentially confine ourselves \textit{at
least} to (non-degenerate) \textit{reduced} FTS's $\mathfrak{F}=\mathfrak{F}%
(J_{3})$, for which a $4D/5D$ special coordinates' symplectic frame can be
defined.

A generic element $x$ of the reduced FTS $%
\mathfrak{F}$ splits as%
\begin{equation}
x=\left( x^{0},x^{i},x_{0},x_{i}\right) ^{T}\equiv\left(
p^{0},p^{i},q_{0},q_{i}\right) ^{T},  \label{split}
\end{equation}%
where the second renaming pertains to the identification of $x$ with a
dyonic charge configuration in $D=4$ (super)gravity, where $p$'s and $q$'s
are magnetic and electric charges, respectively; within the standard
convention in supergravity, $p^{0}$, $p^{i}$, $q_{0}$ and $q_{i}$ will
usually be called $D6$, $D4$, $D2$, $D0$ (brane) charges, respectively.

In the canonical basis the  symplectic product of two generic elements $x$ and $y$ in $%
\mathfrak{F}$ reads%
\begin{equation}
\left\{ x,y\right\}\equiv x^T \Omega y  =-x^{0}y_{0}-x^{i}y_{i}+x_{0}y^{0}+x_{i}y^{i}
\label{sympl-prod-z}
\end{equation}
where $\Omega$ is a symplectic matrix.

\textit{At least} within (non-degenerate) reduced FTS's, the quartic
polynomial invariant $I_{4}= 2\Delta(x) $ of  $\Aut \left( \mathfrak{F}\left( J_{3}\right)
\right) \approx Conf\left( J_{3}\right) $ can be written\footnote{%
Recall that $d_{ijk}=d_{(ijk)}$ and $d^{ijk}=d^{(ijk)}$ throughout.} as
follows\footnote{\textit{At least} in all reduced FTS's,  $\Aut \left(
\mathfrak{F}(J_{3}\right) )$ is \textquotedblleft of type $E_{7}$" \cite%
{Brown:1969}, and the ring of invariant polynomials is one-dimensional, and
finitely generated (\textit{i.e.}, with no syzygies) by $I_{4}$ \cite{Kac-80}.%
} (\textit{cfr. e.g.} \cite{Sato-Kimura,FG-1,CFM-1}; ($i=1,...,N-1$, dim$%
\mathfrak{F}=2N$)):%
\begin{eqnarray}
 \Delta(x)\equiv I_{4}\left( x\right)  &=& -\left(
p^{0}q_{0}+p^{i}q_{i}\right) ^{2}+4q_{0}I_{3}(p)-4p^{0}I_{3}(q)+4\left\{
I_{3}(p),I_{3}(q)\right\}   \label{I4-1} \\
&=&-\left( p^{0}q_{0}+p^{i}q_{i}\right) ^{2}+\frac{2}{3}%
q_{0}d_{ijk}p^{i}p^{j}p^{k}-\frac{2}{3}%
p^{0}d^{ijk}q_{i}q_{j}q_{k}+d_{ijk}d^{ilm}p^{j}p^{k}q_{l}q_{m},  \label{I4-2}
\end{eqnarray}%
where 
\begin{eqnarray}
I_{3}(p) &\equiv&\frac{1}{3!}d_{ijk}p^{i}p^{j}p^{k},\\
I_{3}(q)&\equiv&\frac{1}{3!}%
d^{ijk}q_{i}q_{j}q_{k}, \\
\left\{ I_{3}(p),I_{3}(q)\right\}  &\equiv&\frac{\partial I_{3}(p)}{\partial
p^{i}}\frac{\partial I_{3}(q)}{\partial q_{i}}.
\end{eqnarray}%

The symmetric quantities $d^{ijk}, d_{ijk}$ follow
the so-called \textit{adjoint identity} of the Jordan algebra $J_{3}$ underlying the reduced FTS $\mathfrak{F}$ (\textit{cfr. e.g.} \cite%
{Krutelevich:2004,mccrimmon,CFM-1} and Refs. therein), which reads
\begin{equation}
d_{(ij|k}d_{l|mn)}d^{klp}=\frac{4}{3}\delta _{(i}^{p}d_{jmn)}.
\label{adj-id}
\end{equation}%

The triple product map $T(x,y,z)$ reads (\textit{up to contributions} $\in $
$\mathfrak{F}_{w}^{\perp }=\mathfrak{F}/\mathfrak{F}_{w}$)%
\begin{equation}
T(x,y,z)_{M}=
\frac{\partial I_{4}\left( x,y,z,w\right) }{\partial
w^{M}}=K_{MNPQ}x^{N}y^{P}z^{Q},  \label{T}
\end{equation}%
where the capital Latin indices span the entire FTS $\mathfrak{F}$, and $%
K_{MNPQ}=K_{(MNPQ)}$ is the rank-$4$ completely symmetric tensor
characterizing $\mathfrak{F}$ \cite{Brown:1969,Exc-Reds,FKM-Deg}. Note that, from
its very definition (\autoref{T}), $T(x,y,z)$ is completely symmetric in all its
arguments \cite{Brown:1969}.

Then, by using $\Omega $ to raise the symplectic indices, one can compute%
\begin{equation}
T(x,y,z)^{M}=\Omega ^{MN}T(x,y,z)_{N}.
 \label{contra}
\end{equation}%
By
direct computation, one gets%
\beq
\frac{\partial
I_{4}\left( x\right) }{\partial p^{0}}&=&-2\left( p^{0}q_{0}+p^{i}q_{i}\right)
q_{0}-\frac{2}{3}d^{ijk}q_{i}q_{j}q_{k}; \\[0.0cm]
\frac{\partial
I_{4}\left( x\right) }{\partial p^{i}}&=&-2\left( p^{0}q_{0}+p^{j}q_{j}\right)
q_{i}+2q_{0}d_{ijk}p^{j}p^{k}+2d_{ijk}d^{klm}p^{j}q_{l}q_{m}; \\[0cm]
\frac{\partial
I_{4}\left( x\right) }{\partial q_{0}}&=&-2\left( p^{0}q_{0}+p^{i}q_{i}\right)
p^{0}+\frac{2}{3}d_{ijk}p^{i}p^{j}p^{k}; \\[0cm]
 \frac{\partial
I_{4}\left( x\right) }{\partial q_{i}}&=&-2\left( p^{0}q_{0}+p^{j}q_{j}\right)
p^{i}-2p^{0}d^{ijk}q_{j}q_{k}+d_{jkl}d^{lin}p^{j}p^{k}q_{n}.%
\eeq
In
\autoref{JJexplicit3} we present some explicit expressions for the triple product
and other maps.

\subsection{\label{sec:T^3} The $T^{3}$ Supergravity model and
 $ \mathfrak{F}(\R )$}

The so-called $T^{3}$ model of $\mathcal{N}=2$, $D=4$ supergravity
 is the smallest model in which the plane $\mathfrak{F}_{y\perp x}$
can be defined; such a model is comprised within all models
based on (non-degenerate) reduced FTS's (\textit{cfr. e.g.} (\autoref{tt1})).
In this model, it holds that ($i=1$, and $p^{1}\equiv T$)%
\begin{equation}
\frac{1}{3!}d_{ijk}p^{i}p^{j}p^{k}=T^{3}\Leftrightarrow d_{111}=6.
\end{equation}%
In the usual normalization of $d$-tensors used in supergravity literature%
\footnote{%
Which, however, is not the one used \textit{e.g.} in \cite{Exc-Reds}.}, it
holds that (\textit{cfr. e.g.} \cite{CFM-1})%
\begin{equation}
d^{111}=\frac{2}{9}.
\end{equation}

In this case we have ($N=2$, $i=1$, dim$ \mathfrak{F}=4$)):%
\begin{eqnarray}
x &=& \left ( p^0,p^1,q_0,q_1\right )^T,\\
I_{4}\left( x\right)  &=&\Delta \left( x\right)
=-\left( p^{0}q_{0}+p^{1}q_{1}\right) ^{2}
+4 q_{0}(p^{1})^3-\frac{4}{27} p^{0}(q_{1})^3+\frac{4}{3}(p^{1} q_{1})^2.
\label{ee00340}
\end{eqnarray}%
By
direct computation, one gets%
\beq
\frac{\partial
I_{4}\left( x\right) }{\partial p^{0}}&=&-2\left( p^{0}q_{0}+p^{1}q_{1}\right)
q_{0}-\frac{4}{27}q_{1}^3; \\[0.0cm]
\frac{\partial
I_{4}\left( x\right) }{\partial p^{1}}&=&-2\left( p^{0}q_{0}+p^{1}q_{1}\right)
q_{1}+\frac{4}{9}q_{0}(p^{1})^2+\frac{24}{9}p^{1}q_{1}^2; \\[0cm]
\frac{\partial
I_{4}\left( x\right) }{\partial q_{0}}&=&-2\left( p^{0}q_{0}+p^{1}q_{1}\right)
p^{0}+4(p^{1})^3; \\[0cm]
 \frac{\partial
I_{4}\left( x\right) }{\partial q_{1}}&=&-2\left( p^{0}q_{0}+p^{1}q_{1}\right)
p^{1}-\frac{4}{9}p^{0}q_{1}^2+\frac{12}{9}(p^{1})^2 q_{1}.%
\eeq
which allows to compute the dual components by \autoref{ee00122}.

Since the $T^{3}$ model pertains to the unique reduced FTS for which
$N=2$ (\textit{cfr.} \autoref{tt1}), for this model $\dim \FTS= 2 N=4$ and
 the plane $\mathfrak{F}_{y\perp x}$ coincides with the whole space $\left\{ ,\right\} $-orthogonal
to the Freudenthal plane $\mathfrak{F}_{x}$ :%
\begin{equation}
N=2\Rightarrow \mathfrak{F}_{y\perp x}=\mathfrak{F}_{x}^{\perp }.
\end{equation}%
Thus, the FTS $\mathfrak{F}_{T^{3}}\equiv \mathfrak{F}(\R )$,
sitting in the spin-$\frac{3}{2}$ irrepr. $\mathbf{4}$ of  $\Aut (\mathfrak{F}%
_{T^{3}})\approx Conf\left( J_{3}=\R \right) =SL(2,\R )$, gets
decomposed as follows :%
\begin{equation}
\mathfrak{F}_{T^{3}}\equiv \mathfrak{F}(\R )\equiv \underset{%
Conf\left( J_{3}=\R \right) =SL(2,\R )\longrightarrow
SO(1,1)_{KK}}{\mathbf{4}=\mathbf{1}_{-3}+\mathbf{1}_{-1}+\mathbf{1}_{1}+%
\mathbf{1}_{3}},  \label{ddec}
\end{equation}%
where $SO(1,1)_{KK}$ is related to the radius of the $S^{1}$ in the
dimensional reduction from minimal ($\mathcal{N}=2$) $D=5$ \textquotedblleft
pure" supergravity down to $D=4$ (giving rise to the $T^{3}$ model).


Let us start first with a particular configuration with {$\Delta (x)<0$}.
Specifying (\autoref{1-z}) and (\autoref{ee00122})-(\autoref{2}) for the $T^{3}$ model, one
has ($\epsilon= \sgn (x^0 x_0)$)
\begin{eqnarray}\label{825}
x &=&\left( x^{0},0,x_{0},0\right) ^{T}; \\
\widetilde{x} &=&
\epsilon \left( -x^{0},0,x_{0},0\right) ^{T},
\end{eqnarray}%
with%
\begin{equation}
\Delta_0\equiv \Delta (x)=\Delta (\widetilde{x})=-\left( x^{0}x_{0}\right) ^{2}<0.
\end{equation}%
Then, for a generic GFT transformation on $x$
$$x_F=ax+b\widetilde{x}\in \mathfrak{F}_{x},$$
it
holds that%
\begin{equation}
\Delta (x_F)=-\left( a^{2}-b^{2}\right) ^{2}\left( x^{0}x_{0}\right)
^{2}=\left( a^{2}-b^{2}\right) ^{2}\Delta (x)<0,
\end{equation}%
and therefore $\mathfrak{F}_{x}$ lies completely in the rank-$4$ $\Delta <0$
orbit of  $\Aut \left( \mathfrak{F}\left( J_{3}=\R \right) \right) $.

Analogously, specifying (\autoref{y}) and (\autoref{jjj}) for the $T^{3}$ model, one
obtains%
\beq
y&=&\left( 0,y^{1},0,y_{1}\right) ^{T}
\eeq
and, according to \autoref{ee00340},
\beq
\Delta (y)&=&
\frac{1}{3}\left(y^{1}y_{1}\right) ^{2}>0,
\label{jazzz}
\eeq
where the strict inequality holds, because we assume $y$ to be of maximal($=4
$) rank in $\mathfrak{F}_{T^{3}}$. Note that, while $x$ and $\widetilde{x}$
lie in the $\Delta <0$ orbit of  $\Aut (\mathfrak{F}(\R ))$, $y$
belongs to the other rank-$4$ orbit\footnote{%
As pointed out above, there is a unique $\Delta >0$ orbit in the $T^{3}$
model.}.

Starting from the decomposition (\autoref{ddec}), the Freudenthal plane $%
\mathfrak{F}_{x}$ related to $x$ (\autoref{825}) and the $\left\{ ,\right\} $%
-orthogonal plane $\mathfrak{F}_{y\perp x}=\mathfrak{F}_{x}^{\perp }$ can
respectively be identified as follows:%
\begin{eqnarray}
\mathfrak{F}_{x} &=&\underset{x^{0}}{D6}\oplus \underset{x_{0}^{~}}{D0}=%
\mathbf{1}_{-3}+\mathbf{1}_{3};  \label{jj-1} \\
\mathfrak{F}_{y\perp x} &=&\mathfrak{F}_{x}^{\perp }=\underset{y^{1}}{D4}%
\oplus \underset{y_{1}^{~}}{D2}=\mathbf{1}_{-1}+\mathbf{1}_{1}.  \label{jj-2}
\end{eqnarray}%
Nicely, within the interpretation of $SO(1,1)_{KK}$ as a non-compact
analogue of $D=4$ helicity of a would-be spin-$\frac{3}{2}$
(Rarita-Schwinger) particle, the Freudenthal plane $\mathfrak{F}_{x}$
pertains to the two massless helicity modes.

Let us recall that, while $\mathfrak{F}_{x}$ (\autoref{jj-1}) is a quadratic
sub-FTS of $\mathfrak{F}_{T^{3}}$ (as discussed in Sec. 3.4), $\mathfrak{F}%
_{y\perp x}=\mathfrak{F}_{x}^{\perp }$ (\autoref{jj-2}) is \textit{not} a
sub-FTS of $\mathfrak{F}$ (as discussed in Sec. 5). This can be explicitly
checked by relying on the treatment of \autoref{sec:closure}; in fact, for
the $T^{3}$ model, $\mathfrak{F}_{y\perp x}=\mathfrak{F}_{x}^{\perp }$ (\ref%
{jj-2}) is not closed under $T$. Out of the four cases 1-4 listed at the end
of \autoref{sec:closure}, only the last one (4) is to be considered: in
this case, the condition of closure of $\mathfrak{F}_{y\perp x}=\mathfrak{F}%
_{x}^{\perp }$ (\autoref{jj-2}) under $T$ is that $y^{1}$ is rank$<3$ in $J_{3}=%
\R $ and $y_{1}$ is rank$<3$ in $\overline{J}_{3}=\R $, namely%
\begin{equation}
y_{1}=0=y^{1}\Leftrightarrow y=0\in \mathfrak{F}.
\end{equation}%
Thus, the condition of closure of $\mathfrak{F}_{y\perp x}=\mathfrak{F}_{x}^{\perp }$
(\autoref{jj-2}) under $T$ implies, in the case of the $T^{3}$
model, an \textit{absurdum,} namely that the rank-$4$ element
$y\in\mathfrak{F}_{x}^{\perp }$ be the \textit{null element} of the FTS $%
\mathfrak{F}_{T^{3}}\equiv \mathfrak{F}(\R )$. Therefore,
$\mathfrak{F}_{y\perp x}=\mathfrak{F}_{x}^{\perp }$ (\autoref{jj-2}) is \textit{%
not} closed under $T$.

In other words, as also pointed out above, in order
for $y=\left( 0,y^{1},0,y_{1}\right)^T \in \mathfrak{F}_{x}^{\perp }$ 
 to be rank-$4$ (as assumed throughout), it must have both components
non-vanishing; from (\autoref{jazzz}) one can observe that in the $T^{3}$ model $%
y$ belongs to the rank-4, $\Delta >0$,  $\Aut (\mathfrak{F}_{T^{3}})=SL(2,\mathds{%
R})$ orbit, unless $y^{1}=0$ \textit{and/or} $y_{1}=0$, in which case it has
rank$<4$. Therefore an element $y$ of the form
$$y=\left( 0,y^{1},0,y_{1}\right)^T \in \mathfrak{F}_{x}^{\perp }$$
is rank-$4$ (and necessarily in the unique
$\Delta>0$ orbit) \textit{iff} $y^{1}\neq 0$ \textit{and} $y_{1}\neq 0$. $%
\blacksquare $

Furthermore, we are interested in the  behaviour of the quartic invariant $\Delta$ on the  $D4\oplus D2$
\fxy\ plane. General results
are presented in \autoref{secab}, in particular in \autoref{ee00527} and \autoref{ee00527b}
which can be used here. According to these results
\footnote{
We can explicitly write
 the $\Upsilon$ map (see \autoref{Ipsilon})
\beq
\left( \Upsilon _{x}(y)\right) ^{0}&=&0;
\left( \Upsilon _{x}(y)\right) ^{1}=\frac{1}{2}x^{0}x_{0}y^{1};
\left( \Upsilon _{x}(y)\right) _{0}=0;
\left( \Upsilon _{x}(y)\right) _{1}=-\frac{1}{2}x^{0}x_{0}y_{1},%
\label{Ipsilon-T3}
\eeq
}
\beq
\Delta\left(\Upsilon_x(y)\right)&=&\Delta(x)^2 \Delta(y)
 \eeq
 The sign of $\Delta\left(\Upsilon_x(y)\right)$ depends only on the sign of $\Delta(y)$
 implying that $\Upsilon _{x}(y)$ belongs to the same rank-$4$ ($\Delta >0$)
 $\Aut (\mathfrak{F}(\R ))$-orbit as $y$.
 Explicitly in this case
 \beq
\Delta\left(\Upsilon_x(y)\right)&=&\Delta(x)^2 \Delta(y)
=\frac{1}{3}\left( x^{0}x_{0}\right)
^{4}\left( y^{1}y_{1}\right) ^{2}>0
 \eeq
 For a generic element $r$
$$r=ay+b\Upsilon _{x}(y)\in \mathfrak{F}_{y\perp x}=\mathfrak{F}_{x}^{\perp } \quad   (a,b\in \R ),$$
 one gets (\autoref{ee00527} and \autoref{ee00527b})
\begin{equation}
\Delta (r)=\left( a^{2}-\left\vert \Delta (x)\right\vert b^{2}\right) ^{2}\Delta (y)\geqslant 0.
\label{jjj-1-tris}
\end{equation}%
Thus, following the general behaviour explained in
 appendix B, $r\in \mathfrak{F}_{y\perp x}$ 
is not
of the same (maximal $=4$) rank orbit as $y$ (and $\Upsilon _{x}(y)$) only
when%
\begin{equation}
\Delta (r)=0\Leftrightarrow a^{2}-\left( x^{0}x_{0}\right)
^{2}b^{2}=0\Leftrightarrow a^{2}=\left\vert \Delta (x)\right\vert%
b^{2}.
\end{equation}%


%

The conditions for
$r\in \mathfrak{F}_{y\perp x}\subset \mathfrak{F}_{x}^{\perp }$
 to lie in the rank-3, rank-2 or even rank-1 orbits
might be easily studied using expressions
\autoref{ee00527} and \autoref{ee00527b}).


Let us study now
the family of configurations with $D4-D0$ charges.
This family includes
configurations with  both {$\Delta (x)>0$} and
$\Delta (x)<0$ possibilities.
Let us take \footnote{This can be seen as an special case of
 (\autoref{1-bis}) and (\autoref{1-bis-tilde})-(\autoref{1-bis-tilde-2})
for the $T^{3}$ model.},
\begin{eqnarray}
x &=&\left( 0,x^{1},x_{0},0\right) ^{T},
\end{eqnarray}
then one obtains
\begin{equation}
\Delta (x)=\Delta (\widetilde{x})=4 x_{0}\left( x^{1}\right) ^{3}.
\end{equation}%
Thus the sign of $\Delta(x)$ equals the sign of $x_0 x^1$:
\beq
\sgn \Delta(x) &=& \sgn (x_0 x^1).
\eeq
For a positive sign,  $\sgn (x_0 x^1)>0$,  the dual is a $D6-D2$ configuration, it  reads
\begin{eqnarray}
\widetilde{x} &=&
\frac{1}{\sqrt \Delta_0} \left( -\left( x^{1}\right) ^{3}
,0,0, 3x_{0}\left( x^{1}\right) ^{2}
\right),
\end{eqnarray}%
 $x$ and $\widetilde{x}$ belong to the same (rank-$4$, $\Delta >0$)
orbit of  $\Aut ((\mathfrak{F}=J_{3}))$, which is unique in this model (\textit{%
cfr.} \cite{small-orbits}, and Refs. therein). For a generic element
$x_F=ax+b%
\widetilde{x}\in \mathfrak{F}_{x}$, it holds that
\begin{equation}
\Delta (x_F)=\left( a^{2}+b^{2}\right) ^{2}\Delta (x)>0,
\end{equation}%
implying that $\mathfrak{F}_{x}$ lies completely in the unique rank-4 $%
\Delta >0$ orbit of  $\Aut (\mathfrak{F}(\R ))$.

Then, let us  pick a   rank-$4$ element $y\in \mathfrak{F}_x^\perp$,
that means which is  $\left\{ ,\right\} $-orthogonal to $x$  
and
$\widetilde{x}$ , one can show that
  the most general element of this kind is given by the charge
configuration:
\begin{eqnarray}
y &=&\left( \frac{x^{1}}{x_{0}}y_{1},y^{1},-3\frac{x_{0}}{x^{1}}%
y^{1},y_{1}\right) ,
\end{eqnarray}
whose quartic invariant is given by
\beq
\Delta (y) &=&-\frac{8}{3}\left( y^{1}\right) ^{2}y_{1}^{2}-12\frac{x_{0}}{%
x^{1}}\left( y^{1}\right) ^{4}-\frac{4}{27}\frac{x^{1}}{x_{0}}y_{1}^{4},\\  
  &=& -\frac{4}{27( x_0 x^1)}
\left (9 x_0 (y^1)^2+  x^1 (y_1)^2 \right )^2
\eeq
thus the signs of $\Delta(x)$ and $\Delta(y)$ are opposite
\beq
\sgn \Delta(y) &=& -\sgn(x_0 x^1) =-\sgn \Delta(x).
\eeq
In the case of $(x_0 x^1)=0$ then $\FTS_x^\perp\sim D6\oplus D4$ (for $x_0=0$) and
$\FTS_x^\perp\sim D2\oplus D4$ (for $x^1=0$)
Moreover, according to
\autoref{ee00527} and \autoref{ee00527b}
\beq
\Delta\left(\Upsilon_x(y)\right)&=&\Delta(x)^2 \Delta(y).
 \eeq
 The sign of $\Delta\left(\Upsilon_x(y)\right)$ depends only on the sign of $\Delta(y)$. Both of them
 are negative in our current case.
  For a generic element $r$
$$r=ay+b\Upsilon _{x}(y)\in \mathfrak{F}_{y\perp x}=\mathfrak{F}_{x}^{\perp }$$
($a,b\in \R $), one gets (\autoref{ee00527} and \autoref{ee00527b})
\begin{equation}
\Delta (r)=\left( a^{2}+\Delta (x) b^{2}\right) ^{2}\Delta (y).
\end{equation}%
 implying that $\Upsilon _{x}(y)$ and for the case any $\Upsilon _{x}(r)$
 lies in the same (maximal rank)  $\Aut (\mathfrak{F}(\R ))$-orbit as $y$.
\footnote{ Explicitly,  from (\autoref{Ipsilon-bis}), one obtains%
\beq
\left( \Upsilon _{x}(y)\right) ^{0}&=&-6\left( x^{1}\right) ^{2}y^{1};
\left( \Upsilon _{x}(y)\right) ^{1}=\frac{2}{3}\left( x^{1}\right) ^{2}y_{1};
\left( \Upsilon _{x}(y)\right) _{0}=-2x_{0}x^{1}y_{1};
\left( \Upsilon _{x}(y)\right) _{1}=-6x_{0}x^{1}y^{1},%
\eeq
yielding
\begin{eqnarray}
\Delta \left( \Upsilon _{x}(y)\right) &=&-\frac{4}{3}x_{0}^{2}\left(
x^{1}\right) ^{6}\left( y^{1}\right) ^{2}y_{1}^{2}-6x_{0}^{3}\left(
x^{1}\right) ^{5}\left( y^{1}\right) ^{4}-\frac{2}{27}x_{0}\left(
x^{1}\right) ^{7}y_{1}^{4} .
\end{eqnarray}%
}

\subsection{
General  $D0-D6$/$D2-D4$ sectors
}
\label{sec:D0D6}


Let us consider a particular configuration with only $D0-D6$ charges with an arbitrary number
of them.
We start by identifying $x$ with the rank-$4$, strictly regular element of
the FTS $\mathfrak{F}$ given by the $D0-D6$ brane charge configuration
\begin{equation}
\xdodo\equiv \left( p^{0},0^{i},q_{0},0_{i}\right) ^{T}\in \mathfrak{F},
\label{1-z}
\end{equation}
for any element of this configuration we have
\footnote{This characterizes $\mathfrak{F}$ as a
reduced \cite{Brown:1969} FTS.}
\begin{equation}
\Delta (\xdodo)=-\left( p^{0}q_{0}\right) ^{2}<0.
 \label{jazz-sat}
\end{equation}%
One can compute the Freudenthal dual $\widetilde{x}$.
Using the expressions
\beq
\frac{\partial I_{4}\left( x\right) }{\partial p^{0}}&=&-2\left( p^{0}q_{0}\right) q_{0}, \\[0.0cm]
\frac{\partial I_{4}\left( x\right) }{\partial p^{1}}&=&0,   \\[0cm]
\frac{\partial I_{4}\left( x\right) }{\partial q_{0}}&=&-2\left( p^{0}q_{0}\right) p^{0},\\[0cm]
 \frac{\partial I_{4}\left( x\right) }{\partial q_{1}}&=& 0,
\eeq
which allows to compute the dual components using \autoref{ee00122}.
We arrive to ($\epsilon=\sgn (p^0 q_0)$)
\beq
\widetilde{\xdodo}&\equiv& \left( \widetilde{p}^{0},0^{i},\widetilde{q}_{0},0_{i}\right) ^{T}\in \mathfrak{F},
\label{1-tilde-z}\\
                              &=&\epsilon \left(-p^{0},0^{i},q_{0},0_{i}\right) .
                              \label{2}
\eeq
%

%
%
%
%
%
%
%
%
Thus, depending on the sign of $p^{0}q_{0}$
\footnote{%
The result (\autoref{2}) defines a $\mathds{Z}_{4}$ symmetry in the $2$-dim.
Freudenthal plane $\mathfrak{F}_{x}\subset \mathfrak{F}$, spanned by $%
\xdodo$ (\autoref{1-z}) and its Freudenthal dual $\widetilde{\xdodo}$ (\ref%
{2}) (or, equivalently, in the Darboux canonical basis, by the magnetic and
electric charges $p^{0}$ and $q_{0}$ of the $5D\rightarrow 4D$ KK Abelian
vector - see below - ) : \textit{e.g.}, starting from $p^{0}$ and $q_{0}$
both positive (denoted by \textquotedblleft $\left( +,+\right) $"), the
iteration of Freudenthal duality yields%
\begin{equation*}
\left( +,+\right) \overset{\sim }{\rightarrow }\left( -,+\right) \overset{%
\sim }{\rightarrow }\left( -,-\right) \overset{\sim }{\rightarrow }\left(
+,-\right) \overset{\sim }{\rightarrow }\left( +,+\right) .
\end{equation*}%
This provides a simple realisation of the $\mathds{Z}_{4}$ symmetry
characterizing every Freudenthal plane, as a consequence of the
anti-involutivity of Freudenthal duality itself.}
Note that%
\begin{equation}
\Delta (x)=\Delta (\widetilde{x})<0,
\end{equation}%
and thus $\widetilde{x}$ belongs to the same (unique) rank-$4$ $\Delta <0$
orbit of  $\Aut (\mathfrak{F}(J_{3}))$ as $x$.
Namely, when $p^{0}q_{0}>0$, the action of Freudenthal duality on $\xdodo$
amounts to flipping $p^{0}$ only, whereas when $p^{0}q_{0}<0$, the action of
Freudenthal duality on $\xdodo$ amounts to flipping $q_{0}$ only.\medskip

Associated to a GFT transformation on $x$, one defines the Freudenthal plane $\mathfrak{F}_{x}\subset \mathfrak{%
F}$ (dim$\mathfrak{F}_{x}=2$), spanned by $x$
and $\widetilde{x}$
, whose generic element is
$$x_F=ax+b\widetilde{x}\in\mathfrak{F}_{x} \quad (a,b\in \R ).$$
Within the choice above, $\mathfrak{F}_{x}$ is  coordinatized by the
charges of $D0$ and $D6$ branes,
respectively being the electric and magnetic charges $x_{0}$ and $x^{0}$ of
the KK Abelian vector in the reduction $D=5\rightarrow 4$.
In other words,within the position  (which does not imply any loss of generality
for reduced FTS's), the Freudenthal plane $\mathfrak{F}_{x}$ is spanned (in a
canonical Darboux symplectic frame - see below -) by the electric and
magnetic charges $x_{0}$ and $x^{0}$ of the $D=5\rightarrow 4$ Kaluza-Klein
Abelian vector (which is the $D=4$ graviphoton in the $\mathcal{N}=2$
supersymmetric interpretation).

Note that $\mathfrak{F}_{x}$ lies completely in the (unique) rank-$4$ $%
\Delta <0$ orbit of  $\Aut (\mathfrak{F}(J_{3}))$, because (\autoref{ee003}) %
\beq
\Delta (X_F)&=&-\left( ax^{0}+b\widetilde{x}^{0}\right) ^{2}\left( ax_{0}+b%
\widetilde{x}_{0}\right) ^{2}\\
&=&-\left( a^{2}-b^{2}\right) ^{2}\left(
x^{0}x_{0}\right) ^{2}=\left( a^{2}-b^{2}\right) ^{2}\Delta (x)<0.
\label{jazz-sat-2}
\eeq
This implies that $s$
belongs to the same maximal ($=4$) rank, $\Delta <0$  $\Aut (\mathfrak{F}%
(J_{3}))$-orbit as $x$ and $\widetilde{x}$, \textit{unless }$a^{2}=b^{2}$.
This observation actually yields interesting consequences for multi-centered
black hole physics, as briefly discussed in \autoref{sec9}.

\subsubsection*{The $D2-D4$ orthogonal space $\mathfrak{F}_{x}^{\perp }$, and the plane $%
\mathfrak{F}_{y\perp x}\subset \mathfrak{F}_{x}^{\perp }$}

One can choose a rank-$4$ element $y\in \mathfrak{F}$ which is $%
\left\{ ,\right\} $-orthogonal to the generic $D0-D6$ element $x$ defined before and its dual
 and $\widetilde{x}$.  
A  possible, particular, choice is provided by a $D2-D4$ brane charge
configuration:%
\begin{eqnarray}
y &\equiv &\left( 0,y^{i},0,y_{i}\right) ^{T}\in \mathfrak{F},  \label{y} \\
\Delta (y) &=&-\left( y^{i}y_{i}\right)
^{2}+d_{ijk}d^{ilm}y^{j}y^{k}y_{l}y_{m}\gtrless 0.  \label{jjj}
\end{eqnarray}%
In (\autoref{jjj}), the case of vanishing $\Delta $ has been excluded because $y$
is chosen to be of maximal($=4$) rank in $\mathfrak{F}$. By recalling (\ref%
{sympl-prod-z}), one can immediately check that (dim$\mathfrak{F}_{x}^{\perp
}=2N-2$)%
\begin{equation}
\left\{ x,y\right\} =0=\left\{ \widetilde{x},y\right\} \Leftrightarrow y\in
\mathfrak{F}_{x}^{\perp }\equiv \mathfrak{F}/\mathfrak{F}_{x}.
\end{equation}

One can compute the components of $%
\Upsilon _{x}(y)^{M}$ (\autoref{Ipsilon}) as given by%
\beq
\left( \Upsilon _{x}(y)\right) ^{0}&=&0; \\
\left( \Upsilon _{x}(y)\right) ^{i}&=&x^{0}x_{0}y^{i}; \\
\left( \Upsilon _{x}(y)\right) _{0}&=&0; \\
\left( \Upsilon _{x}(y)\right) _{i}&=&-x^{0}x_{0}y_{i};%
\label{Ipsilon-z}
\eeq
thus, $\Upsilon _{x}(y)$ is still given by a rank-$4$ $D2-D4$ brane charge
configuration, and it holds that%
\begin{equation}
\left\{ x,\Upsilon _{x}(y)\right\} =0=\left\{ \widetilde{x},\Upsilon
_{x}(y)\right\} \Leftrightarrow \Upsilon _{x}(y)\in \mathfrak{F}_{x}^{\perp
}\equiv \mathfrak{F}/\mathfrak{F}_{x}.  \label{jazz-2}
\end{equation}

Consequently, one can define the $2$-dim. plane $\mathfrak{F}_{y\perp
x}\subset \mathfrak{F}_{x}^{\perp }$, spanned by $y$ and $\Upsilon _{x}(y)$,
whose generic element is $r=ay+b\Upsilon _{x}(y)\in \mathfrak{F}_{y\perp x}$
($a,b\in \R $, in our classical/supergravity treatment .
In particular,
note that $\Upsilon _{x}(y)$ belongs to the same  $\Aut (\mathfrak{F}(J_{3}))$%
-orbit as $y$, because
(consistent with the general  \autoref{ee00527} and \autoref{ee00527b}), it holds that%
\begin{equation}
\Delta \left( \Upsilon _{x}(y)\right) =\left( x^{0}x_{0}\right)^{4}\Delta (y)
=\left( \Delta (x)\right) ^{2}\Delta (y)\gtrless 0.
\end{equation}
whose sign depends only on the sign of $\Delta(y)$.

It is worth remarking that $\Upsilon _{x}(y)$
automatically satisfies (\autoref{jazz-2}) for every pair $y^{i}$ and $y_{i}$,
with $i=1,...,N-1$. In fact, regardless of $d_{ijk}$ and $d^{ijk}$, when
only a pair $y^{i}$ and $y_{i}$ for a \textit{fixed} $i$ is non-vanishing
(among all $y^{i}$'s and $y_{i}$'s), then $y$ is non-trivially of rank-$4$
in $\mathfrak{F}$, because generally $\Delta (y)\neq 0$, since at least the
term $-\left( y^{i}y_{i}\right) ^{2}$ is present (\textit{cfr.} (\autoref{I4-1}%
)-(\autoref{I4-2})). Therefore, one can define $N-1$ \textit{distinct} planes $%
\left( \mathfrak{F}_{y\perp x}^{\perp }\right) _{i}$'s, orthogonal to the
Freudenthal plane $\mathfrak{F}_{x}$,
cfr. \autoref{sec:Darboux}.

Moreover, note that $\mathfrak{F}_{y\perp x}$ intersects  at least
three orbits of $\Aut(\mathfrak{F})$. Indeed, it holds that, using \autoref{ee00527} and \autoref{ee00527b},
( $c\equiv x^{0}x_{0} b $)
\begin{eqnarray}
\Delta (r) &=&\left( a+c\right) ^{2}\left( a-c\right) ^{2}\Delta (y)
=\left( a^{2}-c^2\right) ^{2}\Delta (y)  \label{jjjj-1} \\
&=&\left( a^{2}-\left\vert \Delta (x)\right\vert b^{2}\right)
^{2}\Delta (y)\gtreqless 0,  \label{jjj-1-bis}
\end{eqnarray}%
which implies $r\in \mathfrak{F}_{y\perp x}\subset \mathfrak{F}_{x}^{\perp }$
to be not of the same (maximal $=4$) rank as $y$ (and $\Upsilon _{x}(y)$)
only when (recall (\autoref{jjj}))%
\begin{equation}
\Delta (r)=0\Leftrightarrow a^2-c^2=0\to a^{2}-\left( x^{0}x_{0}\right)
^{2}b^{2}=0.
\label{jjjj-2}
\end{equation}%
The conditions for $r\in \mathfrak{F}_{y\perp x}\subset \mathfrak{F}_{x}^{\perp }$
 to lie in the rank-3, rank-2 or even rank-1 orbits
may be easily inferred.

\subsubsection*{\label{sec:closure}Closure of the $D2-D4$ $\mathfrak{F}_{y\perp x}$ under $T$}

The plane $\mathfrak{F}_{y\perp x}$ is not generally closed under the triple map $T$ (or,
equivalently, under Freudenthal duality $\sim $), see \autoref{sec3}.

 Within the framework under
consideration, namely within the $4D/5D$ special coordinates' symplectic
frame of reduced FTS's and within the choice given by \autoref{y}  of
the rank-$4$ element $x\in \mathfrak{F}$ (with $\Delta (x)<0$) and of the
rank-$4$ element $y\in \mathfrak{F}_{x}^{\perp }=\mathfrak{F}/\mathfrak{F}%
_{x}$ (with $\Delta (y)\gtrless 0$), we study now more in
detail the condition of closure of the plane $\mathfrak{F}_{y\perp x}$ under
$T$.

In order to determine the condition of closure of $\mathfrak{F}_{y\perp x}$
under $T$, we have to explicitly compute $T(r)\equiv T(r,r,r)$ for a generic
element $r=ay+b\Upsilon _{x}(y)\in \mathfrak{F}_{y\perp x}$, and
for any $D2-D4$ configuration $y$.
This is given
by, ( see \autoref{JJexplicit3}) ($c\equiv x^{0}x_{0} b $, $T(y)=T(y,y,y)$).
\begin{eqnarray}
T(r)_{M} &:&\left\{
\begin{array}{l}
T(r)_{0}=\left( a-c \right)^{3}T(y)_{0}; \\  \\
T(r)_{i}=(a^2-c^2)(a-c) T(y)_{i}; \\ \\
T(r)^{0}=\left( a+ c \right)^{3}T(y)^{0}; \\ \\
T(r)^{i}=(a^2-c^2) (a+c) T(y)^{i}.%
\end{array}%
\right.  \notag \\
&&  \label{M-1}
\end{eqnarray}%
Then $T(r)^{M}=\Omega ^{MN}T(r)_{N}$. Thus, the plane $\mathfrak{F}_{y\perp x}$ is closed under $T$ \textit{iff}%
\begin{equation}
T(r)^{0}=0=T(r)_{0}\Leftrightarrow \left\{
\begin{array}{l}
\left( a-c \right)^3 d^{ijk}y_{i}y_{j}y_{k}=0; \\  
\\
\left( a+c \right) ^{3}d_{ijk}y^{i}y^{j}y^{k}=0.%
\end{array}%
\right.  \label{sys}
\end{equation}%
There are various cases, as follows:

\begin{enumerate}
\item $y^{i}$ is rank-$3$ in $J_{3}$ and $y_{i}$ is rank-$3$ in $\overline{J}%
_{3}$, namely%
\begin{equation}
d^{ijk}y_{i}y_{j}y_{k}\neq 0,~d_{ijk}y^{i}y^{j}y^{k}\neq 0.
\end{equation}%
In this case, no solutions exist to the system (\autoref{sys}), and $\mathfrak{F}%
_{y\perp x} $ is \textit{not} closed under $T$.

\item $y^{i}$ is rank$<3$ in $J_{3}$ and $y_{i}$ is rank-$3$ in $\overline{J}%
_{3}$, namely%
\begin{equation}
d^{ijk}y_{i}y_{j}y_{k}\neq 0,~d_{ijk}y^{i}y^{j}y^{k}=0.
\end{equation}%
In this case, $T(r)_{0}=0$ is automatically satisfied, while $T(r)^{0}=0$
has solution $a=x^{0}x_{0}b$. However, for a fixed $x$, this
solution is a line in $\F ^{2}=\R ^{2}$ spanned by $\left(
a,b\right) $, and thus is codimension-$1$ in $\mathfrak{F}_{y\perp x}$.
Therefore, only the $x$-dependent $1-$dimensional \textit{locus}
$a=x^{0}x_{0}b$ in $\mathfrak{F}_{y\perp x}\subset \mathfrak{F}_{x}^{\perp }$%
, and \textit{not} $\mathfrak{F}_{y\perp x}$ itself, is closed under $T$.

\item $y^{i}$ is rank-$3$ in $J_{3}$ and $y_{i}$ is rank$<3$ in $\overline{J}%
_{3}$, namely%
\begin{equation}
d^{ijk}y_{i}y_{j}y_{k}=0,~d_{ijk}y^{i}y^{j}y^{k}\neq 0.
\end{equation}%
In this case, $T(r)^{0}=0$ is automatically satisfied, while $T(r)_{0}=0$
has solution $a=-x^{0}x_{0}b$. However, for a fixed $x$, this
solution is a line in $\F ^{2}=\R ^{2}$ spanned by $\left(
a,b\right) $, and thus is codimension-$1$ in $F_{y\perp x}$. Therefore, only
the $x$-dependent $1-$dimensional \textit{locus} $a=-x^{0}x_{0}b$
in $F_{y\perp x}\subset F_{x}^{\perp }$, and \textit{not} $F_{y\perp x}$
itself, is closed under $T$.

\item $y^{i}$ is rank$<3$ in $J_{3}$ and $y_{i}$ is rank$<3$ in $\overline{J}%
_{3}$, namely%
\begin{equation}
d^{ijk}y_{i}y_{j}y_{k}=0=d_{ijk}y^{i}y^{j}y^{k}.  \label{cond-1}
\end{equation}%
In this case, the system (\autoref{sys}) is automatically satisfied $\forall
a,b\in \R $, and $\mathfrak{F}_{y\perp x}$ is therefore closed%
\footnote{It could have also  characterized as a 2-dimensional sub-FTS of
$\mathfrak{F}$.} under $T$.
Note that the
condition (\autoref{cond-1}) is not inconsistent with the assumption of $y$ (\ref%
{y}) to be a rank-$4$ element of $\mathfrak{F}$. In fact, if both $y^{i}$
and $y_{i}$ are rank-$2$ elements in $J_{3}$ resp. $\overline{J}_{3}$, then $%
\Delta (y)$ (\autoref{jjj}) is still generally non-vanishing, with the second
term vanishing \textit{iff} $d_{ijk}d^{ilm}y^{j}y^{k}y_{l}y_{m}=0$ (in this
latter case, when non-vanishing, $\Delta (y)<0$, and $y$ \ - and $\Upsilon
_{x}(y)$ as well - would lie in the same $\Delta <0$  $\Aut (\mathfrak{F}%
(J_{3}))$-orbit as $x$ and $\widetilde{x}$). On the other hand, if $y^{i}$
\textit{and/or} $y_{i}$ are rank-$1$ elements in $J_{3}$ resp. $\overline{J}%
_{3}$, still $y$ can be a rank-$4$ element of $\mathfrak{F}(J_{3})$, because
$\Delta (y)=-\left( y^{i}y_{i}\right) ^{2}\leqslant 0$ in this case, and
thus (when the inequality strictly holds), $y$ - and $\Upsilon _{x}(y)$ as
well - would lie, as above, in the same $\Delta <0$  $\Aut (\mathfrak{F}%
(J_{3})) $-orbit as $x$ and $\widetilde{x}$.
\end{enumerate}


\subsection{
The general $D0-D4$ sector }
\label{sec:D0D4}


\subsubsection*{The Freudenthal plane $\mathfrak{F}_{x}$}

We start by identifying $x$ with the rank-$4$ element of the FTS $\mathfrak{F%
}$ given by the
\footnote{We might have as well started with a $D2-D6$ configuration, and perform an
equivalent treatment (obtaining, as evident from the treatment given below
and from the anti-involutivity of Freudenthal duality, a $D0-D4$
configuration as Freudenthal-dual of the starting $D2-D6$ configuration).} $%
D0-D4$ brane charge configuration :%
\begin{equation}
\xdodc\equiv \left( 0,x^{i},x_{0},0_{i}\right) ^{T}\in \mathfrak{F},
\label{1-bis}
\end{equation}%
and we further impose that $\xdodc$ belongs to (one of) the $\Delta >0$ $%
\Aut(\mathfrak{F})$-orbit(s) (see \autoref{sec:preservation}).
\begin{equation}
\Delta (\xdodc)=\frac{2}{3}x_{0}d_{ijk}x^{i}x^{j}x^{k}>0.
\end{equation}%
From the definition (\autoref{ee00122}) (note that $\epsilon =1$ in this case),
one can compute that the Freudenthal-dual $\widetilde{\xdodc}$ of the $%
D0-D4$ configuration (\autoref{1-bis}) is a $D2-D6$ configuration, namely%
\footnote{%
This has been computed for the $STU$ model in Example 1 of Sec. V.A of \cite%
{Duff-FD} (\textit{cfr.} (198)-(200) therein); see also the treatment of
 \autoref{sec:preservation}.}
  The dual is
\begin{equation}
\widetilde{x}\equiv \left( \widetilde{x}^{0},0^{i},0,\widetilde{x}_{i}\right)
^{T}\in \mathfrak{F},  \label{1-bis-tilde}
\end{equation}%
with (\autoref{ee00122})
\begin{eqnarray}
\widetilde{x}^{0}&=&-\frac{1}{3\sqrt{\Delta (x)}}d_{jkl}x^{j}x^{k}x^{l}; \\
\widetilde{x}_{i}&=&\frac{1}{\sqrt{\Delta (x)}}x_{0}d_{ijk}x^{j}x^{k}.
\label{1-bis-tilde-2}
\end{eqnarray}%
By exploiting the \textit{adjoint identity} (see \cite{Krutelevich:2004,mccrimmon,CFM-1} and Refs. therein)
of the Jordan algebra $%
J_{3}$ underlying the reduced FTS $\mathfrak{F}$ \autoref{adj-id}
one can also check that $\Delta $ is invariant under Freudenthal duality :%
\begin{equation}
\Delta (\xdodc)=\Delta (\widetilde{\xdodc})>0,
\end{equation}%
and thus that $\widetilde{\xdodc}$ would lies in the same  $\Aut (\mathfrak{F}%
(J_{3}))$-orbit as $\xdodc$.

Thus, one can define the Freudenthal plane $\mathfrak{F}_{x}$ (dim$\mathfrak{F%
}_{x}=2$), spanned by $x$ (\autoref{1-bis}) and $\widetilde{x}$ (\ref%
{1-bis-tilde}), whose generic element is $x_F=ax+b\widetilde{x}\in \mathfrak{F}%
_{x}$, $a,b\in \R $. By using (\autoref{adj-id}) again, one can also
compute that
\begin{equation}
\Delta (x_F)=\left( a^{2}+b^{2}\right) ^{2}\Delta (x)>0,
\end{equation}%
implying that $\mathfrak{F}_{x}$ lies completely in the rank-$4$ $\Delta >0$
orbit\footnote{%
Again, as pointed out above, some subtleties may arise, related to the
splitting of the $\Delta >0$ locus of $\mathfrak{F}$ (\textit{cfr. }Sec. \ref%
{sec:preservation}, and\textit{\ e.g.} \cite%
{Shukuzawa,small-orbits,small-orbits-maths}, and Refs. therein).} of  $\Aut (%
\mathfrak{F}(J_{3}))$.

\subsubsection*{The orthogonal space $\mathfrak{F}_{x}^{\perp },$ and the plane $%
\mathfrak{F}_{y\perp x}\subset \mathfrak{F}_{x}^{\perp }$}

Then, one can pick another rank-$4$ element $y\in \mathfrak{F}$ which is $%
\left\{ ,\right\} $-orthogonal to $x$ (\autoref{1-bis}) and $\widetilde{x}$ (\ref%
{1-bis-tilde}); the most general element of this kind is given by the charge
configuration:%
\begin{equation}
y\equiv \left( y^{0},y^{i},y_{0},y_{i}\right) ^{T}\in \mathfrak{F},  \label{y-bis}
\end{equation}%
constrained%
\begin{equation}
\left\{ x,y\right\} =0=\left\{ \widetilde{x},y\right\} \Leftrightarrow
\left\{
\begin{array}{l}
y^{0}=\frac{1}{x_{0}}x^{i}y_{i}; \\
\\
y_{0}=-3x_{0}\frac{d_{ijk}y^{i}x^{j}x^{k}}{d_{lmn}x^{l}x^{m}x^{n}}.%
\end{array}%
\right.
\end{equation}%
One can also compute that%
\begin{eqnarray}
&&%
\begin{array}{l}
\Delta (y)=-\left( -3\frac{d_{klm}y^{k}x^{l}x^{m}}{d_{qrs}x^{q}x^{r}x^{s}}%
x^{j}y_{j}+y^{j}y_{j}\right) ^{2}-2x_{0}\frac{d_{klm}y^{k}x^{l}x^{m}}{%
d_{qrs}x^{q}x^{r}x^{s}}d_{ijt}y^{j}y^{j}y^{t} \\
-\frac{2}{3}\frac{1}{x_{0}}%
x^{j}y_{j}d^{klm}y_{k}y_{l}y_{m}+d_{ijk}d^{ilm}y^{j}y^{k}y_{l}y_{m}\gtrless
0.%
\end{array}
\notag \\
&&  \label{jjj-bis-z}
\end{eqnarray}%
In (\autoref{jjj-bis-z}), the case of vanishing $\Delta $ has been excluded
because $y$ is chosen to be of rank-$4$ in $\mathfrak{F}$.

One can compute the components of $%
\Upsilon _{x}(y)^{M}$ (\autoref{Ipsilon}) as given by%
\beq 
\left( \Upsilon _{x}(y)\right) ^{0}&=&-d_{ijk}x^{i}x^{j}y^{k}; \\
\left( \Upsilon _{x}(y)\right) ^{i}&=&-d_{klm}d^{mij}x^{k}x^{l}y_{j}+2x^{j}y_{j}x^{i}; \\
\left( \Upsilon _{x}(y)\right) _{0}&=&-2x^{j}y_{j}x_{0}; \\
\left( \Upsilon _{x}(y)\right) _{i}&=&2x_{0}d_{ijk}x^{j}y^{k}-3x_{0}\frac{d_{klm}y^{k}x^{l}x^{m}}{d_{qrs}x^{q}x^{r}x^{s}}d_{ijt}x^{j}x^{t},
\label{Ipsilon-bis2}
\eeq
which for this case can be written as
\beq
\left( \Upsilon _{x}(y)\right) ^{0}&=&-d_{ijk}x^{i}x^{j}y^{k}; \\
\left( \Upsilon _{x}(y)\right) ^{i}&=&-d_{klm}d^{mij}x^{k}x^{l}y_{j}+y^0 x_0 x^{i}; \\
\left( \Upsilon _{x}(y)\right) _{0}&=&-2y^0 x_{0}^2; \\
\left( \Upsilon _{x}(y)\right) _{i}&=&2x_{0}d_{ijk}x^{j}y^{k}+y_0d_{ijt}x^{j}x^{t}.%
\label{Ipsilon-bis}
\eeq

By exploiting (\ref{adj-id}), one can then check that $\Upsilon _{x}(y)$ (%
\autoref{Ipsilon-bis}) automatically satisfies%
\begin{equation}
\left\{ x,\Upsilon _{x}(y)\right\} =0=\left\{ \widetilde{x},\Upsilon
_{x}(y)\right\} \Leftrightarrow \Upsilon _{x}(y)\in \mathfrak{F}_{x}^{\perp
}\equiv \mathfrak{F}/\mathfrak{F}_{x},
\end{equation}%
for every pair $y^{i}$ and $y_{i}$, with $i=1,...,N-1$. In fact, regardless
of $d_{ijk}$ and $d^{ijk}$, when only a pair $y^{i}$ and $y_{i}$ for a
\textit{fixed} $i$ is non-vanishing (among all $y^{i}$'s and $y_{i}$'s),
then $y$ is non-trivially of rank-$4$ in $\mathfrak{F}$, because generally $%
\Delta (y)\neq 0$, since at least the term $-\left( y^{i}y_{i}\right) ^{2}$
is present (\textit{cfr.} (\autoref{I4-1})-(\autoref{I4-2})). Therefore, one can
define $N-1$ \textit{distinct} planes $\left( \mathfrak{F}_{y\perp x}^{\perp
}\right) _{i}$'s, orthogonal to the Freudenthal plane $\mathfrak{F}_{x}$,
cfr.   \autoref{sec:Darboux}.

Moreover, (consistently with
\autoref{ee00527} and \autoref{ee00527b}) one can compute that
\begin{eqnarray}
\Delta \left( \Upsilon _{x}(y)\right) &=&-\frac{8}{9}x_{0}^{2}\left(
d_{abc}x^{a}x^{b}x^{c}\right) ^{2}\left[
\begin{array}{l}
\frac{1}{2}\left( -3\frac{d_{klm}y^{k}x^{l}x^{m}}{d_{qrs}x^{q}x^{r}x^{s}}%
x^{j}y_{j}+y^{j}y_{j}\right) ^{2}+x_{0}\frac{d_{klm}y^{k}x^{l}x^{m}}{%
d_{qrs}x^{q}x^{r}x^{s}}d_{ijt}y^{j}y^{j}y^{t} \\
\\
+\frac{1}{3}\frac{1}{x_{0}}%
x^{j}y_{j}d^{klm}y_{k}y_{l}y_{m}+d_{ijk}d^{ilm}y^{j}y^{k}y_{l}y_{m}\gtrless
0.%
\end{array}%
\right]  \notag \\
&=&\left( \Delta (x)\right) ^{2}\Delta (y)\gtrless 0,
\end{eqnarray}%
implying that $\Upsilon _{x}(y)$ lies in the same  $\Aut (\mathfrak{F}(J_{3}))$%
-orbit as $y $.

The same holds for a generic element $r$ $=ay+b\Upsilon _{x}(y)\in \mathfrak{%
F}_{y\perp x}$ ($a,b\in \R $), which belongs to the same  $\Aut (%
\mathfrak{F})$-orbit as $y$ : indeed it can be
checked that (consistently with \autoref{ee00527} and \autoref{ee00527b})
\begin{eqnarray}
\Delta (r) &=&-2\left( a^{2}+\frac{2b^{2}}{3}x_{0}d_{abc}x^{a}x^{b}x^{c}%
\right) ^{2}\left[
\begin{array}{l}
\frac{1}{2}\left( -3\frac{d_{klm}y^{k}x^{l}x^{m}}{d_{qrs}x^{q}x^{r}x^{s}}%
x^{j}y_{j}+y^{j}y_{j}\right) ^{2}+x_{0}\frac{d_{klm}y^{k}x^{l}x^{m}}{%
d_{qrs}x^{q}x^{r}x^{s}}d_{ijt}y^{j}y^{j}y^{t} \\
\\
+\frac{1}{3}\frac{1}{x_{0}}%
x^{j}y_{j}d^{klm}y_{k}y_{l}y_{m}+d_{ijk}d^{ilm}y^{j}y^{k}y_{l}y_{m}\gtrless
0.%
\end{array}%
\right]  \notag \\
&=&\left( a^{2}+\Delta (x)b^{2}\right) ^{2}\Delta (y)\gtrless 0.
\label{jjj-1-bis-2}
\end{eqnarray}


\subsubsection*{\label{sec:Darboux}The canonical Darboux symplectic frame}

We recall that in the $4D/5D$ special coordinates' symplectic frame a
generic element $Q$ of the reduced FTS $\mathfrak{F}(J_{3})$ splits as given
by (\autoref{split}), while the $2N\times 2N$ symplectic metric is given by (\ref%
{Omega-z}). By a simple re-ordering of rows and columns (amounting to a
relabelling of indices
, one can switch to a \textit{canonical Darboux
symplectic frame}\footnote{%
For some applications of the canonical Darboux frame to supergravity, see
\textit{e.g.} \cite{Ferrara-Macia, Mohaupt, Klemm}, and Refs. therein.}  (in
which the $4D/5D$ covariance is still manifest), in which $x$ (\autoref{split})
splits as follows:%
\begin{equation}
x=-\left( p^{0},q_{0},p^{1},q_{1},...,p^{N-1},q_{N-1}\right) ^{T},
\label{Q2}
\end{equation}%
and in which the symplectic metric (\autoref{Omega-z}) acquires the following
form
\begin{equation}
\Omega =-\mathbf{1}\otimes \varepsilon ,  \label{Omega2}
\end{equation}%
where $\varepsilon $ is the $2\times 2$ symplectic metric of the defining
irrepr. $\mathbf{2}$ of $Sp(2)\approx SL(2)$ defined by (\autoref{epsilon!}).

At a glance, in a physical (Maxwell-Einstein) framework (\autoref{Q2}) suggests
that the choice of the (canonical) Darboux symplectic frame defined by (\ref%
{Q2}) (or, equivalently, by (\autoref{Omega2} and (\autoref{epsilon!})), amounts to
making manifest the splitting of the electric-magnetic fluxes of the Abelian
$2$-form field strengths, grouped, within the symplectic vector $Q$ (\autoref{Q2}%
), into the KK vector's fluxes (magnetic $p^{0}$ and electric $q_{0}$), and
into the fluxes (magnetic $p^{i}$ and electric $q_{i}$, $i=1,...,N-1$) of
each of the $N-1$ Abelian vectors with a $D=5$ origin. When specifying such
a generic (supersymmetry-independent) interpretation for minimal $D=5$
supergravity dimensionally reduced down to $\mathcal{N}=2$, $D=4$
supergravity, $p^{0}$ and $q_{0}$ are the magnetic resp. electric charges of
the $D=4$ graviphoton (the Abelian vector in the $\mathcal{N}=2$ gravity
multiplet), whereas each of the $N-1 $ pairs $\left( p^{i},q_{i}\right) $
denote the magnetic resp. electric charges of the Abelian vector belonging
to each of the $N-1$ vector supermultiplets coupled to the gravity one
(these all have a $D=5$ origin, thereby comprising the $D=5$ graviphoton, as
well).

Thus, the $\left(
2N-2\right) $-dim. space $\mathfrak{F}_{x}^{\perp }$, $\left\{ ,\right\} $%
-orthogonal to the $2$-dim. Freudenthal plane $\mathfrak{F}_{x}$, gets
decomposed into $N-1$ $2$-dim. subspaces, all mutually orthogonal with
respect to the symplectic product $\left\{ ,\right\} $ defined by (\ref%
{Omega2}) : each of them corresponds to the electric-magnetic flux degrees
of freedom of a vector supermultiplet in the corresponding $\mathcal{N}=2$, $%
D=4$ supergravity, or, more generally, to the electric-magnetic fluxes of a $%
D=4$ Abelian vector fields with a five-dimensional origin.

\subsection{\label{sec:preservation} F-duality preserves the $\Delta >0$
$\Aut(\FTS)$-orbits: The $STU$ model}
\label{sec:84}

\textit{At least} for (non-degenerate) reduced FTS's, $\Aut(\mathfrak{F})$
has a transitive action on elements with a given $\Delta <0$. Thus, the $\Delta <0$ locus 
 corresponds to a  one-parameter family of
$\Aut(\mathfrak{F})$-orbits
\footnote{We will equivalently use $\Aut(\mathfrak{F})$, electric-magnetic duality, or
U-duality, even if the first terminology is mathematical, the second more
pertaining to Maxwell gravity, and the third one deriving from string
theory.}; consequently, Freudenthal duality   trivially preserves
the orbit structure for $\Delta <0$.

The story is more complicated for the $\Delta >0$ locus, which, again
\textit{at least} for (non-degenerate) reduced FTS's,   generically (with
the unique exception of the $T^{3}$ model) has two or more $\Aut(%
\mathfrak{F})$-orbits. However, the existence of the Freudenthal rotations presented in \autoref{f-rot} ensures that $x$ and $\tilde{x}$ always lie in the same orbit. Here we explicitly present the
non-trivial orbit structure of the $\Delta >0$ locus and its properties under Freudenthal duality for the $STU$ model using only the (discrete) U-duality invariants characterising the orbits.

The $STU$ model, introduced independently in \cite{Senvafa,triality}, provides an  interesting subsector of string compactification to four dimensions. This model has a low energy limit which is described by $\N = 2$ supergravity coupled to three vector multiplets interacting through the special K\"ahler manifold  $[SL(2, \R)/SO(2)]^3$.  (In the version of \cite{Senvafa}, the discrete $SL(2, \Z)$ are replaced by a subgroup denoted $\Gamma_0(2)$).  The three complex scalars are denoted by the letters $S, T$ and $U$, hence the name of the model \cite{triality,stublack}. The remarkable feature that distinguishes it from generic $\N=2$ supergravities coupled to vectors \cite{Kounas} is its $S-T-U$ triality  \cite{triality}.  There are three different versions with two of the $SL(2)$s perturbative symmetries of the Lagrangian and the third a non-perturbative symmetry of the equations of motion. In a  fourth version all three are non-perturbative \cite{triality,stublack}. All four are on-shell equivalent. If there are in addition four hypermultiplets, the $STU$ model is self-mirror. Even though the simplest reduced FTS exhibiting more than one $\Delta >0$ $%
\Aut(\mathfrak{F})$-orbit is given by $\mathfrak{F}(%
\R \oplus \R)$, which corresponds  to the  $ST^{2}$
model of $\mathcal{N}=2$, $D=4$ supergravity \cite{Bellucci:2008sv}, we will explicitly treat 
 $\mathfrak{F}(\R \oplus \R \oplus \R)$
corresponding  to the slightly larger $STU$ model, because it can be considered
as a genuine truncation of all (non-degenerate) reduced FTS (with  the
exception of the $T^{3}$ and $ST^{2}$ models, which are however particular
\textquotedblleft degenerations" of the $STU$ model itself), thus covering all such cases.

As determined in \cite{Bellucci:2008sv} (see also the treatment in Sec. F.1 of \cite%
{small-orbits}), in the $STU$ model there are two orbits with $\Delta >0$,
one supersymmetric and one non-supersymmetric (the one with vanishing
central charge at the horizon : $Z_{H}=0$), and their coset expressions are
isomorphic (even if they are $SL(2,\R )^{\times 3}$-disjoint orbits) :%
\begin{equation}
\mathcal{O}_{\Delta >0,BPS}\cong \frac{SL(2,\R )^{\times 3}}{%
U(1)^{\times 2}}\cong \mathcal{O}_{\Delta >0,non-BPS,Z_{H}=0}.
\end{equation}

Following the treatment of \cite{CFM-1}, one can consider a $D0-D4$
representative (also in the FTS representation \cite{Duff-FD}) of the orbits
$\mathcal{O}_{\Delta >0,BPS}$ and $\mathcal{O}_{\Delta >0,non-BPS,Z_{H}=0}$ :%
\begin{eqnarray}
x_{D0-D4,\Delta >0} &=&\left( 0,p^{1},p^{2},p^{3},q_{0},0,0,0\right)
^{T}=\left(
\begin{array}{cc}
-q_{0} & (p^{1},p^{2},p^{3}) \\
\left( 0,0,0\right) & 0%
\end{array}%
\right) ,  \label{D0-D4-pos-pre} \\
\Delta \left( x_{D0-D4,\Delta >0}\right) &=&4q_{0}p^{1}p^{2}p^{3}>0,
\label{D0-D4-pos}
\end{eqnarray}%
and determine the further $SL(2,\R )^{\times 3}$-invariant
constraints on the sign of $q_{0}^{0}$, $p^{1}$, $p^{2}$ and $p^{3}$
defining the orbits $\mathcal{O}_{\Delta >0,BPS}$ and $\mathcal{O}_{\Delta
>0,non-BPS,Z_{H}=0}$ within the $D0-D4$ configuration (\ref{D0-D4-pos-pre})-(%
\autoref{D2-D6-pos}) with $\Delta >0$ (all other sign combinations pertain to
the unique non-BPS $Z_{H}\neq 0$ orbit $\mathcal{O}_{\Delta <0}$) :%
\begin{itemize}
\item $\frac{1}{2}$-BPS~conditions~(triality invariant):
\begin{equation}
(q_{0} , p^{1} , p^{2} , p^{3})=(+ , + , + , + )\ \text{or}\  (- , - , - , - ).
 \label{conds-1}
 \end{equation}
 \item
non-BPS~$Z_{H}=0$~{conditions~(up to triality) }:
\begin{equation}
(q_{0} , p^{1} , p^{2} , p^{3})=(+ , + , - , - )\ \text{or}\  (- , - , + , + ).
 \label{conds-2}
 \end{equation}

\end{itemize}

Analogously, we can consider (also in the FTS representation \cite{Duff-FD})
a $D2-D6$ representative of the orbits $\mathcal{O}_{\Delta >0,BPS}$ and $%
\mathcal{O}_{\Delta >0,non-BPS,Z_{H}=0}$ :%
\begin{eqnarray}
x_{D2-D6,\Delta >0} &=&\left( p^{0},0,0,0,0,q_{1},q_{2},q_{3}\right)
^{T}=\left(
\begin{array}{cc}
0 & (0,0,0) \\
\left( q_{1},q_{2},q_{3}\right) & p^{0}%
\end{array}%
\right) ,  \label{D2-D6-pos-pre} \\
\Delta \left( x_{D2-D6,\Delta >0}\right) &=&-4p^{0}q_{1}q_{2}q_{3}>0.
\label{D2-D6-pos}
\end{eqnarray}%
From the treatment of \cite{Bellucci:2008sv}, we can write down the further $%
SL(2,\R )^{\times 3}$-invariant constraints on the sign of $p^{0}$, $%
q_{1}$, $q_{2}$ and $q_{3}$ defining the orbits $\mathcal{O}_{\Delta >0,BPS}$
and $\mathcal{O}_{\Delta >0,non-BPS,Z_{H}=0}$ within the $D2-D6$
configuration (\autoref{D2-D6-pos-pre})-(\autoref{D2-D6-pos}) with $\Delta >0$ (all
other sign combinations pertain to the unique non-BPS $Z_{H}\neq 0$ orbit $%
\mathcal{O}_{\Delta <0}$) :
\begin{itemize}
\item $\frac{1}{2}$-BPS~conditions~(triality invariant):
\begin{equation}
(q_{0} , p^{1} , p^{2} , p^{3})=(+ , - , - , - )\ \text{or}\  (- , + , + , + ).
 \label{conds-3}
 \end{equation}
 \item
non-BPS~$Z_{H}=0$~{conditions~(up to triality) }:
\begin{equation}
(q_{0} , p^{1} , p^{2} , p^{3})=(+ , - , + , + )\ \text{or}\  (- , + , - , - ).
 \label{conds-4}
 \end{equation}

\end{itemize}

As computed in Example 1 of Sec. V.A of \cite{Duff-FD} (\textit{cfr.}
(198)-(200) therein), the Freudenthal dual of the $D0-D4$ element (\ref%
{D0-D4-pos-pre})-(\autoref{D0-D4-pos}) yields a particular element of the type $%
D2-D6$ type (\autoref{D2-D6-pos-pre})-(\autoref{D2-D6-pos}) (up to triality; $%
c(m,n)>0$) :%
\begin{equation}
x_{D0-D4,\Delta >0}=\left(
\begin{array}{cc}
-n & (n,-m,-m) \\
\left( 0,0,0\right) & 0%
\end{array}%
\right) \overset{\sim }{\longrightarrow }\widetilde{x_{D0-D4,\Delta >0}}%
=c(m,n)\left(
\begin{array}{cc}
0 & (0,0,0) \\
\left( m,-n,-n\right) & -m%
\end{array}%
\right) .  \label{F-dual-map}
\end{equation}%
By using (\autoref{F-dual-map}), one can then compute that the action of
Freudenthal duality on the $D0-D4$ element (\autoref{D0-D4-pos-pre})-(\ref%
{D0-D4-pos}) maps the conditions (\autoref{conds-1}) and (\autoref{conds-2})
respectively into the conditions (\autoref{conds-3}) and (\autoref{conds-4}) : by
using the commutativity of  $\Aut (\mathfrak{F}(\R \oplus \mathds{R%
}\oplus \R ))=SL(2,\R )^{\times 3}$ and Freudenthal duality $%
\sim $ \cite{Duff-FD}, one can consequently conclude that the orbit
stratification of the $\Delta >0$ locus of $\mathfrak{F}(\R %
\oplus \R \oplus \R )$ is \textit{preserved} under Freudenthal
duality $\sim $.

\section{Linear realisations of general Freudenthal transformations}
\label{sec:90}

As consequence of its definition, Freudenthal duality $\sim $
can only be consistently defined in the \textit{locus} $\Delta \neq 0$ of
the FTS $\mathfrak{F}$ itself. In general, the group $\Aut(\mathfrak{F})$ has
a non-transitive action over such a \textit{locus}, which undergoes a,
(\textit{at least}) twofold stratification, into a (always unique) $\Delta <0$
orbit and into a $\Delta >0$ \textit{sub-locus}, which may in turn further
stratify into $\Aut(\mathfrak{F})$-orbits.

While F-duality is a non-linear operation, as discussed in \cite{Duff-FD} its action  can be
realised  by finite ``local/gauged'' U-duality
transformations $U: \FTS \rightarrow \Aut(\FTS)$, namely, as we will
understand throughout the following treatment, that depend on the element of
$\mathfrak{F}$ they are applied onto.

More generally, Freudenthal duality, and so GFT,
 can be mimicked by
finite transformations of at least three different kinds, the first two of which are not contained in the U-duality group:

\begin{enumerate}
\item Gauged anti-symplectic transformations, as we will discuss, within
non-degenerate, reduced FTS's, for the $\Delta <0$ orbit in \autoref{F-Dual-D0-D6};

\item  Gauged $Sp(2N,\R )$ transformations, where $%
\dim _{\R }\mathfrak{F}=2N$. We will discuss these, within the $T^{3}$
model ($N=2$; \textit{cfr.} (\autoref{tt1})), for the $\Delta <0$ orbit in
  \autoref{F-Dual-D0-D6-sympl};

\item Gauged $\Aut(\mathfrak{F})$ transformations, as we
will discuss, within the $T^{3}$ model, for the $\Delta <0$ and $\Delta >0$
orbit in Secs. \autoref{F-Dual-D0-D6-U} and \autoref{F-Dual-D0-D4}, respectively.
\end{enumerate}

\subsection{Anti-symplectic realisation: $\Delta <0$}
 \label{F-Dual-D0-D6}

\textit{At least} within (non-degenerate) reduced FTS's, the quartic
polynomial invariant $\Delta$ of $\Aut\left( \mathfrak{F}\left( J_{3}\right)
\right) \approx Conf\left( J_{3}\right) $ can be written  as
in \autoref{I4-2} (see \autoref{secJ3} for this and other expressions)
(\textit{cfr. e.g.} \cite{Sato-Kimura,FG-1,CFM-1}; ($i=1,...,N-1$, dim$%
\mathfrak{F}=2N$)).
Let us consider a $D0-D6$ configuration, studied in \autoref{sec:D0D6}.
As we can seen there
(we refer to \autoref{1-z},\autoref{jazz-sat},\autoref{1-tilde-z} and \autoref{2}),
the action of the Freudenthal duality $\sim $ on $\xdodo$ (\autoref{1-z}) can
be represented by a (maximal-rank) $2N\times 2N$ matrix%
\begin{eqnarray}
\widetilde{x} &=&-\epsilon\mathbf{O}x,
\label{plus-minus} \\
&&  \notag \\
\mathbf{O} &=&\left(
\begin{array}{cccc}
1 &  &  &  \\
& \mathbf{A} &  &  \\
&  & -1 &  \\
&  &  & \mathbf{B}%
\end{array}%
\right) ,
\end{eqnarray}%
where $\mathbf{A}$, $\mathbf{B}\in GL(N-1,\R )$.
The action of a general Freudenthal transformations is then
\beq
x_F &=& a x +b \tilde{x} = \left (\mathbf{1}-\epsilon \mathbf{O} \right ) x.
\eeq
The
transformation $\mathbf{O}\in GL(2N,\R )$ is inherently \textit{not
unique}. Also, apart from the \textquotedblleft $\mp $" branching in (\ref%
{plus-minus}), the whole realisation  of Freudenthal duality does not depend
on $p^{0}$ nor on $q_{0}$, and so it can be (loosely) considered an
\textquotedblleft ungauged" transformation in $\mathfrak{F}$. In particular,
$\mathbf{O}$ is anti-symplectic, namely%
\begin{equation}
\mathbf{O}^{T}\Omega \mathbf{O}=-\Omega \Leftrightarrow \mathbf{A}^{T}%
\mathbf{B}=-\mathbf{1}
\end{equation}%
where $\mathbf{1}$ is the identity matrix in $N-1$ dimensions. Note,
however, that $\mathbf{O}$ is never symplectic (\textit{i.e.}, it always
holds that $\mathbf{O}^{T}\Omega \mathbf{O}\neq \Omega \Leftrightarrow
\mathbf{O}\notin Sp\left( 2N,\R \right) $).

Let us consider a particularly simple anti-symplectic case, for which $%
\mathbf{A}=-\mathbf{1\Rightarrow B}=\mathbf{1}$ :
\begin{equation}
\mathbf{O}^{\prime }=\left(
\begin{array}{cccc}
1 &  &  &  \\
& -\mathbf{1} &  &  \\
&  & -1 &  \\
&  &  & \mathbf{1}%
\end{array}%
\right) .  \label{O'}
\end{equation}%
\textit{At least} in all reduced FTS's $\mathfrak{F}$'s based on simple and
semi-simple rank-$3$ Jordan algebras
(see   \autoref{sec:fts} and  \autoref{tt1} therein), it can be proved
(\textit{cfr.} App. D of  \cite{M-Horizon})
that $\mathbf{O}^{\prime }$ (\autoref{O'}) realizes an \textit{outer%
} automorphism of  $\Aut \left( \mathfrak{F}(J_{3})\right) $, namely that%
\begin{eqnarray}
\mathbf{O}^{-1}\widehat{\mathbf{R}}_{\mathfrak{F}}(g)\mathbf{O} &\subset &%
\widehat{\mathbf{R}}_{\mathfrak{F}}(g), \\
\mathbf{O} &\in &\frac{\Aut\left(\Aut\left( \mathfrak{F}(J_{3})\right)
\right) }{Inn\left(\Aut\left( \mathfrak{F}(J_{3})\right) \right) }%
\equiv Out\left(\Aut\left( \mathfrak{F}(J_{3})\right) \right) ,
\end{eqnarray}%
where $\widehat{\mathbf{R}}_{\mathfrak{F}}(g)$ denotes the $2N\times 2N$
matrix representation of the element $g$ of  $\Aut \left( \mathfrak{F}%
(J_{3})\right) $ acting on $\mathfrak{F}$ itself, and  $\Aut $, $Inn$ and $Out$
respectively denote the automorphism group, and its inner resp. outer
components. Note that $\mathbf{O}^{\prime }$ is an involution:%
\begin{equation}
\mathbf{O}^{\prime 2}=\mathbf{1},
\end{equation}%
where here $\mathbf{1}$ denotes the the identity matrix in $2N$ dimensions.
However, from (\autoref{ee00122})-(\autoref{2}) it follows that the correct iteration
of the Freudenthal duality on the $D0-D6$ configuration $x$ (\autoref{1-z}) is
provided by the application of $\mp \mathbf{O}^{\prime }$ and then
necessarily of $\pm \mathbf{O}^{\prime }$, thus yielding to $-\mathbf{O}%
^{\prime 2}=-\mathbf{1}$ acting on $x$, and correctly implying%
\begin{equation}
\widetilde{\widetilde{x}}=-x.
\end{equation}%
The anti-simplecticity of $\mathbf{O}^{\prime }$ (\autoref{O'}) implies that it
does not preserve the symplectic structure of $\mathfrak{F}$ (as neither
Freudenthal duality does, as well \cite{Duff-FD}). This is consistent with
the fact that $\mathbf{O}^{\prime }$ realizes an \textit{outer} automorphism
of the electric-magnetic U-duality group $\Aut\left( \mathfrak{F}%
(J_{3})\right) $, which in turn is generally realized in a symplectic way
\cite{Gaillard-Zumino,U-duality-revisited,Aschieri-Ferrara-Zumino}:%
\begin{equation}
\widehat{\mathbf{R}}_{\mathfrak{F}}(g)^{T}\Omega \widehat{\mathbf{R}}_{%
\mathfrak{F}}(g)=\Omega ,~\forall g\in \Aut\left( \mathfrak{F}(J_{3})\right) .
\end{equation}

\subsubsection{Anti-symplectic symmetries and parity transformations.}

As observed in \cite{M-Horizon}, \textit{at least} for all automorphism
groups of reduced FTS's over simple or semi-simple rank-$3$ Jordan algebras
it holds that (see e.g. \cite{Murakami})
\begin{equation}
Out\left(\Aut\left( \mathfrak{F}(J_{3})\right) \right) \subseteq \mathds{Z}%
_{2}.
\end{equation}%
Thus, all non-trivial elements of $Out\left(\Aut\left( \mathfrak{F}%
(J_{3})\right) \right) $ are implemented by anti-symplectic transformations.
In \cite{trig} (also cfr. \cite{trig-2}), it was discussed that the global
symmetry of the resulting Maxwell-Einstein (super)gravity contains the
factor $\mathds{Z}_{2}$, which can be offset by a \textit{spatial parity} $%
\mathbf{P}$ transformation. In particular, from Eq. (2.118) of \cite{trig},
it follows that the global symmetry group $G$ of the resulting
Maxwell-Einstein (super)gravity theory is given by%
\begin{equation}
G=G_{0}\times \mathds{Z}_{2}=\left\{ G_{0},G_{0}\cdot \mathbf{p}\right\} ,
\end{equation}%
where $G_{0}$ is the identity-connected, proper electric-magnetic ($U$%
-)duality, $\Aut\left( \mathfrak{F}(J_{3})\right) $-part of $G$, whereas $%
\mathbf{p}$ corresponds to an element of $G$ implemented by an \textit{%
anti-symplectic} transformation.

Interestingly, the above results relate a realisation  (\textit{not} the
unique one, though! - see Secs. \autoref{F-Dual-D0-D6-sympl} and \ref%
{F-Dual-D0-D6-U}) of the Freudenthal duality on the well-defined
representative $D0-D6$ of the unique, non-BPS (non-supersymmetric) rank-$4$ $%
\Delta <0$ orbit in $\mathfrak{F}$, to spatial parity transformations in the
corresponding theory; in fact, anti-symplectic transformations, such as $%
\mathbf{O}^{\prime }$ (\autoref{O'}) are symmetries of the theory, provided that
they are combined with spatial ($3D$) parity $\mathbf{P}$.

\subsection{Gauged symplectic realisation: $\Delta<0$}

\label{F-Dual-D0-D6-sympl}

It is also possible to find a (non-unique) symplectic
transformation realizing the Freudenthal duality transformation (\autoref{ee00122})
on the $D0-D6$ representative of the $\Delta <0$  $\Aut \left( \mathfrak{F}%
\right) $-orbit of $\mathfrak{F}$. However, this will necessarily be
\textquotedblleft gauged" in $\mathfrak{F}$, namely it will depend on the
element of $F$ it acts upon (\textit{i.e.}, in this case, on the $D0-D6$
element (\autoref{1-z})).

Let us exemplify this in the $T^{3}$ model ($N=2$, as from (\autoref{tt1}));
we look for a finite $Sp(4,\R )$ transformation (realized as a $%
4\times 4$ matrix $\mathcal{C}$) such that%
\begin{equation}
\mathcal{C}\left(
\begin{array}{c}
p^{0} \\
0 \\
q_{0} \\
0%
\end{array}%
\right) =sgn\left( p^{0}q_{0}\right) \left(
\begin{array}{c}
-p^{0} \\
0 \\
q_{0} \\
0%
\end{array}%
\right) .
\end{equation}%
Defining $z\equiv q_{0}/p^{0}$, long but straightforward algebra yields to the
following expression (for $sgn\left( p^{0}q_{0}\right) =sgn(z)=\pm 1$) :%
\begin{equation}
\mathcal{C}_{\pm }\left( z;a,b,c,d,e,f,g\right) =\left(
\begin{array}{cccc}
\pm \left( 1+\frac{a}{z}\right) & \mp \left( bc\mp \frac{de}{z}\right) &
\frac{2}{z}+\frac{a}{z^{2}} & \mp \left( bf\mp \frac{dg}{z}\right) \\
\pm bz & e & b & g \\
a & -\left( bcz\mp de\right) & \pm \left( 1+\frac{a}{z}\right) & -\left(
bfz\mp dg\right) \\
d & c & \pm \frac{d}{z} & f%
\end{array}%
\right) _{ef-gc=1}.  \label{C}
\end{equation}%
$\mathcal{C}_{\pm }$ (\autoref{C}) depends on $z$, and it is thus
\textquotedblleft gauged" in $\mathfrak{F}$. Moreover, it depends on $7$
real parameters, with a constraint ($ef-gc=1$) : therefore, it realizes a
particular, $\left( D0-D6\right) $-dependent finite transformation of $Sp(4,%
\R )$, which mimicks the action of Freudenthal duality over the $%
D0-D6 $ element (\autoref{1-z}).

\subsection{Gauged $\Aut\left(\mathfrak{F}\right) $ realisation: $\Delta<0$}
\label{F-Dual-D0-D6-U}

In the $\Delta <0$ locus of $\mathfrak{F}$ (on which the action of $%
\Aut\left( \mathfrak{F}\right) $ is always transitive, thus defining a unique
orbit $\mathcal{O}_{\Delta <0}$ of $\mathfrak{F}$), we deal with the issue
of mimicking the action of Freudenthal duality by an $\Aut\left( \mathfrak{F}%
\right) $  transformation (which will generally be local in $\mathfrak{F}$), and consider the following (commutative) diagram:%
\be\label{diagram-1}
    \xymatrix{
          x  \ar[dd]_{\sim}      \ar[rr]^{U} & &x_{can} \ar[dd]^{\sim} \\
          \\
        *[l]{\tilde{x} =M_{x}x}   \ar[rr]_{U'} & &*[r]{\widetilde{x}_{can}=M_{x_{can}}x_{can}}  \\
        }
            \ee
where $x$, $x_{can}\in \mathcal{O}_{\Delta <0}$. Here,  $x_{can}$ denotes a convenient  \textquotedblleft canonical"
  representative that can be defined in a uniform manner for all relevant FTS as in \cite{Borsten:2011nq}. The corresponding $\Aut(\FTS)$ transformations taking $x$ and $\tilde{x}$ to $x_{can}$ and $\tilde{x}_{can}$, respectively, are denoted $U$ and $U^{\prime}$. Similarly,  $M_{x}$ and $M_{x_{can}}$ are the gauged $\Aut\left( \mathfrak{F%
}\right) $ transformations that send $x$ and $x_{can}$ to $\tilde{x}$ and  $\tilde{x}_{can}$, respectively. Since the square commutes we free to pick a convenient canonical representative.

Generally, \textit{at least} for (non-degenerate) reduced FTS's, the
homogeneous space $\mathcal{O}_{\Delta <0}$ can be written as%
\begin{equation}
\mathcal{O}_{\Delta <0}=\frac{Conf\left( J_{3}\right)}{Str_{0}\left( J_{3}\right) },
\end{equation}%
where we recall that $\Aut\left( \mathfrak{F}\left( J_{3}\right) \right)
\simeq Conf\left( J_{3}\right) $, $Conf\left( J_{3}\right) $ and $%
Str_{0}(J_{3})$ respectively denote the conformal and reduced structure
groups of the cubic Jordan algebra $J_{3}$.\medskip

For simplicity's sake, let us  assume $Str_{0}\left( J_{3}\right)
=Id$ (namely, there is no continuous nor discrete stabiliser for $\mathcal{O}%
_{\Delta <0}$, which thus is a group manifold: $\mathcal{O}_{\Delta
<0}\cong\Aut\left( \mathfrak{F}\right) $). Actually, this
only holds for the $T^{3}$ model of $\mathcal{N}=2$, $D=4$ supergravity,
associated to the simplest example of (non-degenerate) reduced FTS \cite%
{Bellucci:2008sv,small-orbits}. Let us also choose a convenient  representative of $%
\mathcal{O}_{\Delta <0}$. An obvious choice is given by the $D0-D6$
configuration (\autoref{1-z}), $%
x_{can}=x_{D0D6}$,  which makes the $Str_{0}\left( J_{3}\right) $ stabiliser of $\mathcal{O}_{\Delta <0}$  manifest for all reduced FTS.

The assumption $%
Str_{0}\left( J_{3}\right) =Id$ implies $U^{\prime }=U$. From (\autoref{diagram-1}) we have%
\begin{equation}
\begin{array}{ll}
\widetilde{x}_{D0D6}&=\widetilde{Ux} \\
&=U^{\prime }M_{x}x=U^{\prime }\widetilde{x}=\widetilde{%
U^{\prime }x}; \\
\end{array}
\label{ress}
\end{equation}%
where the last step of the second line follows from the fact that $\Aut\left(
\mathfrak{F} \right) $  and Freudenthal duality
commute \cite{Duff-FD}. Thus, applying
Freudenthal duality to (\ref%
{ress}) one obtains%
\begin{equation}
Ux=U^{\prime }x\Rightarrow x=U^{\prime -1}Ux\Leftrightarrow \left.
\begin{array}{r}
U^{\prime -1}U \\
U^{-1}U^{\prime }%
\end{array}%
\right\} \in Stab\left( x\right) =Id\Leftrightarrow U=U^{\prime
}~\blacksquare
\end{equation}
Consequently, any reasoning involving the diagram (\autoref{diagram-1}) is
independent  of the  choice of $x_{can}$; indeed, the $\Aut\left(
\mathfrak{F}\left( J_{3}\right) \right) $ transformation connecting any two
elements of $\mathcal{O}_{\Delta <0}=Aut\left( \mathfrak{F}\left(
J_{3}\right) \right) $ (say $x_{can}$ and $x_{can}^{\prime }$) will be
unique, since $\Aut\left( \mathfrak{F}\left(
J_{3}\right) \right)$ is free on the orbit
$\mathcal{O}_{\Delta <0}$ by assumption.

In order to determine the  gauged  $\Aut \left( \mathfrak{F}%
\left( J_{3}\right) \right) $ transformation $M_{x_{D0D6}}$ mimicking
Freudenthal duality acting on $x_{can}=x_{D0D6}$, \textit{i.e.} such that $%
\widetilde{x}_{D0D6}=M_{x_{D0D6}}x_{D0D6}$ (\textit{cfr.} (\autoref{diagram-1}%
)), we \ will use the \textquotedblleft $T^{3}$ degeneration" of the quantum
information symplectic frame of the $STU$ model (see \textit{e.g.} \cite%
{Borsten:2008wd, KL-QIT, GLS, Bellucci:2008sv, small-orbits}). By recalling
that the usual parametrization of an element $x\in \mathfrak{F}\left(
J_{3}\right) $ as a formal $2\times 2$ matrix reads \cite{Duff-FD}%
\begin{equation}
x=\left(
\begin{array}{cc}
\alpha & A \\
B & \beta%
\end{array}%
\right) ,~\alpha ,\beta \in \mathds{R},~A\in J_{3},~B\in \overline{J_{3}},
\label{FTS-p}
\end{equation}%
the relation between the quantum information symplectic frame $x=a$, with $%
a_{abc}=a_{(abc)}$ ($a,b,c=0,1$), the $4D/5D$ special coordinates'
symplectic frame $x=\left( p^{0},p^{1},q_{0},q_{1}\right) ^{T}$ and the FTS
parametrization (\autoref{FTS-p}) for the $T^{3}$ model ($J_{3}=\mathds{R}$)
reads as follows :%
\begin{equation}
\begin{array}{lllll}
\text{QIT~frame:} & a_{000} & a_{111} & a_{001} & a_{110}, \\
4D/5D~\text{frame:} & p^{0} & q_{0} & -p^{1} & q_{1}/3, \\
\text{FTS~frame:} & \beta & -\alpha & -A & B/3,%
\end{array}%
\end{equation}%
which is consistent with the \textquotedblleft $T^{3}$ degeneration" of
(175) and (183), Table VII of \cite{Duff-FD}.

Thus, we are searching for a $2\times 2$ matrix $\mathbf{M}$ such that
\begin{equation}
\widetilde{a_{(abc)D0D6}}=\mathbf{M}_{a}^{a^{\prime }}\mathbf{M}%
_{b}^{b^{\prime }}\mathbf{M}_{c}^{c^{\prime }}a_{(a^{\prime }b^{\prime
}c^{\prime })D0D6}\equiv M_{x_{D0D6}}\left(
\begin{array}{cc}
-q_{0} & 0 \\
0 & p^{0}%
\end{array}%
\right).  \label{condd-1}
\end{equation}%
Straightforward algebra yields the unique solution:%
\begin{equation}
\mathbf{M}(z)=sgn\left( z\right) \left(
\begin{array}{cc}
0 & -z^{-1/3} \\
z^{1/3} & 0%
\end{array}%
\right), \qquad z = q_0/p^0.
\label{M-bold}
\end{equation}%
Note that if and only if $z=\pm 1\Leftrightarrow p^{0}=\pm q_{0}$,
the matrix $\mathbf{M}(z=\pm 1)$ (\autoref{M-bold}) (and thus, through (\ref%
{condd-1}), $M_{x_{D0D6}}$) belongs to $SL(2,\mathds{Z})$ :%
\begin{equation}
\mathbf{M}(z=\pm 1)=\pm \left(
\begin{array}{cc}
0 & -1 \\
1 & 0%
\end{array}%
\right) =\pm \varepsilon \in SL(2,\mathds{Z}),  \label{jazz!}
\end{equation}%
where $\varepsilon $ is nothing but the symplectic $2\times 2$ metric $%
\Omega _{2\times 2}$ :%
\begin{equation}
\varepsilon \equiv \left(
\begin{array}{cc}
0 & -1 \\
1 & 0%
\end{array}%
\right) \equiv \Omega _{2\times 2}.
\end{equation}%
Note that for $p^{0}, q_0 =\pm1$  the black hole is
 \emph{projective}, \textit{cfr.} the treatment in the quantized
charge regime presented in \cite{Duff-FD}).
The integral automorphism group  $\Aut (\FTS_{\mathds{Z}})$ acts transitively on projective
charge configurations.

With the generalisation to arbitrary reduced FTS in mind, it is useful to reexpress $%
M_{x_{D0D6}}(z)$, as defined by (\autoref{condd-1}), through the elementary
$\Aut(\mathfrak{F})$ transformations
defined  in \cite{Seligman:1962, Brown:1969} for generic reduced FTS
\begin{equation}
M_{x_{D0D6}}(z)=\psi \left( -|z|^{-1/3} \right) \circ \varphi \left( |z|^{1/3} \right) \circ \psi \left( -|z|^{-1/3} \right), \label{Krut-j-1}
\end{equation}%
where
\be
\label{ftstrans}
{
\small
\begin{split}
\varphi(C):\pmtwo{\alpha}{A}{B}{\beta}&\mapsto \pmtwo{\alpha+(B,C)+(A,C^\sharp)+\beta N(C)}{A+\beta C}{B+A\times C +\beta C^\sharp}{\beta};\\[4pt]
\psi(D):\pmtwo{\alpha}{A}{B}{\beta}&\mapsto \pmtwo{\alpha}{A+B\times D +\alpha D^\sharp}{B+\alpha D}{\beta+(A,D)+(B,D^\sharp)+\alpha N(D)};\\[4pt]
T(\tau):\pmtwo{\alpha}{A}{B}{\beta}&\mapsto \pmtwo{\lambda^{-1}\alpha}{\tau A}{^t\tau^{-1} B}{\lambda\beta};
\end{split}
}
\ee
where $C,D\in J_3$ and $\tau\in Str(J_3)$ s.t. $N(\tau A)=\lambda N(A)$.  In this form it is  straightforward to generalise to arbitrary cubic Jordan algebra as follows. Consider  the F-dual pair given by
\begin{equation}
w=\left(
\begin{array}{cc}
\alpha & 0 \\
0 & \beta%
\end{array}%
\right),\qquad
\tilde{w}=\sgn(\alpha\beta)\left(
\begin{array}{cc}
\alpha & 0 \\
0 & -\beta%
\end{array}%
\right).
\label{FTS-dualnonBPS}
\end{equation}%
For base point $E$ and $c,d\in\R$  we find
\be
w' = \psi \left( dE \right) \circ \varphi \left(cE \right) \circ \psi \left( dE \right) (w)
\ee
is given by
\be
{
\small
w'=\left(
\begin{array}{cc}
 (c d+1)^3\alpha+ c^3 \beta & (c d+1) \left(\alpha d^2 (c d+2)^2+\beta c (c d+1)\right)E \\
 (c d+1) \left(\beta c^2+\alpha d \left(c^2 d^2+3 c d+2\right)\right)E &  d^3 (c d+2)^3 \alpha +  (c d+1)^3\beta \\
\end{array}
\right).
}
\ee
Hence, on setting
$$d=-\sgn(\alpha\beta)(\beta/\alpha)^{1/3}=-|\beta/\alpha|^{1/3}$$
and
$$c=\sgn(\alpha\beta)(\alpha/\beta)^{1/3}=|\beta/\alpha|^{1/3}$$
we obtain
\be
\label{Fdualexample}
w'=\sgn(\alpha\beta)\left(
\begin{array}{cc}
\alpha & 0 \\
0 & -\beta%
\end{array}\right)=\tilde{w},
\ee
as required. Using \autoref{diagram-1} this gives an explicit realisation of a U-duality relating   any F-dual pair $x, \tilde{x}$ with $\Delta(x)<0$. Explicitly, setting $x_{can}=w$ using $U$ and then applying $M_w= \psi \left( dE \right) \circ \varphi \left(cE \right) \circ \psi \left( dE \right)$ we have
\be
\tilde{x} = U'^{-1} M_w U x.
\ee
Note, since any $x$ in a $\Delta(x)<0$ orbit has stabiliser $Str_0(J_3)$ the U-duality transformations are generically   non-unique,
\be
\tilde{x} = S_{\tilde{x}} U'^{-1} S_{\tilde{w}} M_w S_w U S_x x
\ee
for arbitrary $S_y\in Stab(y)\subset\Aut(\FTS)$.

It is also straightforward to define a $\hat{w}\in\FTS$ such that
\be
\exp\left [\frac{\pi}{2}\Upsilon_{\hat{w}}\right]w = \tilde{w}.
\ee
where $\hat{w}$ is determined by $w$.
Recall, for $x=(\alpha, \beta, A, B),\; y=(\gamma, \delta, C,D)$,  the Freudenthal product
\[
\wedge:\FTS\otimes\FTS\rightarrow\text{Hom}(\FTS)
\]
is defined by,
\be
x\wedge y\equiv \Phi_{(\phi, X, Y, \nu)}, \quad\textrm{where}\quad
\left\{ \begin{array}{lll}
\phi    &=&-\frac{1}{2}(A\vee D+C\vee B),\\[4pt]
X           &=&-\frac{1}{4}(B\times D-\alpha C- \gamma A),\\[4pt]
Y           &=&\frac{1}{4}(A\times C-\beta D-\delta B),\\[4pt]
\nu     &=&\frac{1}{8}(\Tr(A,D)+\Tr(C,B)-3(\alpha\delta+\gamma\beta)),\\
\end{array}\right.
\ee
and $A\vee B\in\mathfrak{Str}{J}$ is defined by
$$(A\vee B)C\equiv \frac{1}{2}\Tr(B,C)A+\frac{1}{6}\Tr(A,B)C-\frac{1}{2} B\times(A\times C).$$
The action of $\Phi:\FTS  \rightarrow\FTS$ is given by
\be\label{eq:ftslieaction}
\Phi_{(\phi,X,Y,\nu)}
\begin{pmatrix}
\alpha&A\\
B&\beta
\end{pmatrix}=
\begin{pmatrix}
\alpha\nu+(X, B)&\phi A-\frac{1}{3}\nu A+Y\times B +\beta X \\
-^t\phi B+\frac{1}{3}\nu B+X\times A+\alpha Y&-\beta\nu+(Y, A)
\end{pmatrix}.
\ee
The set of all such homomorphisms yields the automorphism Lie algebra,
\be
\mathfrak{Aut}(\mathfrak{F})=\{\Phi(\phi,X,Y,\nu)\in \Hom_{\R}(\mathfrak{F})|\phi\in\mathfrak{Str}_{0}(\mathfrak{J}), X,Y\in\J, \nu\in\R\},
\ee
where the Lie bracket
\be
[\Phi(\phi_1,X_1,Y_1,\nu_1),\Phi(\phi_2,X_2,Y_2,\nu_2)]=\Phi(\phi,X,Y,\nu)
\ee
is given by
\be
\begin{split}
\label{aut_commutators}
\phi&=[\phi_1,\phi_2]+2(X_1\vee Y_2-X_2\vee Y_1),\\
X&=(\phi_1+\frac{2}{3}\nu_1)X_2-(\phi_2+\frac{2}{3}\nu_2)X_1,\\
Y&=({}^t\phi_2+\frac{2}{3}\nu_2)Y_1-(^t\phi_1+\frac{2}{3}\nu_1)Y_2,\\
\nu&=\Tr(X_1,Y_2)-\Tr(Y_1,X_2).\\
\end{split}
\ee

Noting  that
\be
 \label{exptrans}
\varphi(C) = \exp\left[\Phi{(0,C,0,0)}\right], \qquad
\psi(D) = \exp\left[\Phi{(0,0,D,0)}\right]
\ee
and, from \autoref{aut_commutators}, that
\be
e\equiv \Phi{(0,E,0,0)}, \quad f\equiv \Phi{(0,0, E,0)},\quad h\equiv \Phi{(0,0,0,3)},
\ee
generates an $\mathfrak{sl}(2, \R)$ subalgebra,
\be
[e,f]=h,\quad  [h, e]=2e,\quad [h, f]=-2f,
\ee
we find
\be
\psi(dE)\circ\varphi(-E/d)\circ\psi(dE) = \exp\left [\frac{\pi}{2}\Phi{(0,-E/d, dE,0)}\right],
\ee
which on setting  $d=-|\beta/\alpha|^{1/3}$ gives the F-dual transformation \autoref{Fdualexample}.

Using
\be
4x\wedge x (y) = 3T(x,x,y)+\{x, y\}x\ee
and for $\hat{w}=2(0, 0, \sqrt{d}E, E/\sqrt{d})$
\be
\hat{w}\wedge w = \Phi{(0,-1/dE, dE,0)}
\ee
we have
\be
\exp\left [\frac{\pi}{2}\Upsilon_{\hat{w}}\right]w = \tilde{w}.
\ee

\subsection{Gauged  $\Aut \left(\mathfrak{F}\right)$ realisation: $\Delta>0$}
\label{sec:Delta>0-transf}
\label{F-Dual-D0-D4}
\label{sec:94}

As for the $\Delta (x)<0$ case treated above, when $\Delta (x)>0$ the action
of Freudenthal duality on $x$ can generally be realised by linear
 gauged $Sp(2N,\mathds{R})$  (recalling
that $\dim _{\mathds{R}}\mathfrak{F}=2N$) or    $
\Aut(\mathfrak{F})$ transformations. For simplicity's sake, we confine
ourselves here to the study of the  gauged
$\Aut(\mathfrak{F})$ transformations
\footnote{\textit{At least} in (non-degenerate) reduced FTS's, a similar treatment along the
non-BPS $Z_{H}=0$ orbit ($\Delta >0$) can be given. Within
(non-degenerate) reduced FTS's, the smallest model exhibiting a non-BPS $%
\Delta >0$ orbit is the $ST^{2}$ model \protect\cite{Bellucci:2008sv}
}.

Concerning the $\Delta >0$ locus of $\mathfrak{F}$, it is generally
stratified in two or more orbits under the non-transitive action of  $\Aut (%
\mathfrak{F})$. In the following treatment,  we will
consider the particularly simple case of the $T^{3}$ model, in which such a
stratification does not take place, and thus the $\Delta >0$ locus of $%
\mathfrak{F}(\mathds{R})$ corresponds to a unique  $\Aut (\mathfrak{F}%
)=SL(2,\mathds{R})$ supersymmetric ($1/2$-BPS) orbit \cite{Bellucci:2008sv,
small-orbits} :%
\begin{equation}
\mathfrak{F}(\mathds{R})_{\Delta >0}=\mathcal{O}_{\Delta >0}\cong SL(2,%
\mathds{R})/\Z_3.
\end{equation}%
The  non-trivial, discrete, stabiliser of the $\mathcal{O}_{\Delta >0}$ orbits for the
$T^{3}$ model is up to conjugation given by the  $\mathds{Z}%
_{3}\subset SO(2)\subset SL(2,\mathds{R})$, generated by \cite{small-orbits}%
\begin{equation}
\hat{M}\equiv \frac{1}{2}\left(
\begin{array}{cc}
-1 & \sqrt{3} \\
-\sqrt{3} & -1%
\end{array}%
\right),~~~\text{det}\hat{M}=1,~~~\hat{M}^{-1}=\hat{M}^{T}.  \label{M-tilde}
\end{equation}%
.

As computed  below, the Freudenthal dual of
the $D0-D4$ representatives of the orbits $\mathcal{O}_{\Delta >0}$
are given by  $D2-D6$ elements
\begin{equation}
\widetilde{x}_{D0D4}= \frac{1}{\sqrt{q_{0}p^{3}}}
\left( -p^{3},0,0,3q_{0}p^{2}\right) .
 \label{F-dual-D0-D4}
\end{equation}%
In order to determine the  gauged  $\Aut \left( \mathfrak{F}%
\left( J_{3}\right) \right) $ transformation $M_{x_{D0D4}}$ mimicking, along
$\mathcal{O}_{\Delta >0}$, Freudenthal duality acting on $x_{can}=x_{D0D4}$,
\textit{i.e.} such that $\widetilde{x}_{D0D4}=M_{x_{D0D4}}x_{D0D6}$ (\textit{%
cfr.} (\autoref{diagram-1})), we \ will again use the \textquotedblleft $T^{3}$
degeneration" of the quantum information symplectic frame of the $STU$
model, namely we search for a $2\times 2$ matrix $\mathds{M}$ such that
\begin{equation}
\widetilde{a}_{(abc)D0D4}=\mathds{M}_{a}^{a^{\prime }}\mathds{M}%
_{b}^{b^{\prime }}\mathds{M}_{c}^{c^{\prime }}a_{(a^{\prime }b^{\prime
}c^{\prime })D0D4}\equiv M_{x_{D0D4}}x_{D0D4},  \label{condd-2}
\end{equation}%
where $x_{D0D4}$ in the FTS parametrization (following the conventions of
\cite{small-orbits}) reads%
\begin{equation}
x_{D0D4}=\left(
\begin{array}{cc}
-q_{0} & p \\
0 & 0%
\end{array}%
\right) .  \label{1-zz}
\end{equation}%
Long but straightforward algebra yields the twofold solution ($y=q_0/p>0$):%
\begin{equation}
\mathds{M}_{\pm }(y)=\left(
\begin{array}{cc}
\pm \frac{\sqrt{3}}{2} & \frac{1}{2}y^{-1/2} \\
-\frac{1}{2}y^{1/2} & \pm \frac{\sqrt{3}}{2}%
\end{array}%
\right).  \label{M-mathbold}
\end{equation}%
  $%
M_{x_{D0D4}}(y)$ (defined by (\autoref{condd-2})) can be
realized in terms of the  $\Aut (\mathfrak{F}(\mathds{R}))$ transformations \autoref{ftstrans} as
follows:
\begin{equation}
M_{x_{D0D4}}(y)=T\left( \pm \frac{\sqrt{3}}{2}\right) \circ \varphi \left(\mp\frac{\sqrt{3}}{4}y^{1/2}\right) \circ \psi \left( \pm\frac{1}{\sqrt{3}}y^{-1/2}\right).
\label{Krut-2}
\end{equation}%

It is interesting to note that for $y=1\Leftrightarrow q_{0}=p$, the matrix $%
\mathds{M}_{\pm }(y=1)$ (\autoref{M-mathbold}) (and thus, through (\autoref{condd-2}%
), $M_{x_{D0D4}}$) does not belong to $SL(2,\mathds{Z})$:%
\begin{equation}
\mathds{M}(y=1)=\left(
\begin{array}{cc}
\pm \frac{\sqrt{3}}{2} & \frac{1}{2} \\
-\frac{1}{2} & \pm \frac{\sqrt{3}}{2}%
\end{array}%
\right) \notin SL(2,\mathds{Z}).
\end{equation}%
However, remarkably, only for $y=1$ another solution to (\autoref{condd-2}) can
be found (the subscript \textquotedblleft $add$" stands for additional;
recall (\autoref{jazz!})):%
\begin{equation}
\mathds{M}_{add}^{y=1}=\left(
\begin{array}{cc}
0 & -1 \\
1 & 0%
\end{array}%
\right) =\varepsilon =\mathbf{M}(z=1)\in SL(2,\mathds{Z}),  \label{M-add}
\end{equation}%
with $\varepsilon $ defined by (\autoref{epsilon!}).  Due to the existence of the additional
solution (\autoref{M-add}), the integral (projective) case is obtained for $y=1$
(by further setting $p\in \mathds{Z}$) (\textit{cfr.} the
treatment in the quantized charge regime, presented in \cite{Duff-FD}%
).

Recall, since we are assuming $\Delta(x)>0$, we  can use the Freudenthal rotation in   $\Aut (\FTS)$,
\be
\exp \left[\frac{\pi}{2}\overline{\Upsilon}_x\right] x = \tilde{x}.
\ee
Specialising to a generic
\footnote{Generically, this is actually a larger class of charge configurations, but in the
special cases of the $T^3, ST^2, STU$ models it is precisely the $D0D4$ subsector.
For $\mathcal{N}=8$,  imposing $A$ is diagonal restricts to the $D0D4$ subsector.}
 element in the $D0D4$ system
\be
w= \begin{pmatrix}\alpha&A\\0&0\end{pmatrix}, \qquad \tilde{w}= \frac{1}{\sqrt{-\alpha N(A)}}\begin{pmatrix}0&0\\\alpha A^\sharp& N(A)\end{pmatrix},
\ee
and using
\be
\overline{\Upsilon}_x = \frac{4}{3\sqrt{\Delta(x)}} x \wedge x,
\ee
we have
\be
\overline{\Upsilon}_w = \frac{1}{3\sqrt{-\alpha N(A)}} \Phi(0, \alpha A, A^\sharp, 0),
\ee
so that
\be
\exp \left[\frac{\pi}{6\sqrt{-\alpha N(A)}} \Phi(0, \alpha A, A^\sharp, 0)\right] w = \tilde{w}.
\ee
Further restricting to the $T^3$ model and setting $\alpha=-q_0, A=p$ we find,
\be
\frac{\pi}{6\sqrt{-\alpha N(A)}} \Phi(0, \alpha A, A^\sharp, 0) = \frac{\pi}{6} \Phi(0, -y^{1/2}, y^{-1/2}, 0)
\ee
and
\be
\exp \left[\frac{\pi}{6} \Phi(0, -y^{1/2}, y^{-1/2}, 0)\right] = T\left(  \frac{\sqrt{3}}{2}\right) \circ \varphi \left(-\frac{\sqrt{3}}{4}y^{1/2}\right) \circ \psi \left( \frac{1}{\sqrt{3}}y^{-1/2}\right)
\ee
as given in \autoref{Krut-2}.

\subsection{Non-trivial orbit stabilizers}
\label{General-Treatment-Mimicking}

Here, generalizing the reasoning at the end of   \autoref{F-Dual-D0-D6-U}, we
want to reconsider the diagram (\autoref{diagram-1}), and generalize the
treatment to the case in which (regardless of the sign of $\Delta (x)$) the $%
\Aut\left( \mathfrak{F}\right) $-orbit to which $x\in \mathfrak{F}$ belongs
is endowed with a non-trivial stabilizer $\mathcal{H}$, such that the
corresponding homogeneous (generally non-symmetric) manifold can be written
as%
\begin{equation}
x\in \mathcal{O}=\frac{Aut\left( \mathfrak{F}\right) }{\mathcal{H}}\subset
\mathfrak{F}.
\end{equation}%
Let us deal with the issue of mimicking the action of Freudenthal duality by
an  $\Aut \left( \mathfrak{F}\right) $ finite transformation (which will
generally be \textquotedblleft gauged" in $\mathfrak{F}$), and consider
again the (commutative) diagram (\autoref{diagram-1}). In general\footnote{%
If $U$ and $U^{\prime }$ are more general finite transformations, such as
\textquotedblleft gauged" symplectic (not belonging to $\Aut(\mathfrak{F})$)
or anti-symplectic transformations, things become more complicated, since,
for instance, finite transformations of the pseudo-Riemannian non-compact
non-symmetric coset $Sp(2n,\mathds{R})/\Aut(\mathfrak{F})$ do not generally
preserve the $\Aut(\mathfrak{F})$-orbit structure of $\mathfrak{F}$, and thus
they generally do not commute with Freudenthal duality.} ($U,U^{\prime }\in \frac{Aut\left(
\mathfrak{F}\right) }{\mathcal{H}}$), it holds that%
\begin{equation}
\begin{array}{l}
\widetilde{x}_{can}=\widetilde{Ux}; \\[4pt]
\widetilde{x}_{can}=U^{\prime }\widetilde{x}=\widetilde{U^{\prime }x},%
\end{array}
\label{jazz!!}
\end{equation}%
where in the last step of the second line we used the commutativity of
Freudenthal duality and  $\Aut \left( \mathfrak{F}\right) $
. Thus, (\autoref{jazz!!}) implies%
\be
 x= U^{\prime-1}Ux ~~\Leftrightarrow ~~
 U^{\prime -1}U=Z_x\in Stab(x)
  \label{res-1}
\ee
On the other hand:%
\be
 \widetilde{x}=\widetilde{U^{-1}x_{can}}, \qquad
\widetilde{x}=U^{\prime -1}\widetilde{x}_{can}=\widetilde{U^{\prime
-1}x_{can}},%
\label{jazz!!!}
\ee
where again in the last step  we used the commutativity of
Freudenthal duality and  $\Aut \left( \mathfrak{F}\right) $. Thus, (\ref%
{jazz!!!}) implies%
\be
x_{can}=U^{\prime }U^{-1}x_{can} ~~\Leftrightarrow ~~
U^{\prime }U^{-1}= Z_{x_{can}} \in Stab(x_{can}).
 \label{res-2}
\ee
Let us observe that%
\begin{equation}
Stab(x)=Stab(\widetilde{x}),  \label{Th}
\end{equation}%
since for $Z_{x}\in Stab(x)$ and $Z_{\widetilde{x}}\in Stab(%
\widetilde{x})$ we have
\be
Z_{\widetilde{x}}\widetilde{x}=\widetilde{x}=\widetilde{Z_{x}x}=Z_{x}%
\widetilde{x},%
\label{blues!}
\ee
and vice versa,  implying (\autoref{Th}).

By virtue of results (\autoref{res-1}), (\autoref{res-2}) and (\autoref{Th}), one can
write%
\begin{equation}
U^{\prime }x=Z_{x_{can}}UZ_{x}x.
\end{equation}

Moreover, it is here worth commenting that, in presence of non-trivial $%
\mathcal{H}$, any reasoning involving the diagram (\autoref{diagram-1}) is
actually dependent from the actual choice of $x_{can}$; indeed, the $%
Aut\left( \mathfrak{F}\left( J_{3}\right) \right) $ transformation
connecting any two elements of $\mathcal{O}$ (say $x_{can}$ and $%
x_{can}^{\prime }$) will not be unique, for the reasons highlighted above.

\section{Summary, concluding and further remarks}
\label{sec9}

The purpose of this work has been  to present, extend and clarify old and new results concerning
general Freudenthal transformations, which generalise  Freudenthal duality, while  filling in necessary details not yet appearing in the mathematics or physics literature.

We begin with   a detailed and self-contained  treatment of FTS, groups of type $E_7$, F-duality and GFT in 
\autoref{sec:2}-\autoref{sec:5}, laying the groundwork for the subsequent sections that apply the formalism to Einstein-Maxwell-Scalar (super)gravity theories.

In \autoref{sec:pure}  we  study the entropy properties of
 $\N=2$, $D=4$ pure supergravity from the point of view of FTS formalism, where is it given by a non-reduced
 FTS with  positive-definite  quartic invariant.
 In this case,   general Freudenthal transformations are nothing other than the familiar $U(1)$ electromagnetic  duality and Freudenthal duality is an anti-involutive duality transformation.

In \autoref{sec:axion} we  considered the axion-dilaton model, an $\N=2, d=4$ supergravity
 minimally coupled to
one vector multiplet, which can be considered a consistent truncation of $\N=4$ supergravity.
The mathematical structure is in this case a two dimensional FTS with negative-semi-definite quartic invariant.
 The corresponding extremal BH  is non-BPS.
 A $SO_0(1,1)$  subgroup of the GFTs leave invariant the entropy.
 Freudenthal duality reduces  in this case to a  U-duality transformation.
This model, as the pure supergravity studied before, cannot be uplifted to $D=5$, which corresponds directly to the fact that the associated  FTS is not reduced.

In \autoref{sec:reduced}
we proceeded to the analysis of $\N=2,D=4$ supergravities with a  $D=5$ origin.  The mathematical
structure of these models is that of a  FTS derived from a cubic Jordan Algebra.
In first place we study the  $T^3$ model,  or in Freudenthal terminology a  $\FTS(J_3=\R)$
structure. This model is represented by the unique reduced FTS with $\dim \FTS=4$ as vector
space. The automorphism group is a four dimensional representation of $SL(2, \R)$  which can be
decomposed into $SO(1,1)$  representations, making the $D=5$ origin manifest.  The full space is split into two two-dimensional planes  orthogonal (with respect to the symplectic bilinear form), which can be identified respectively with purely $D6-D0$ or $D4-D2$ configurations. Other identifications
are, however, possible, for example a $D4-D0/D6-D2$ decomposition. General Freudenthal transformations
leave invariant the entropy for configurations chosen in each of the planes for
every decomposition.
Any $D6-D0$ configuration has a negative value for the quartic invariant, corresponding to
extremal non-BPS BHs. Meanwhile, the  $D4-D2$ configurations correspond to BHs with either  positive (BPS or non-BPS) or negative (non-BPS)
quartic invariant. Each non-BPS $D6-D0$ BH can be put in correspondence with a BPS
$D4-D2$ one, with the same entropy. Mathematically this correspondence is performed
by a $\Upsilon_x$ map, which exponentiates to an element of  $\Aut (\FTS)$.
It is also possible to consider an  initial family of only $D4-D0$ configurations. This family
includes both BPS and non-BPS  BHs (respectively configurations with positive and negative
quartic invariants). The application of general Freudenthal transformations to any member
of this family generates a two-parameter subset of the full $D6-D4-D0-D2$ configuration space.
The configurations with $D6$ or $D4$ charges alone are ``small'' ($\Delta=0$) BH. They are
related among themselves by general Freudenthal transformations forming a $D6-D4$ plane, which
is $\Upsilon$-mapped to a $D2-D0$ plane.
Similar results are obtained in general theories beyond $T^3$ containing an arbitrary number
of charges, they are studied in full detail in  \autoref{sec:D0D6} and \autoref{sec:D0D4}.

General Freudenthal transformations, and in particular  Freudenthal duality $\sim $,
 will  preserve the orbit structure for $\Delta <0$. This is trivially true since,
at least for (non-degenerate) reduced FTS,  $\Aut(\mathfrak{F})$
has a transitive action on elements with a given $\Delta <0$. 
The situation is more complicated for the locus of $\Delta>0$ configurations, for which, in general
(with known exceptions)
 are stratified in two or more automorphism orbits.
In \autoref{sec:84} we study this question and conclude
 that the orbit
stratification of the $\Delta >0$ locus of $\mathfrak{F}(\R %
\oplus \R \oplus \R )$ is \textit{preserved} under Freudenthal
duality $\sim $, and by extension by General Freudenthal transformations.
This result generalises to all cases with (non-degenerate)
reduced FTS, as they can either be invectively embedded in, or truncated to, the  $STU$ model.

In \autoref{sec:90} we show, in different examples,
how the action of General Freudenthal transformations, and, in particular,
Freudenthal duality can be
mimicked/undone by  finite U-duality  ($\Aut(\mathfrak{F})$),
transformations which are gauged in   that they depend on the element of
$\mathfrak{F}$ they are applied onto.
We restrict to  two situations. First, for configurations within the   $\Delta <0$ locus of $\mathfrak{F}$, 
the action of  $\Aut \left( \mathfrak{F}\right) $ is always transitive on elements of a given $\Delta$, thus defining a unique one-parameter family of
orbits $\{\mathcal{O}\}_{\Delta(x) <0}$ of $\mathfrak{F}$. The
 $\Delta>0$ locus of $\FTS$  is generally stratified in two or more orbits for any fixed $\Delta>0$ under the
 non-transitive action of  $\Aut(\FTS)$. However it is possible to find particular
 cases where this stratification does not take place. For example,  the  $T^3$ model has 
 a unique orbit for all $\Delta>0$ under the $SL(2, \R)$ automorphism group. Hence, for the $T^3$ model all BH with $\Delta>0$ are $(1/2)-$BPS. This
 example is studied in \autoref{sec:94}.

The entropy of a linear superposition of  configurations $x,y$  is given by
\beq
\frac{S_{BH}^2(x+y)}{4\pi} &=&\mid \Delta(x+y)\mid,
\eeq
which is not the sum of the entropies of the individual constituents  
\beq
\Delta(x+y) &=& \Delta(x)+\Delta(y)+3\{x',y\}+3\{x,y'\}+\{y,\Upsilon_x(y)\}.
\eeq
However for linear combinations of the form given by a General Freudenthal transformation,
$x_F= a x +b\tilde{x}$,
the entropy  of the composite object is simply related to that one of $x$
\beq
\frac{S_{BH}^2(x_F)}{4\pi}  &=&
 \mid a^2+\epsilon b^2 \mid
\frac{S_{BH}^2(x)}{4\pi} .
\eeq
Thus there is a family of configurations for which the entropy is the same to the
entropy of $x$, those with $a^2+\epsilon b^2=\pm1$, with $\epsilon=\sgn \Delta(x)$.

 We can use these results to show that it is possible to construct
  asymptotically ``small'' (zero entropy)  interacting black holes from an initial non-trivial configuration. First, when $\Delta (x)>0$ $ (\epsilon =1$),
the  elements defined by   (see also  \autoref{ee00101})
$x_{F\pm}=x\pm \tilde{x}$
are rank-4 element of $\mathfrak{F}_{x}\subset \mathfrak{F}$
(see \autoref{ee00102} and   \autoref{tt5}).
However, when $\Delta(x)<0$ ($\epsilon =-1$)
   it follows that $x_{F\pm}$ are null elements
\footnote{
The elements $x_{F\pm}$ are actually  rank-1 element of
$\mathfrak{F}_{x}\subset \mathfrak{F}$.
In this case (\autoref{ee00102}, \autoref{tt5})
\beq
(x_{F\pm})'\equiv( x\pm \tilde{x})' &=&0,\\
 \Upsilon_{x_{F\pm}} (y)&=&0,~\forall y\in \mathfrak{F}.
\eeq
In particular it does not have a well defined \~{F}-dual.
}
\begin{equation}
\Delta(x_{F\pm})\equiv\Delta (x\pm \tilde{x})=0.
\label{r-1}
\end{equation}%
with vanishing  Bekenstein-Hawking entropy
\beq
S_{BH}(x\pm \tilde{x}) &=&0.
\eeq
This suggests the  existence of a class of ``two-centered
black hole solutions'' where each centre
is \textquotedblleft large" non-BPS
($\Delta (x)=\Delta (\pm \tilde{x})<0$),
they are interacting since
$\{\tilde{x},x\}\not=0$, yet asymptotically their Bekenstein-Hawking
entropy
vanishes, so the  total system (\textit{%
before} crossing a line of marginal stability) belongs to a
 small \textit{nilpotent}\footnote{
In the $D=3$ language; \textit{cfr.} \textit{e.g.} the nilpotent orbits of $%
\mathfrak{so}(4,4)$ acting on its adjoint irrep. $\mathbf{28}$ in
\protect\cite{Bossard:2011kz}.} orbit. The physical or geometric significance of such configurations remains unclear.

Alternatively,  small BH solutions can constructed by the application of the
properties of
$  \Upsilon_x\in \autFF$ maps.
  The behaviour of $S_{BH}$ (or $\Delta$) on the $\FTS$ or \fxy\  planes is similar
  (see \autoref{sec:41}).
The null elements of $\FTS_x$ and any \fxy\ are ``aligned''. The locus of null entropy,  is
 given by the same \autoref{e3001} which it is independent of $y$.
For $\Delta(x)<0, y\in \fxt$,  any element of the form
\beq
z_\pm&=&\sqrt{\mid\Delta(x)\mid} y \pm \Upsilon_x(y)
\eeq
is null, $\Delta(z_\pm)=0$.  This describes another class of two-centred black hole
configurations, which are interacting since
$\{y,\Upsilon_x{y}\}\not=0$, yet asymptotically their Bekenstein-Hawking
entropy
vanishes
\beq
S_{BH} \left(y \pm \overline\Upsilon_x(y)\right)&=&0.
\eeq
Further work on such small BHs will be presented elsewhere.

The extension of  these results to systems with quantized charges is challenging.  In this case the requirement that
the set of charge vectors $x_f= a x+b \tilde{x}$ belongs to the charge lattice is extremely restrictive.
Let us recall that for the case of Freudenthal duality, demanding that $x,\tilde{x}$ are integers restrict us to a open subset of black holes
where the entropy is necessarily an integer multiple of $\pi$.
The complete characterisation of discrete U-duality invariants,  which may or may not  also be F- and GFT invariant, remains an open question and, hence, so does  the F-dual invariance  of higher order corrections to the entropy.

Let us just present  a simple result. In the case of $\N=8, D=4$, the automorphism group is $E_{7(7)}(\mathds{Z})$ and  $\Delta(x)$ is quantised \cite{
Borsten:2009zy},
$\Delta(x)=0_{[4]}$ or $\Delta(x)=1_{[4]}$, where $n_{[4]}\equiv n \mod4$. The  requirement  that both $x,\tilde{x}$ are integer restricts us  to the subset
of black holes for which $\pm \Delta$ is a perfect square among other conditions \cite{
Borsten:2009zy}.
 Let us explore what happens for integral
GFT. Noting that the F-dual scales linearly $\tilde{(n x)}= n \tilde{x},\ n\in \mathds{Z}$, according to \autoref{ee003b}, we find, 
\beq
\Delta(n x+m \txx) &=& \left(n^2+\epsilon m^2\right)^2\Delta(x),\quad n,m\in\mathds{Z}.\nonumber
\eeq
From the structure of this expression it is obvious that the requirement the entropy being a
 perfect square is automatically preserved under a
GFT.

Finally, we note that the charges of five dimensional stringy black holes may be described in the context of (cubic) Jordan algebras.
The cubic norm defining the relevant algebra determines the BH entropy to lowest order.
The Jordan dual (introduced in \cite{Borsten:2009zy}) is related to the Freudenthal dual of the corresponding 4D model.
The generalization of this 4D/5D correspondence from a general Freudenthal transformation to the corresponding putative 
``General Jordan Transformations'' (GJT), of which the Jordan dual is a particular case, will be treated  elsewhere
\cite{futuretorrente}.

\vspace{0.5cm}
\section*{Acknowledgments}

 The work of LB is supported by a Schr\"odinger Fellowship.
The work of MJD  was supported in part by the STFC under rolling grant ST/P000762/1.
The work of AM has been partially supported by a fellowship from Fundacion Seneca, CARM, Spain.
AM would like to thank the hospitality of the FISPAC research group at the Universidad de Murcia, Spain, where
part of this work has been conducted.
The work of JJFM has been supported by the Spanish Ministry of Education FPU fellowship AP2008-00919 and grants E024-018 and FIS2011-24924.
The work of ETL has been supported in part by the Ministerio de Educaci\'on y Ciencia, grants FIS2011-24924, FIS2015-28521,  Universidad de Murcia project E024-018 and Fundacion Seneca (21257/PI/19 and 20949/PI/18).
ETL would like thank the hospitality of the CERN TH Division, Harvard Physics Department and
Yukawa Institute (Kyoto University) where part of the research has been conducted, he would also like to thank ESV for unvaluable support. LB is grateful to Sergio Ferrara for hospitality at the CERN TH Division, where part of this project was initiated. LB and MJD are grateful to Philip Candelas for hospitality at the Mathematical Institute, University of Oxford. MJD is grateful to Marlan Scully for his hospitality in the Institute for Quantum Science and Engineering, Texas A\&M University. MD acknowledges the Leverhulme Trust for an Emeritus Fellowship and the Hagler Institute for Advanced Study at Texas A\&M for a Faculty Fellowship. 

\appendix

\vspace{0.3cm}
\section{Freudenthal triples: Assorted properties.}
\label{sec:properties}

A summary of some FTS definitions, notation and properties used through
this work.
See Ref.\cite{Brown:1969} for additional ones and proofs.
\begin{align}
\{T(x,y,z), w\} &=2 \Delta (x,y,z,w),\\
x'&\equiv T(x,x,x)\equiv T(x),\\
\Delta(x) &\equiv \Delta(x,x,x,x),\\
3\{T(x,x,y),y'\}&=2\{x,y\}\Delta(x,y,y,y),\\
6\{T(x,w,y),y'\}&=\{x,y\}2\Delta(w,y,y,y)+\{w,y\}2\Delta(x,y,y,y),\\
6\{T(x,y,y),y'\}&=\{x,y\}2\Delta(y),\\
9T\left((T(x,y,y),y,y\right)&=
-2\Delta(x,y,y,y)y-\Delta(y)x-\{x,y\}y',\\
T(x',x,x)       &=-\tfrac{1}{3}\Delta(x) x,\\
T(x',x',x)    &=\tfrac{1}{3}\Delta(x) x',\\
(x')' \equiv T(x',x',x')                &=-\Delta^2(x)x,\\
\Delta(x',x,x,x)&=0,\\
\Delta(x',x',x,x)   &=\tfrac{1}{3}\Delta^2(x),\\
\Delta(x',x',x',x)&=0,\\
\Delta(x',x',x',x')            &=\Delta^3(x).
\end{align}


Some additional properties related to \autoref{e9001}:
\begin{align}
For \quad \Delta(x) =0:& \quad T(a x+b x') = a^3 x',\label{e8001}\\
                &,\quad T( x') = 0,\\
        \Delta(x) <0:        &,\quad T(\sdx x\pm  x') = 0,\\
\Delta(x) >0:&\quad T(\spdx x\pm  x') = \mp 2\Delta(x)^{3/2}\left (\spdx x\mp x'\right).
\end{align}
 Moreover
\begin{align}
 \Delta(x) <0:       &\quad T(x\pm  \txx) = 0,\\
\Delta(x) >0:&\quad T(x\pm  \txx) =
\mp 2\spdx\left ( x\mp \txx\right).
\end{align}

\section{More \FT-plane properties}
\label{secab}

\subsection{Complexification of real \FTS's and \FT-planes}
\label{sec:b1}

The  square of the $\Upsilon_x$ map, for a fixed $x$, is proportional to the identity
($\Delta_x\equiv \Delta(x),\epsilon=\sgn \Delta(x)$)
$$\Upsilon_x^2(y)=-2\Delta_x y$$
for
 any $y\in\fxy$  (see \autoref{ee120}).
 For $\epsilon=1$ this map defines then a local (depending on the point) complex structure
 \footnote{It defines a split complex structure in the  $\epsilon=-1$ case.}, allowing one to
 endow, the originally real, \FTS with the structure of a complex vector space. We may define
 its complexification by extension to
 \beq
 \FTS^{\mathds C}=\FTS\otimes_{\R }\mathds{C}.
 \eeq
 since the algebraic closure of $\mathds{C}$, the map $\Upsilon_x$ is guaranteed to have eigenvalues, proportional to $\pm i$ in the
 $\epsilon=1$ case..
 The space \fxy\  can be then split into one dimensional
 eigenspaces of  $\Upsilon_x$. We write
$$\fxy=(\fxy)_-\oplus(\fxy)_+$$
where, for any $y_\pm\in (\fxy)_\pm$,
\beq
\Upsilon_x( y_\pm) &=& \pm i \sqrt{\Delta_x} y_\pm.
\eeq
Let us choose normalized eigenvectors (the \fxy\ ``light-cone" or ``null'' basis)
as follows
\beq
y_\pm &=& \pm i \sqrt{\Delta_x} y +\ups{y}.
\eeq
It is clear that $\{ y_-,y_+\}=0$.
In addition we have the following properties
(see \cite{Brown:1969},Eqs.(18,19)) ,
\beq
\Delta(y_\pm)&=&0,\label{e90103}\\
\Delta(y_-,y_+,y_+,y_+)&=&\Delta(y_+,y_-,y_-,y_-)=0,\label{e90102}\\
\{ (y_\pm)',y_\mp\}&=&0.
\eeq
Any element  $u\in \fxy$,  such that $u=a y +b\ups{y}$, can be written as
$u=\alpha_+ y_++\alpha_-y_-$, with the coordinates in both
basis related by  $ 2\alpha_\mp=b\pm i a/\sqrt{\Delta(x)}$.
This null basis will be useful in what follows.



We will study now the behaviour of the quadrilinear map $\Delta$ and the
antisymmetric form in this plane \fxy.

For any two vectors $u_1,u_2\in \fxy$ we have the following properties
($u_i=a_i y+b_i \ups{y}=\alpha_{i+} y_{+}+\alpha_{i-}y_-, i=1,2$)
 ($ 2\alpha_{i\mp}=b_i\pm i a_i/\sqrt{\Delta(x)}$),
\beq
\{u_1,u_2\}&=& \left (b_1 a_2-a_1 b_2\right) \{\ups{y}, y\}\\
     &=&\left (b_1 a_2-a_1 b_2\right) 6 \Delta(x,x,y,y),\\
\{\ups{u_1},\ups{u_2}\}&=& \Delta(x)\{u_1, u_2\},\label{e90301}\\
\Delta(u)
&=& \frac{1}{4\Delta_x^2}(a^2+b^2\Delta_x)^2\Delta(y_-,y_-,y_+,y_+),\label{e90101}\\
\Delta(y_-,y_-,y_+,y_+) &=& 4 \Delta(x)^2 \Delta(y)= 4 \Delta( \ups{y})\label{e90104}.
 \eeq
The first two expressions are obtained by direct computation.
To get \autoref{e90101}, we note that
\beq
\Delta(u)&=& a_+^4 \Delta(y_+)+a_-^4\Delta(y_-)+
     3\alpha_+^3\alpha_- \Delta(y_-,y_+,y_+,y_+)+
    \\
    && 3\alpha_+\alpha_-^3\Delta(y_+,y_-,y_-,y_-)+
     +4 \alpha_+^2\alpha_-^2\Delta(y_-,y_-,y_+,y_+)\\
    &=& 4 \alpha_+^2\alpha_-^2\Delta(y_-,y_-,y_+,y_+)\\
&=& \frac{1}{4\Delta_x^2}(a^2+b^2\Delta_x)^2\Delta(y_-,y_-,y_+,y_+)
\eeq
where we have used  ( \autoref{e90103}, \autoref{e90102}).
Finally \autoref{e90104}   is a  particular  case of \autoref{e90101}.

 \autoref{e90301}   can also be written as (for $\Delta(x)\neq 0$),
\beq
\left\{\frac{\ups{u_1}}{\sqrt{\mid \Delta(x)\mid}},\frac{\ups{u_2}}{\sqrt{\mid \Delta(x)\mid}}\right\}&
=& \epsilon\{u_1, u_2\}.
 \eeq
As consequence, for $\epsilon=1$, the map
$$\tilde \Upsilon_x\equiv \frac{1}{\surd \Delta_x} \Upsilon_x$$
preserves the bilinear antisymmetric form $\{,\}$ in each  \fxy\ plane.
This implies necessarily that $\tilde\Upsilon_x$ is a symplectic transformation on the plane.


\subsection{Maximal rank $\FT$-Planes are disjoint}
\label{ss32}

Let us  have two non-degenerate (generated by maximal rank elements) planes $\FT_{x_0}$,$\FT_{x_1}$
generated by
distinct elements $x_0,x_1$ ($x_0\not= x_1$).
We will show that the two planes are, or the same, or disjoint.

Suppose we can find a common element $y\in \FT_{x_0}\cap \FT_{x_1}$ ,
this implies that a) the signs of $\Delta(x_0),\Delta(x_1)$ are the same and
b) also
  $y'\equiv T(y)\in \FT_{x_0}\cap \FT_{x_1}$. Then, we can find coefficients $a_i,b_i,\alpha_i,\beta_i$ such that
\beq
y&=& a_0 x_0+b_0 x_0'=a_1 x_1+b_1 x_1',\\
y'&=& \alpha_0 x_0+\beta_0 x_0'=\alpha_1 x_1+\beta_1 x_1'.
\label{e8002}
\eeq
The coefficients $\alpha_i,\beta_i$ are given  in terms of the $a_i,b_i$
by \autoref{ee001}. Inserting the values of these coefficients we can see that these equations are invertible  as long as
$$ \Delta(y)/\Delta(x_1)= \Delta(y)/\Delta(x_0)=(a_0^2+b_0^2\Delta_0)^2=(a_1^2+b_1^2\Delta_1)^2\not= 0.$$
If $\Delta(y)\not=0$ the equations are invertible:
 we can write any pair ($x_0,x_0')$ or $(x_1,x_1')$ as linear
combinations of $y,y'$. The existence of such linear combinations shows
that $\FT_{y}=\FT_{x_0}=\FT_{x_1}$.

If the common element is not of maximal rank then the situation is different.
If  $\Delta(y)=0$,  the equations are not invertible but then necessarily both $\Delta(x_0)<0,\Delta(x_1)<0$.
$y$ is a ``null'' or ``light-cone'' element, it can be seen it is then  proportional to one of the two vectors
  $$\pm \sdxx{0} x_0+x_0'$$
 and one of the two
    $$\pm \sdxx{1} x_1+x_1'.$$
 We can use these
last expressions to arrive to:
\beq
\pm \sdxx{0} x_0+x_0' &\propto &\pm \sdxx{1} x_1+x_1'.
\eeq
 ``Some'' elements of $\FT_{x_0}$ (a one dimensional subset of them) can be written as linear
 combination of those of $\FT_{x_1}$.

In the case of degenerate planes (generated by non-maximal rank elements), one can show in the same way that the
T-''lines'' $\FT_{x_0},\FT_{x_1}$,
corresponding to
elements $x_0,x_1$ with $\Delta(x_0)=\Delta(x_1)=0$    are
either identical (if  $x_0'=x_1'=0$ or $x_0'\propto x_1'$) or disjoint (if $x_0'\not\propto x_1'$).

\subsection{Rank on the \FT/$\FTS$-plane}

Let it be a generic vector $x_0$ of a given rank and
its associated (possible degenerate)  $\FT$-plane $\FT_{x_0}\equiv\FT_0$ ( we write
 $\Delta_0=\Delta(x_0)$).
We are interested in studying what it can be said about the rank of any element $x\in \FT_0$ on the
 plane. Let us consider the different cases.

Let us assume then a nonzero rank and consider the following possibilities:

\begin{itemize}
\item[A)] If $\Delta_0>0$ then, by \autoref{ee003}, $\Delta(x)>0$, and $Rank(x)=Rank(x_0)=4,\forall x\in \FT_0$.

\item[B)] If $\Delta_0<0$ then, by \autoref{ee003}, we have two possibilities:  $\Delta(x)<0$
and then $Rank(x)=4$  or $\Delta(x)=0$  .

 If $\Delta(x)=0$ then $x$ is a null vector, it belongs to the light-cone
 generated by $ x_\pm=\pm \sdxx{0} x_0+x_0'$, then, by \autoref{ee001}, $x'=T(x)=0$.
For the sake of concreteness, let us take $x=x_+$ .
We can find at least one element, its associated element $x_-$, such that
$$\Upsilon_{x_{+}}(x_{-})\propto -8\mid \Delta_0\mid \sdxx{0} x_{+}\neq 0;$$
similarly would happen if we start by assuming $x=x_-$. We arrive to the conclusion that $Rank(x)=2$
in this case.

\item[C)]If $\Delta_0=0$ ($Rank(x_0)=1,2,3$)
 then, by \autoref{ee003}, $\Delta(x)=0$. We are confronted with two self-excluding possibilities:

C1)If $x_0'\neq 0$ ($Rank(x_0)=3$) then $x'= a^3 x_0'\neq 0$ ($a\neq 0$) and $Rank(x)=3$.
If $a=0$ then $T(x')=0$ and the rank of $x$ is $R=1,2$.

C2)If $x_0'=0$ ($Rank(x_0)=1,2$) then
$x'\propto x_0' = 0,x\propto x_0$.   $\Upsilon_x(y)\propto\Upsilon_{x0}(y)$ and
  $Rank(x)=Rank(x_0)$ in this case.

\end{itemize}

  The results are summarized in  \autoref{tt5}. Similar results are obtained for the $\FTS$-plane.
  \begin{table}[h]
{
\begin{tabular}{|c|c|c|l|c|}\hline
  $Rk(x_0)$ &                    &  Cases                       & &Rank(x) \\ \hline
              4 &$\Delta_0>0$ & $\Delta(x)>0$   & &$4$ \\
              4 &$\Delta_0<0$ & $\Delta(x)<0$   &  &$4$ \\
                 &                    & $\Delta(x)=0$   &
{\small $x'=0$ ,  $x\propto x_\pm$, $\Upsilon_{x+}(x_-)\neq 0$},&$2$\\
3        &$\Delta_0=0,x_0'\neq0$ & $x\not\propto x_0'$ &   $x'\propto x_0'$, & $3$\\
          &                                    &  $x\propto x_0'$ &   $T(x')=0$, & $1,2$\\
1,2     &$\Delta_0=0,x_0'=0$ &   $\Delta(x)=0$  &  $x'=0,x\propto x_0$,
$\Upsilon_x\propto\Upsilon_{x0}$, &$={\small Rank(x_0)}$
 \\
                 \hline
  \end{tabular}
  }
  \centering
\caption{Relation of  rank$(x)$, the rank of any element $x$ in the \FT-plane
$\FT_0$ generated by $x_0$.}
\label{tt5}
\end{table}

\subsection{The \FT-  or $\FTS$-plane as a (quadratic) sub-FTS system: Euclidean and hyperbolic planes}
\label{ss3a}

Let us fix a maximal rank element $x_0$ ($\Delta(x_0)\not=0$) and its
\FT-plane $\FT_{x_0}\equiv \FT_0$.
As a consequence of  \autoref{ee003}, the restriction of the quartic  map $\Delta$ to $\FT_0$
can be written in terms of a symmetric bilinear form $I_2$ as follows
\beq
\Delta_{\FT_0}(x)&=&\frac{\epsilon_0}{4}I_2^2(x), \quad (\forall x\in \FT_0),
\eeq
where $\epsilon_0=\sgn  \Delta(x_0)$ (we call $\epsilon_0$
 the "signature of $\FT_0$", it does not depend on the plane
maximal rank element  chosen as generator).

Choosing a basis on the plane and coordinates with respect it
  ($x=x^I e_I, (I=1,2)$, $e_1\equiv x_0, e_2\equiv x_0'$), the bilinear form is given by
\beq
I_2(x,y)&\equiv& x^I g_{IJ} y^J=
2\sqrt{\mid\Delta_0\mid} (x^1 y^1+x^2 y^2 \Delta_0).
\eeq
Then the quadratic form
$$I_2(x)=2\sqrt{\mid\Delta_0\mid} ((x^1)^2+(x^2)^2 \Delta_0).$$
The full quadrilinear map restricted to the plane is then of the
form (for generic vectors $x,y,z,w\in \FT_0$)
{\small
\beq
\Delta_{T_0}(x,y,z,w) &=& \frac{\epsilon_0}{12} \left (
I_2(x,y)I_2(z,w)
+I_2(x,z)I_2(y,w)+
I_2(x,w)I_2(y,z)
\right).
\eeq
}
To arrive to this expression  we have used the multi-linearity of $\Delta$ and the properties
$\Delta(x,x,x,x')=\Delta(x,x',x',x')=0$ (see properties in \autoref{sec:properties}).
In the case of $\Delta_0=0$ the expressions reduce trivially to
$I_2\equiv 0$.

We can quickly convince us that the $\FT_0$ plane is  itself a two dimensional FTS
of quadratic (or degenerate)  type (see \cite{torrentemelgarejoquad})
with a characteristic "signature ",$\epsilon_0$, and whose
symmetric quadrilinear and antisymmetric bilinear forms are those inherited from the original FTS.

Restricted to the $\FT$-plane the quadrilinear and bilinear forms
 acquire simple expressions.
The antisymmetric bracket is
\beq
\{ x,y\}_{\FT_0} &\equiv&  x^I \omega_{IJ} y^J= \omega_{12} \left( x^1 y^2 -x^2 y^1\right ).
\eeq
with
$$\omega_{12}=\{x_0,x_0'\}=-2\Delta_0.$$
We define the matrix $S=(S_{IJ})$ (see \cite{torrentemelgarejoquad}) , such that
$$S^{I}_{J}=(g^{-1}){}^{IK} \omega_{KJ},$$
then
$$S^2=-\epsilon_0 I.$$
The explicit form of the matrix $S$ is
{\small
\beq
(S^{I}_{J})&=& \epsilon_0 \begin{pmatrix}
0 & -\sqrt{\mid \Delta_0\mid} \\
\frac{1}{ \sqrt{\mid \Delta_0\mid}}& 0  \\
\end{pmatrix},
\label{ee95}
\eeq
}


The trilinear $T$ map, the  F-dual  maps restricted to $\FT_0$
are given by, in component form  \cite{torrentemelgarejoquad}
\beq
T(x,y,z) &=& \frac{1}{6}\left( I_2(x,y) \hS( z)+I_2(x,z) \hS(y) +I_2(y,z) \hS(x)\right),\\
x'&=& \frac{1}{2} I_2(x) \hS (x), \label{ee00319}\\
\tilde{x} &=&  \eta(x) \hS( x).
\eeq
where it has been used the linear map  defined by $\hS(x)\equiv S^{I}{}_J x^J e_I$ and where
$$\eta(x)\equiv \sgn (I_2(x))=\sgn \left((x^1)^2+\epsilon_0 (x^2)^2 \mid\Delta_0\mid\right).$$
If $\epsilon_0=1$ then $\eta(x)=1,\ \forall x\in\FT_0$.

In summary any non-degenerate $T$ (or $F$) plane can be considered
as a two-dimensional quadratic FTS system, the three FTS axioms, see  \autoref{sec:fts},
are trivially satisfied by the $\FT_0$-restricted maps
 $\Delta_T,\{\}_T, T_T$ ($\Delta_F,\{\}_F, T_F$).

\subsubsection*{Euclidean, hyperbolic \FT-planes}

 The signature of the $I_2$ bilinear form coincides with the signature of the $\FT_0$
plane, the  sign of $\Delta(x_0)$ with $x_0$ any maximal rank element in the plane.  Thus $I_2$  defines an Euclidean or a
Minkowskian $R^{1,1}$ (hyperbolic or split-complex) structure on the  \FT-plane according to it.

 Let us focus on the second case. Endowed by the metric $I_2$,
$\FT_0$ becomes a Minkowski plane.
 The set of all transformations of the hyperbolic plane which preserve the
$I_2$ form is  the  group $O(1, 1)$. This group consists of the hyperbolic
subgroup $SO^+(1, 1)$,
combined with four discrete reflections given by
$$x \to \pm x, x'\to \pm x' .$$
%
%

Using standard notation,
we say that a non-zero vector $x\in \FT_0$ is spacelike if  $I_2(x)>0$,
lightlike, or null,  if  $I_2(x)=0$ and timelike if $I_2(x)<0$.

The T operation,  $x\to x'$,  change the character of the vector.
The vector $x'$ is  " $I_2$-orthogonal" (or "simultaneous events")
to $x$. One can check that \cite{torrentemelgarejoquad},
\begin{eqnarray}
 I_2(x,x')&=&\frac{1}{2} I_2(x) I_2(x,\hS x)\\
             &=&\frac{1}{2} I_2(x) \{x,x\} = 0.
\label{eq31b}
\end{eqnarray}

The relation $x'=\pm x$ is true if and only if $x$ is lightlike;
 $x'$ is timelike (resp. spacelike) if $x$ is spacelike (resp. timelike):
\begin{eqnarray}
x: lightlike & \longleftrightarrow & x': lightlike,\quad x'=\pm x,\\
x: timelike (spacelike)& \longleftrightarrow & x': spacelike (timelike).
\end{eqnarray}
%
%
%
\subsection{Coordinates in the \FT-plane. Orbits}

On the "Euclidean" $\FTS$-plane, by  \autoref{eq:rot1}, the orbits
of the exponential of the $\Upsilon$ map
for $\epsilon =1$ are
closed circles.
Let $e_{i}=(e,\tilde{e})$ be the orthonormal basis given by $e\equiv x/|\Delta
(x)|^{1/4}$. Then
\begin{equation}
\Delta (e)=\epsilon, \qquad \{e_{i},e_{j}\}=-2\varepsilon _{ij},
\end{equation}
For a generic $u\in\fx$ expressed in this basis,
$u=u^{1}e+u^{2}\tilde{e}$, we arrive to, applying \autoref{eq:Frot1},
\begin{equation}
\Delta (x)>0:e^{\overline{\Upsilon }_{x}}\left( u\right) =(\cos (\theta
)u^{1}-\sin (\theta )u^{2})e+(\sin (\theta )u^{1}+\cos (\theta )u^{2})\tilde{%
e}.
\end{equation}%

Whereas, on a "Minkowskian" plane, with $\epsilon =-1$, the same
orbits  are hyperbolic, they are given by
\begin{equation}
\Delta (x)<0:e^{\overline{\Upsilon }_{x}}\left( u\right) =(\cosh (\theta
)u^{1}-\sinh (\theta )u^{2})e+(-\sinh (\theta )u^{1}+\cosh (\theta )u^{2})%
\tilde{e}.
\end{equation}%
The case $\epsilon =-1$ with $u^{1}=\pm u^{2}$ corresponds to null vectors with orbits made by $\pi /4$ rays:
\begin{equation}
\Delta (x)<0,u^{1}=\pm u^{2}:e^{\overline{\Upsilon }_{x}}\left( u\right)
=\pm e^{\mp \theta}u^{2}(e\pm \tilde{e}).
\end{equation}%

\subsection{``Light-cone'' coordinates in the \FT-plane}

It is of interest to define a basis formed by the vectors
\beq
e_\pm&=& x\pm \tilde{x}.
\label{ee00101}
\eeq
For any $y\in \ff$
we define  coordinates  $\alpha_\pm$ such that
$$y=\alpha_+ e_++\alpha_- e_-.$$
In terms of these coordinates the quadrilinear map is given by
\beq
\Delta(y)=((\alpha_++\alpha_-)^2+\epsilon (\alpha_+-\alpha_-)^2)^2\Delta(x).
\label{ee00102}
\eeq
For $\Delta(x)<0$  ($\epsilon=-1$ in the previous expression)
 the basis thus defined is a null basis ($\Delta(e_\pm)=0$), $\alpha_\pm$ are "null" or "light-cone" coordinates, and then
\beq
\Delta(y)&=&(4\alpha_+\alpha_-)^2\Delta(x),  \quad (\epsilon=-1),\\
\Delta(y)&=&4(\alpha_++\alpha_-)^2\Delta(x),  \quad (\epsilon=+1).
\label{ee00103}
\eeq
Similar coordinates will be defined on a $\FT$-plane.

\subsection{The general exponential map}


Let us consider now the action on the orthogonal complement of a given element $x\in\FTS$.
For any $y\in \FT_x^\perp$ ( that means
$\{x,y\}=0,\{x',y\}=0$, see    \autoref{ee00351}), we have also seen that
$\Upsilon_x(y)\in \FT_x^\perp$.
In conclusion, for any $y\in \FT_x^\perp$ the action of successive applications of $\Upsilon_x$
is restricted to lie on \fxy, $\Upsilon_x^{n}(y)$. The orbit of any $y\in \FT_x^\perp$ under
$\sigma_x(\theta)$ lies completely on \fxy.
This can be seen from   \autoref{ee900} and \autoref{ee901}.
These equations allow to explicitly compute the exponential map
 by summing an exponential series, for any rank-4 $x$,
\beq
(\exp \theta \Upsilon_x)(y)&=&
\sum_{k=0} \frac{(\sqrt{-\Delta_x}\theta)^{2k}}{(2k)!} y+
\frac{1}{\sqrt{-\Delta_x}}\sum_{k=0} \frac{(\sqrt{-\Delta_x}\theta)^{2k+1}}{(2k+1)!} \Upsilon_x(y).
\eeq
A similar series is obtained for $(\exp \theta \Upsilon_x) (\Upsilon_x(y))$.

Summing the series, the exponential of the ``normalized'' map in the orthogonal plane is fully determined
 by the expressions
($\overline{\overline{\Upsilon}}_x\equiv \Upsilon_x/ ( \sqrt{\mid \Delta(x)\mid}$,$\epsilon=\sgn \Delta(x)$)
\beq
\left(\exp \theta \overline{\overline{\Upsilon}}_x\right)\left(y\right)
&=& \cos \left(\sqrt{\epsilon} \theta\right) y+
\epsilon\sqrt{\epsilon}\sin\left(\sqrt{\epsilon} \theta\right) \ooUpsilon_x\left(y\right),\label{ee125}\\
\left(\exp \theta {\ooUpsilon}_x\right)\left({\ooUpsilon}_x \left(y\right)\right)
&=& -{\sqrt{\epsilon}} \sin \left(\sqrt{\epsilon} \theta\right) y+
\cos\left(\sqrt{\epsilon} \theta\right) {\ooUpsilon}_x\left(y\right).
\label{ee126}
\eeq


The geometrical character of the orbits of the exponential of the $\Upsilon_x$ map in the \fxt\  plane
 solely depends on  $\epsilon$, the sign of $\Delta(x)$, and not, for example, on the signature of $y$. They are closed (circles or ellipses) or hyperbolic,
 respectively for $\epsilon=1 $ or $-1$.

 It can be explicitly checked that
 \beq
\{ e^{ \theta \overline{\Upsilon}_x}(y),  e^{ \theta \overline{\Upsilon}_x} (\overline{\Upsilon}_x( y)) \}
&=& \{ y,   \overline{\Upsilon}_x ( y) \} .
\eeq
We have also, according to \autoref{ee00527f}
\beq
\Delta(\tilde\Upsilon_x^n(y)) &=& \Delta(y).
\eeq


Let us compute now, for a fixed element $x$, the exponential map $\exp \theta \overline{\Upsilon}_x$ on a generic FTS element $z$, not necessarily on the
orthogonal complement $\FT_x^\perp$.
For that purpose, first we decompose the element on its $\fx$ parallel and orthogonal components
$$ z=z_\parallel+z_\perp.$$
Without loss of generality  we can assume that
$z_\parallel=x$ (if it is not so, we simply realign
the $\FTS$-plane by choosing $z_\parallel$ as the defining element of the plane:
$\fx\equiv \FTS_{z\parallel}$). Then $z= x+z_\perp$.
The action of any power of $\Upsilon_x$
on $z$ is, by linearity,
\beq
\Upsilon_x^n(x+z_\perp)
            &=& \Upsilon_x^n(x)+\Upsilon_x^n(z_\perp)
\eeq
with $\Upsilon_x^n(x)\in\FTS_x$ and $\Upsilon_x^n(z_\perp)\in \fxt$.

As a consequence, the exponential of the $\Upsilon_x$ (or $\bar\Upsilon_x$)
is of the form ( for $z= x+z_\perp$) :
\beq
(\exp \theta \overline{\Upsilon}_x)(z)
&=&
(\exp \theta \overline{\Upsilon}_x)(x)+
(\exp \theta \overline{\Upsilon}_x)(z_\perp)
\eeq
where any of summands is computed independently
using the corresponding relations ( \autoref{eq:Frot1} and  \autoref{ee115} for $\fx$ , or,  \autoref{ee125} and \autoref{ee126} for $\fxp$ ).
Putting together these relations, one arrives to
\begin{eqnarray}
\label{eq:Froty22}
e^{\theta\overline{\Upsilon }_{x}}(z)
&=&\cos \left(\sqrt{\epsilon}\frac{\theta}{3}\right)z_\perp+3\epsilon\sqrt{\epsilon}\sin \left(\sqrt{\epsilon}\frac{\theta}{3}\right)
{\bar\Upsilon}_{x}\left( z_\perp\right) \notag\\
&&+\cos \left(\sqrt{\epsilon}\theta\right)x+\epsilon\sqrt{\epsilon}\sin \left(\sqrt{\epsilon}\theta\right) \txx .
\end{eqnarray}

In particular, for any $z_\perp$ in the orthogonal space, the vector
\begin{eqnarray}
w &=&z_\perp\pm 3\epsilon\sqrt{-\epsilon} {\bar\Upsilon}_x (z_\perp)
\end{eqnarray}
 of the exponential map (strictly, we have to deal with  a complexified FTS for $\epsilon=-1$, see \autoref{sec:b1}):
\begin{eqnarray}
(\exp \theta {\bar\Upsilon}_x) w
&=&e^{ \mp \sqrt{-\epsilon}\frac{\theta}{3}} w.
\end{eqnarray}


The map $\exp(\theta \overline{\Upsilon}_x)$ over  a generic element $y$ in the  FTS can be obtained
from the previous formula performing a suitable rotation in the $\FTS_x$ plane bringing $x$ to a generic $y_\parallel$, and thus $\tilde{x}$ to $\widetilde{y_{\parallel _{x}}}$ :\begin{eqnarray}
\exp \left( \theta \bar{\Upsilon}_{x}\right) (y) &=&\cos \left( \sqrt{%
\varepsilon }\frac{\theta }{3}\right) y+3\varepsilon \sqrt{\varepsilon }\sin
\left( \sqrt{\varepsilon }\frac{\theta }{3}\right) \bar{\Upsilon}_{x}(y)
\nonumber \\
&&+\left[ \cos \left( \sqrt{\varepsilon }\theta \right) -\cos \left( \sqrt{%
\varepsilon }\frac{\theta }{3}\right) \right] y_{\parallel _{x}}  \nonumber
\\
&&+\sqrt{\varepsilon }\left[ \sin \left( \sqrt{\varepsilon }\theta \right)
-3\eta \sin \left( \sqrt{\varepsilon }\frac{\theta }{3}\right) \right]
\widetilde{y_{\parallel _{x}}}
\end{eqnarray}%
where $\eta =sgn\left( \left\{ \tilde{x},y\right\} ^{2}+\varepsilon \left\{
x,y\right\} ^{2}\right) $.
Explicitly, for $\epsilon=1$, one obtains
\begin{eqnarray}
e^{\theta\overline{\Upsilon }_{x}}(y) &=&\cos \left(\frac{\theta}{3}\right) y+
3\sin \left(\frac{\theta }{3}\right) {\bar\Upsilon} _{x}\left( y\right)  \notag \\
&&+\left[ \cos \left(\theta \right) -\cos \left(\frac{\theta }{3}\right) \right] y_{\parallel x}  +\left[ \sin \left(\theta \right) -3\sin\left( \frac{\theta }{3}\right) \right]\widetilde{y_{\parallel x}}, \label{eq:Froty1}
\end{eqnarray}%
and similarly for $\epsilon=-1$.
Here we have introduced the projections of the generic vector $y$ and its F-dual.
\footnote{\emph Note that, comparing with \protect\autoref{ee0046}-\protect\autoref{ee0048}, {\small
$P_{z}\left( y\right) \equiv y_{\parallel _{z}}=\frac{1}{2\sqrt{\left\vert
\Delta (z)\right\vert }}\left[ \left\{ \tilde{z},y\right\} z-\left\{
z,y\right\} \tilde{z}\right] $, $Q_{z}\left( y\right) \equiv \widetilde{%
y_{\parallel _{z}}}=\frac{\eta }{2\sqrt{\left\vert \Delta (z)\right\vert }}%
\left[ \left\{ \tilde{z},y\right\} \tilde{z}+\varepsilon \left\{ z,y\right\}
z\right] $}.}.
%
By reordering the terms in \autoref{eq:Froty1}, one can also write
\begin{eqnarray}
\label{eq:Froty3}
e^{\theta\overline{\Upsilon }_{x}}(y) &=&\cos \left(\frac{\theta}{3}\right) y_\perp+
3\sin \left(\frac{\theta }{3}\right)\left ( {\bar\Upsilon} _{x}\left( y\right) -\widetilde{y_{\parallel x}}\right) \notag \\
&&+ \cos \left(\theta \right)  y_{\parallel x}  +\sin \left(\theta \right) \widetilde{y_{\parallel x}} .
\end{eqnarray}%


We see that the orbit of any generic element of the FTS under a $\sigma_x(\theta)$, for fixed $x$, lies on a
 4-dimensional hyperplane spanned by  (taking into account that $x'\sim\Upsilon_x(x)$)
$$(x, \Upsilon_x(x), z_\perp,\Upsilon_x(z_\perp))$$
or
\beq
&&\exp \theta \Upsilon_x : \FTS_x\oplus \FTS_{(z_\perp,\Upsilon_x(z_\perp))}
 \to \FTS_x\oplus \FTS_{(z_\perp,\Upsilon_x(z_\perp))}
\eeq
or
\beq
\exp \theta \Upsilon_x :
\begin{cases}
\FTS_x\to \FTS_x,\\
 \FTS_{(z_\perp,\Upsilon_x(z_\perp))} \to \FTS_{(z_\perp,\Upsilon_x(z_\perp))}.
 \end{cases}
\eeq


It is easy to show some explicit examples of the application of \autoref{eq:Froty22}. For $y=x+y_\perp$, $w=\exp \theta {\bar\Upsilon}_x (y)$  ($\epsilon=1$)
where it is obvious the $6\pi$ periodicity behaviour  of the exponential map for this signature:
\begin{eqnarray}
\theta=0 &&,  w= y \\
\theta=3\pi/2 &&, w= -\txx+3 {\bar\Upsilon}_x (y_\perp)\\
\theta=3\pi &&, w= -y\\
\theta=9\pi/2 &&, w= \txx-3 {\bar\Upsilon}_x (y_\perp)\\
\theta=6\pi &&, w= y.
\end{eqnarray}
One can explicitly check that $\Delta(w)=\Delta(y)$ in all the cases.

\section{FTS Darboux canonical form: A foliation on \fxy\ planes}
\label{sec6}

In the previous sections we have seen how it is natural to define
structures on the FTS space as $\FTS$-planes and $\FTS_x$, their orthogonal complement \fxt. Within any \fxt\ it results also natural to define planes
\fxy\ closed under the action of the $\Upsilon_x$ map.
This decomposition of the \fxt\   space can be performed in a systematic
way providing a natural canonical form for any FTS, similar to the Darboux
canonical form of any symplectic space.

The orthogonal space \fxt
 can be further decomposed in 2-dimensional subpaces orthogonal
 with respect to the antisymmetric bilinear form $\{,\}$.

Given a fixed initial element $x_0$ of maximal rank,
let us first define for convenience the shorthand notation
$$\xdot \equiv\Upsilon_{x_0}(x), $$
for the fixed element $x_0$. In particular $\xdot_0=3 x_0'=3 T(x_0)$.
We will  construct on continuation a series of mutually
orthogonal vectors iterating the procedure used before (\autoref{ee0046}) in a sort of modified Gram-Schmidt procedure.
Let us initially assume a number of pairs, formed by some vectors and their transforms,  $(x_0,\xdot_0), (x_1,\xdot_1),...(x_{n-1},\xdot_{n-1})$, which
are already mutually orthogonal, that means (for $i,j=0,n-1$)
\beq
\{ x_i, x_j\} &=&0,\\
 \{ x_i, \xdot_j\} &=&c_{i}\delta_{ij}.
\eeq
Where $c_{i}$ are nonzero constants.
We now extend this set of pairs by iteration. We show that it is possible to find
a pair ($x_n,\xdot_n$) orthogonal to the previous ones.
 Let us take an arbitrary vector $ z$ and decompose in parallel  and orthogonal parts with respect
all these vectors, $z=z_\parallel+z_\perp$.
The parallel  part is easily computed, it is the sum of the parallel parts to each of the individual pairs.  It is  given by
(\autoref{ee0046})
\beq
z_\parallel=\sum_{i=0,n-1} z_{\parallel x_i} &=& \sum_{i=0,n-1} \frac{1}{\{\xdot_i  ,x_i  \}}
\left| \begin{array}{cc} x_i & \xdot_i \\  \{x_i  ,z  \}&\{\xdot_i  ,z  \}  \end{array} \right|.
\eeq
Obviously, the $z_{\parallel }$
defined in this way is on the subspace generated
by $x_i,\xdot_i, (i=0,n-1)$.
The vector $z_\perp=z-z_\parallel$ is  orthogonal by construction to all the subspace,
\beq
\{x_i,z_\perp\}&=&0,\\
\{\xdot_i,z_\perp\}&=&0.
\eeq

It is also straightforward to show that
${(z_\perp)}{}{\dot{}}\equiv\Upsilon_{x_0}( z_\perp)=3 T(x_0,x_0,z_\perp)$ is
also orthogonal to the full set :
\beq
\{\Upsilon_{x_0}(z_\perp), x_i\} &=&\{3 T(x_0,x_0,z_\perp),x_i\}=\{3 T(x_0,x_0,x_i),z_\perp)\}\\
&=&\{\Upsilon_{x_0}(x_i), z_\perp\}=0,\\
\{\Upsilon_{x_0}(z_\perp), \xdot_i\} &=&9\{ T(x_0,x_0,z_\perp),T(x_0,x_0,x_i)\}=-C_i\Delta(x_0) \{x_0,z_\perp\}=0.
\eeq
In the last line $C_0=3,C_i=1 ,\, (i=1,n-1)$.
So $(z_\perp,\dot{(z_\perp)})$ is the pair we were looking for, we redefine
\beq
x_n &\equiv& z_\perp,\\
\xdot_n &\equiv & 3 T(x_0,x_0,z_\perp)=\dot{(z_\perp)}.
\eeq
The process is iterated as long as we exhaust the dimensionality of the vector space ($n=N$)
or  we cannot find vectors with non trivial pairs, $(x_i,\xdot_i)\not=0$.

In this way we reduce the symplectic form to a canonical Darboux form.
In the basis formed by the vectors $(x_0,\xdot_0, x_1,\xdot_1,...)$ the
the symplectic form is expressed by the matrix
{\small
\beq
\Omega =
\begin{bmatrix}
\begin{matrix}0 & \lambda_1\\ -\lambda_1 & 0\end{matrix} & & 0 \\ & \ddots & \\ 0 & &
\begin{matrix}0 & \lambda_N \\ -\lambda_N & 0
\end{matrix}
\end{bmatrix},
\eeq
}
where
$$\lambda_i=\{x_i,\xdot_i\}.$$

\section{The Reduced $\mathfrak{F}(J_{3})$ case: explicit  expressions}
\label{JJexplicit3}
\label{secad}

We present here some explicit formulas used in \autoref{sec8}.
%
By exploiting the results in App. D of \cite{Special-Road}, one can compute
the components of $T(x,y,z)_{M}$ (\autoref{T}) in the $4D/5D$ special
coordinates' symplectic frame, characterizing every \textit{reduced} FTS.
Using obvious notation (see \autoref{sec8}), these components read:
\begin{eqnarray}
T(x,y,z)_{0} &=&-\frac{1}{3}d^{ijk}x_{i}y_{j}z_{k}-\frac{1}{3}\left(x^{0}y_{0}z_{0}+x_{0}y^{0}z_{0}+x_{0}y_{0}z^{0}\right)  \notag \\
&&
-\frac{1}{6}\left[ \left( x^{i}y_{i}+x_{i}y^{i}\right)z_{0}+(x^{i}z_{i}+x_{i}z^{i})y_{0}+(y^{i}z_{i}+y_{i}z^{i})x_{0}\right] ;\label{0-up} \\
T(x,y,z)_{i} &=&
\frac{1}{3}d_{ijk}\left(x_{0}y^{j}z^{k}+x^{j}y_{0}z^{k}+x^{j}y^{k}z_{0}\right)  +\frac{1}{3}d_{ijm}d^{mkl}(x^{j}y_{k}z_{l}+x_{k}y^{j}z_{l}+x_{k}y_{l}z^{j})
\notag \\
&&-\frac{1}{6}\left[ \left( x^{j}y_{j}+x_{j}y^{j}\right) z_{i}+\left(x^{j}z_{j}+x_{j}z^{j}\right) y_{i}+\left( y^{j}z_{j}+y_{j}z^{j}\right) x_{i}
\right]  \notag \\
&&
-\frac{1}{6}\left[ \left( x^{0}y_{0}+x_{0}y^{0}\right) z_{i}+\left(x^{0}z_{0}+x_{0}z^{0}\right) y_{i}+\left( y^{0}z_{0}+y_{0}z^{0}\right) x_{i}\right] ;
  \label{i-up} \\
T(x,y,z)^{0} &=&
\frac{1}{3}d_{ijk}x^{i}y^{j}z^{k}-\frac{1}{3}\left(x^{0}y^{0}z_{0}+x^{0}y_{0}z^{0}+x_{0}y^{0}z^{0}\right)  \notag \\
&&-\frac{1}{6}\left[ \left( x^{i}y_{i}+x_{i}y^{i}\right)z^{0}+(x^{i}z_{i}+x_{i}z^{i})y^{0}+(y^{i}z_{i}+y_{i}z^{i})x^{0}\right] ;\label{0-down} \\
T(x,y,z)^{i} &=&-\frac{1}{3}d^{ijk}\left(x^{0}y_{j}z_{k}+x_{j}y^{0}z_{k}+x_{j}y_{k}z^{0}\right)  +
\frac{1}{3}d_{klm}d^{mij}(x_{j}y^{k}z^{l}+x^{k}y_{j}z^{l}+x^{k}y^{l}z_{j})
\notag \\
&&-\frac{1}{6}\left[ \left( x^{j}y_{j}+x_{j}y^{j}\right) z^{i}+
\left(x^{j}z_{j}+x_{j}z^{j}\right) y^{i}+\left( y^{j}z_{j}+y_{j}z^{j}\right) x^{i}\right]  \notag \\
&&
-\frac{1}{6}\left[ \left( x^{0}y_{0}+x_{0}y^{0}\right) z^{i}+\left(x^{0}z_{0}+x_{0}z^{0}\right) y^{i}+\left( y^{0}z_{0}+y_{0}z^{0}\right) x^{i}\right] .
 \label{i-down}
\end{eqnarray}

Let us recall that in this symplectic frame the $2N\times 2N$ symplectic
metric is given by \autoref{Omega-z}.
In particular, one can compute the various components of the linear map
\footnote{Note that when $y\in \mathfrak{F}_{x}^{\perp }$, then $\left\{ x,y\right\} =0
$, and the second term in (\autoref{Ipsilon}) is missing \cite{Brown:1969}.}\
$\Upsilon _{x}\left( y\right)$ defined in \autoref{Ipsilon}
to read in components%
\begin{eqnarray}
\Upsilon _{x}\left( y\right) ^{0}&=&-3T(x,x,y)^{0}+\left(-x^{0}y_{0}-x^{j}y_{j}+x_{0}y^{0}+x_{j}y^{j}\right) x^{0}; \\
\Upsilon _{x}\left( y\right) ^{i}&=&-3T(x,x,y)^{i}+\left(-x^{0}y_{0}-x^{j}y_{j}+x_{0}y^{0}+x_{j}y^{j}\right) x^{i}; \\
\Upsilon _{x}\left( y\right) _{0}&=&3T(x,x,y)_{0}+\left(-x^{0}y_{0}-x^{j}y_{j}+x_{0}y^{0}+x_{j}y^{j}\right) x_{0}; \\
\Upsilon _{x}\left( y\right) _{i}&=&3T(x,x,y)_{i}+\left(-x^{0}y_{0}-x^{j}y_{j}+x_{0}y^{0}+x_{j}y^{j}\right) x_{i},%
\end{eqnarray}%
where
\begin{eqnarray}
T(x,x,y)^{0} &=&
\frac{1}{3}d_{ijk}x^{i}x^{j}y^{k}-\frac{1}{3}\left[ 2\left(x^{0}\right) ^{2}y_{0}+2x^{0}x_{0}y^{0}\right] -\frac{1}{3}\left[x^{i}x_{i}y^{0}+(x^{i}y_{i}+x_{i}y^{i})x^{0}\right] ;
 \label{j-3} \notag\\
&&  \notag \\
T(x,x,y)^{i} &=&
-\frac{1}{3}d^{ijk}\left(2x^{0}x_{j}y_{k}+x_{j}x_{k}y^{0}\right) +\frac{1}{3}d_{klm}d^{mij}(2x_{j}x^{k}y^{l}+x^{k}x^{l}y_{j})  \notag \\
&&-\frac{1}{3}\left[ x^{j}x_{j}y^{i}+\left( x^{j}y_{j}+x_{j}y^{j}\right)x^{i}\right] -\frac{1}{3}\left[ x^{0}x_{0}y^{i}+\left(x^{0}y_{0}+x_{0}y^{0}\right) x^{i}\right] ;
\label{j-4} \notag\\
&&  \notag \\
T(x,x,y)_{0} &=&
-\frac{1}{3}d^{ijk}x_{i}x_{j}y_{k}-\frac{1}{3}\left[2x^{0}x_{0}y_{0}+\left( x_{0}\right) ^{2}y^{0}\right] -\frac{1}{3}\left[x^{i}x_{i}y_{0}+(x^{i}y_{i}+x_{i}y^{i})x_{0}\right] ;
\label{j-1}\notag \\
&&  \notag \\
T(x,x,y)_{i} &=&
\frac{1}{3}d_{ijk}\left(2x_{0}x^{j}y^{k}+x^{j}x^{k}y_{0}\right) +\frac{1}{3}d_{ijm}d^{mkl}(2x^{j}x_{k}y_{l}+x_{k}x_{l}y^{j})  \notag \\
&&
-\frac{1}{3}\left[ x^{j}x_{j}y_{i}+\left( x^{j}y_{j}+x_{j}y^{j}\right)x_{i}\right] -\frac{1}{3}\left[ x^{0}x_{0}y_{i}+\left(x^{0}y_{0}+x_{0}y^{0}\right) x_{i}\right] .  \label{j-2}\notag
\end{eqnarray}

The following quantity, $x$- and $y$- dependent function, plays an important role
\begin{eqnarray} 
\quad \{y,\Upsilon_x(y)\} &=&3\left\{ y,T(x,x,y)\right\} -\left\{
x,y\right\} ^{2} \\
&=&-3
\left[-y^{0}T(x,x,y)_{0}-y^{i}T(x,x,y)_{i}+y_{0}T(x,x,y)^{0}+y_{i}T(x,x,y)^{i}\right]
 \notag \\
&&-\left( -x^{0}y_{0}-x^{j}y_{j}+x_{0}y^{0}+x_{j}y^{j}\right) ^{2}.
\end{eqnarray}%
where $T(x,x,y)_{0}$, $T(x,x,y)_{i}$, $T(x,x,y)^{0}$ and $T(x,x,y)^{i}$ are
given above.

\subsubsection*{Further expressions}

 Within  the $4D/5D$ special coordinates' symplectic
frame of reduced FTS's and within the choices (\autoref{1-z}) resp. (\autoref{y}) of
the rank-$4$ element $x\in \mathfrak{F}$ (with $\Delta (x)<0$) and of the
rank-$4$ element
$y\in \mathfrak{F}_{x}^{\perp }=\mathfrak{F}/\mathfrak{F}%
_{x}$ (with $\Delta (y)\gtrless 0$),
we present here some further formulas useful for \autoref{sec8}.
In order to determine the condition of closure of $\mathfrak{F}_{y\perp x}$
under $T$, we have to explicitly compute $T(r)\equiv T(r,r,r)$ for a generic
element $r=ay+b\Upsilon _{x}(y)\in \mathfrak{F}_{y\perp x}$, which is given
by%
\begin{eqnarray}
T(r) &\equiv &T(r,r,r)=T(ay+b\Upsilon _{x}(y),ay+b\Upsilon _{x}(y),ay+b\Upsilon
_{x}(y))  \notag \\
&=&a^{3}T\left( y,y,y\right) +3a^{2}bT\left( y,y,\Upsilon _{x}(y)\right)
+3a^{2}bT\left( \Upsilon _{x}(y),\Upsilon _{x}(y),y\right )\notag\\
&&+b^{3}T\left(
\Upsilon _{x}(y),\Upsilon _{x}(y),\Upsilon _{x}(y)\right) .  \notag
\end{eqnarray}%
Let us then start and compute the components of $T(r)_{M}$ .
In first place, the $T(y,y,y)_{M}$ quantities are given by
\begin{eqnarray}
T(y,y,y)_{0}&=&-\frac{1}{3}d^{ijk}y_{i}y_{j}y_{k}; \\
T(y,y,y)_{i}&=&\left(d_{mij}d^{mkl}y^{j}y_{k}y_{l}-y^{j}y_{j}y_{i}\right) ; \\
T(y,y,y)^{0}&=&\frac{1}{3}d_{ijk}y^{i}y^{j}y^{k}; \\
T(y,y,y)^{i}&=&\left(d_{klm}d^{mij}y_{j}y^{k}y^{l}-y^{j}y_{j}y^{i}\right) ;
\end{eqnarray}
similarly
\begin{eqnarray}
T(y,y,\Upsilon _{x}(y))_{0}&=&\frac{1}{3}x^{0}x_{0}d^{ijk}y_{i}y_{j}y_{k}; \\
T(y,y,\Upsilon _{x}(y))_{i}&=&-\frac{1}{3}x^{0}x_{0}\left(d_{mij}d^{mkl}y^{j}y_{k}y_{l}-y^{j}y_{j}y_{i}\right) ; \\
T(y,y,\Upsilon _{x}(y))^{0}&=&\frac{1}{3}x^{0}x_{0}d_{ijk}y^{i}y^{j}y^{k}; \\
T(y,y,\Upsilon _{x}(y))^{i}&=&\frac{1}{3}x^{0}x_{0}\left(d_{klm}d^{mij}y_{j}y^{k}y^{l}-y^{j}y_{j}y^{i}\right) ;%
\end{eqnarray}
and
\begin{eqnarray}
T\left( \Upsilon _{x}(y),\Upsilon _{x}(y),y\right) _{0}&=&-\frac{1}{3}\left(x^{0}x_{0}\right) ^{2}d^{ijk}y_{i}y_{j}y_{k}; \\
T\left( \Upsilon _{x}(y),\Upsilon _{x}(y),y\right) _{i}&=&-\frac{1}{3}\left(x^{0}x_{0}\right) ^{2}\left(d_{mij}d^{mkl}y^{j}y_{k}y_{l}-y^{j}y_{j}y_{i}\right) ; \\
T\left( \Upsilon _{x}(y),\Upsilon _{x}(y),y\right) ^{0}&=&\frac{1}{3}\left(x^{0}x_{0}\right) ^{2}d_{ijk}y^{i}y^{j}y^{k}; \\
T\left( \Upsilon _{x}(y),\Upsilon _{x}(y),y\right) ^{i}&=&-\frac{1}{3}\left(x^{0}x_{0}\right) ^{2}\left(d_{klm}d^{mij}y_{j}y^{k}y^{l}-y^{j}y_{j}y^{i}\right) ;%
\end{eqnarray}
finally
\begin{eqnarray}
T\left( \Upsilon _{x}(y),\Upsilon _{x}(y),\Upsilon _{x}(y)\right) _{0}&=&\frac{1}{3}\left( x^{0}x_{0}\right) ^{3}d^{ijk}y_{i}y_{j}y_{k}; \\
T\left( \Upsilon _{x}(y),\Upsilon _{x}(y),\Upsilon _{x}(y)\right) _{i}&=&\left( x^{0}x_{0}\right) ^{3}\left(d_{mij}d^{mkl}y^{j}y_{k}y_{l}-y^{j}y_{j}y_{i}\right) ; \\
T\left( \Upsilon _{x}(y),\Upsilon _{x}(y),\Upsilon _{x}(y)\right) ^{0}&=&\frac{1}{3}\left( x^{0}x_{0}\right) ^{3}d_{ijk}y^{i}y^{j}y^{k}; \\
T\left( \Upsilon _{x}(y),\Upsilon _{x}(y),\Upsilon _{x}(y)\right) ^{i}&=&-\left( x^{0}x_{0}\right) ^{3}\left(d_{klm}d^{mij}y_{j}y^{k}y^{l}-y^{j}y_{j}y^{i}\right) .%
\end{eqnarray}

Therefore, putting together all the computations, the various components of $T(r)_{M}$ reads%
\begin{eqnarray}
T(r)_{0}&=&
\left( a-x^{0}x_{0}b\right)^{3}
T(y,y,y)_{0};
\notag\\
T(r)_{i}
&=&
\left( a-x^{0}x_{0}b\right)^{2}\left( a+x^{0}x_{0}b\right)
T(y,y,y)_{i};
\notag\\
T(r)^{0}&=&
\left( a+x^{0}x_{0}b\right)^{3}
T(y,y,y)^{0}; \notag\\
T(r)^{i}
&=&
\left( a-x^{0}x_{0}b\right)\left( a+x^{0}x_{0}b\right)^{2}
 T(y,y,y)^{i}.
\end{eqnarray}%
Then
\begin{equation}
T(r)^{M}=\Omega ^{MN}T(r)_{N}=\left(
-T(r)^{0},
-T(r)^{i},
T(r)_{0},
T(r)_{i}%
\right)^T ,
\end{equation}%
where $T(r)^{0}$, $T(r)^{i}$, $T(r)_{0}$ and $T(r)_{i}$ are given by (\ref%
{M-1}). Final expressions are given in the text, see \autoref{sec8}.


\end{document}